\documentclass[aps,twocolumn,superscriptaddress]{revtex4-1}  

\usepackage{natbib}
\usepackage{graphicx}
\usepackage{amsmath}
\usepackage[inline]{enumitem}
\usepackage{textcomp} 
\usepackage{gensymb} 
\usepackage[mathscr]{euscript}

\usepackage{setspace} 
\usepackage{xcolor}

\usepackage{xr} 

\def\p{_\mathit{p}}

\def\rout{\mathbf{r_{out}}}
\def\rin{\mathbf{r_{in}}}
\def\xout{x_\mathrm{out}}
\def\xin{x_\mathrm{in}}
\def\uout{\mathbf{u_{out}}}

\def\tin{\theta_\mathrm{in}}
\def\kout{k_\mathrm{out}}
\def\kin{k_\mathrm{in}}
\def\Rut{\mathbf{R}_{\mathbf{u}\theta}}
\def\Rxx{\mathbf{R}_{xx}}

\def\Rrr{\mathbf{R}_{\mathbf{rr}}}

\def\in{_\text{in}}
\def\out{_\text{out}}
\def\xout{x_\mathrm{out}}
\def\xin{x_\mathrm{in}}
\def\rout{\mathbf{r}\out}
\def\rin{\mathbf{r}\in}
\def\uout{\mathbf{u}\out}

\def\tin{\theta\in}
\def\kout{k_\mathrm{out}}
\def\kpout{k^{\prime}_\mathrm{out}}
\def\kin{k_\mathrm{in}}
\def\ut{_{\mathbf{u}\theta}}
\def\Rxx{\mathbf{R}_{\mathbf{x}\mathbf{x}}\left(z_i\right)}
\def\Rut{\mathbf{R}\ut}

\def\Rrr{\mathbf{R}_{\mathbf{r}\mathbf{r}}}

\newcommand*{\laura}[1]{\textcolor{black}{#1}}
\newcommand*{\alex}[1]{\textcolor{black}{#1}}

\begin{document}

\title{Distortion matrix approach for ultrasound imaging of random scattering media}
\author{William Lambert}
\affiliation{Institut Langevin, ESPCI Paris, CNRS UMR 7587, PSL University, 1 rue Jussieu, 75005 Paris, France}
	\affiliation{SuperSonic Imagine,
Les Jardins de la Duranne, 510 Rue Ren\'{e} Descartes, 13857 Aix-en-Provence, France}
\author{Laura A. Cobus}
\affiliation{Institut Langevin, ESPCI Paris, CNRS UMR 7587, PSL University, 1 rue Jussieu, 75005 Paris, France}
\author{Thomas Frappart}
	\affiliation{SuperSonic Imagine,
Les Jardins de la Duranne, 510 Rue Ren\'{e} Descartes, 13857 Aix-en-Provence, France}
\author{Mathias Fink}
\affiliation{Institut Langevin, ESPCI Paris, CNRS UMR 7587, PSL University, 1 rue Jussieu, 75005 Paris, France}
\author{{Alexandre Aubry}}
\email{alexandre.aubry@espci.fr}
\affiliation{Institut Langevin, ESPCI Paris, CNRS UMR 7587, PSL University, 1 rue Jussieu, 75005 Paris, France}

\date{\today}


\begin{abstract}
Focusing waves inside inhomogeneous media is a fundamental problem for imaging. Spatial variations of wave velocity can strongly distort propagating wavefronts and degrade image quality. Adaptive focusing can compensate for such aberration, but is only effective over a restricted field of view. Here, we introduce a full-field approach to wave imaging based on the concept of the distortion matrix. This operator essentially connects any focal point inside the medium with the distortion that a wavefront, emitted from that point, experiences due to heterogeneities. A time-reversal analysis of the distortion matrix enables the estimation of the transmission matrix that links each sensor and image voxel. Phase aberrations can then be unscrambled for any point, providing a full-field image of the medium with diffraction-limited resolution. Importantly, this process is particularly efficient in random scattering media, where traditional approaches such as adaptive focusing fail. Here, we first present an experimental proof-of-concept on \alex{a tissue mimicking phantom\laura{, and then apply the method to} 
\textit{in vivo} imaging of human soft tissues}. \laura{While introduced here in the context of acoustics, t}his approach can also be extended to optical microscopy, radar or seismic imaging. 
\end{abstract}

\maketitle

{L}ight travelling through soft tissues, ultrasonic waves propagating through the human skull, or seismic waves in the Earth's crust are all examples of wave propagation
through inhomogeneous media. Short-scale inhomogeneities of the refractive index, referred to as \textit{scatterers}, cause incoming waves to be reflected. These backscattered echoes are those which enable reflection imaging; this is the principle of, for example, ultrasound imaging in acoustics, optical coherence tomography for light or reflection seismology in geophysics. However, wave propagation between the sensors and a focal point inside the medium is often degraded by: (\textit{i}) wavefront distortions (aberrations) induced by long-scale heterogeneities of the wave velocity, or (\textit{ii}) multiple scattering if scatterers are too bright and/or concentrated. Because both phenomena can strongly degrade the resolution and contrast of the image, they constitute the most fundamental limits for imaging in all domains of wave physics.

\begin{figure}[t]
	\centering
	\includegraphics[width=\columnwidth]{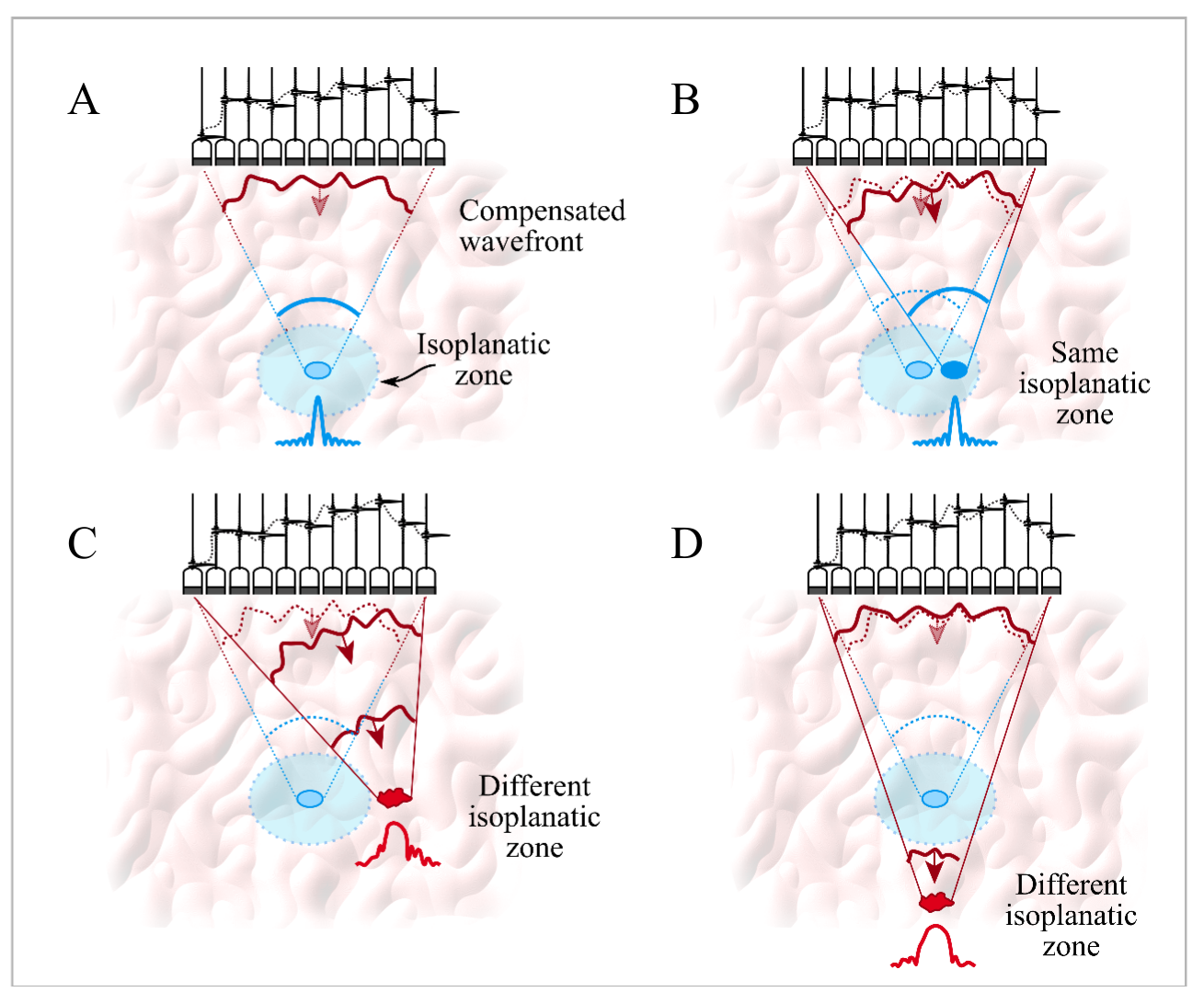}
	\caption{\textbf{Adaptive focusing in ultrasound imaging.} (
	(\textbf{A}) Adaptive focusing consists in adjusting the time delays added to each emitted and/or detected signal in order to focus at a certain position inside the medium. (\textbf{B}) 
	Tilting the same adaptive phase law allows the focal spot to be scanned over the vicinity of the initial focal point. The area over which adaptive focusing remains effective is called an isoplanatic patch. (\textbf{C}-\textbf{D}) Beyond this zone, the correction no longer works.}	
	\label{focusfig}
\end{figure}

Astronomers were the first to deal with aberration issues in wave imaging. 
Their approach to improve image quality was to measure and compensate for the wavefront distortions induced by the spatial variations of the optical index in the atmosphere; this is the concept of adaptive optics, proposed as early as the 1950s~\cite{Roddier1999}. Subsequently, ultrasound imaging~\cite{Angelsen2000} and optical microscopy~\cite{Booth2007} have also drawn on the principles of adaptive optics to compensate for the aberrations induced by uneven interfaces or tissues’ inhomogeneities. In ultrasound imaging, for instance, arrays of transducers are employed to emit and record the amplitude and phase of broadband wavefields. Wavefront distortions can be compensated for by adjusting the time-delays added to each emitted and/or detected signal in order to focus at a certain position inside the medium {(see Fig.~\ref{focusfig}A)}.

Conventional adaptive focusing methods generally require the presence of a dominant scatterer (guide star) from which the signal to be optimized is reflected. 
While it is possible in some cases to generate an artificial guidestar, the subsequent optimization of focus will nevertheless be imperfect for a heterogeneous medium. This is because a wavefront returning from deep within a complex biological sample is composed of a superposition of echoes coming from many unresolved scatterers (resulting in a speckle image), and its interpretation is thus not at all straightforward. A first alternative to adaptive focusing, derived from stellar speckle interferometry~\cite{Labeyrie1970}, is to extract the aberrating phase law from spatial/angular correlations of the reflected wavefield~\cite{ODonnell1988,Mallart1994,Masoy2005,Montaldo2011,Kang2017,Kim2019}. A second alternative is to correct the aberrations not by measuring the wavefront, but by simply optimizing the image quality, i.e. by manipulating
the incident and/or reflected wavefronts in a controlled manner in order to converge towards an optimal image~\cite{Muller1974,Booth5788,Nock1989,Debarre2009,Ji2010,Ji22,Adie7175}. However, both methods generally imply a time-consuming iterative focusing process. More importantly, these alternatives rely on the hypothesis that aberrations do not change over the entire field of view. This assumption of spatial invariance is simply incorrect at large imaging depths for biological media~\cite{Dahl2005,Judkewitz2015}. High-order aberrations 
\laura{induced by small-scale variations in the speed of sound of the medium} are only invariant over small regions \laura{of the image}, often referred to as isoplanatic patches in the literature {(see Fig.~\ref{focusfig}B-D)}. Conventional adaptive focusing methods thus suffer from a very limited field of view at large depths, which severely limits their performance for in-depth imaging. 
\laura{Recently, however, several acoustic imaging groups have demonstrated convincing approaches for heterogeneous media, whether by mapping the speed-of-sound distribution in the medium and using it to reconstruct an image~\cite{Ali2018,Rau2019}, or via estimation of (and compensation for) time delays for a \alex{local} correction of aberrations~\cite{Jaeger2015SosMap,Chau2019,PFM2019}.
Each of these methods leverages the multi-element capabilities of ultrasonic transducers to} \alex{extract spatial coherence or travel time difference between signals recorded by each array element.} In this paper, we \alex{propose a more general solution to optimize the information} 
\laura{offered by transducer arrays -- a universal matrix approach for wave imaging. We develop a rigorous mathematical formalism for our approach, and apply the theoretical results to aberration correction for in vivo imaging of the human body.}  

Historically, the matrix approach for imaging was inspired by the advent of multi-element technology in acoustics, and by the insight that a 
matrix formalism is a natural way to describe ultrasonic wave propagation between arrays of transducers~\cite{Prada1994,Tanter2000,Derode2003}. In the 2010s, the emergence of spatial light modulators allowed the extension of this transmission matrix approach to optics~\cite{Popoff2010}. Experimental access to the transmission matrix then enabled researchers to take advantage of multiple scattering for optimal light focusing~\cite{Popoff2010,Kim2012} and communication across a diffusive layer~\cite{Popoff2010b} or multi-mode fibre~\cite{Cizmar2012}. However, a transmission configuration is not adapted to non-invasive and/or \laura{\textit{in vivo}} imaging of biological media, motivating the development of approaches in an epi-illumination configuration.

The reflection matrix has already been shown to be a powerful tool for focusing in multi-target media~\cite{Prada1994,Borcea2002,Popoff2011}, target detection~\cite{Shajahan2014,Badon2016,Blondel2018} and energy delivery~\cite{Choi2013,Jeong2018} in \laura{scattering} media.  A few studies have also looked at reflection imaging under a matrix formalism~\cite{Varslot2004,Robert2008,Kang2015,Kang2017}; however, as with most conventional adaptive focusing methods, their effectiveness is limited to a single isoplanatic patch (Fig.~\ref{focusfig}). 

\alex{Spatially-distributed aberrations have not been addressed under a matrix approach until very recently \cite{Kim2019,Badon2019,PFM2019}. Inspired by the pioneering work of Robert and Fink~\cite{Robert2008},}  the concept of the \textit{distortion matrix}, $\mathbf{D}$, has been introduced in optical imaging~\cite{Badon2019}. Whereas the reflection matrix $\mathbf{R}$ holds the wavefronts which are reflected from the medium, $\mathbf{D}$ contains the \textit{deviations} from an ideal reflected wavefront which would be obtained in the absence of inhomogeneities. In addition, while $\mathbf{R}$ typically contains responses between inputs and outputs in the same basis, e.g. responses between individual ultrasonic transducer elements~\cite{Prada1996,Aubry2009} or between focal points inside the medium~\cite{Lambert2019}, $\mathbf{D}$ is concerned with the `dual basis' responses between a set of incident plane waves~\cite{Montaldo2009} and a set of focal points inside the medium~\cite{Robert2008}. \alex{In optical imaging, Badon \textit{et al.}~\cite{Badon2019} recently showed that, for a large specular reflector, the matrix $\mathbf{D}$ exhibits long-range correlations in the focal plane. Such spatial correlations can be taken advantage of to decompose the field-of-view (FOV) into a set of isoplanatic modes and their corresponding wavefront distortions in the far-field. The Shannon entropy $\mathcal{H}$ of $\mathbf{D}$ is also shown to yield an effective rank of the imaging problem, i.e the number of isoplanatic patches in the FOV. This decomposition was then used to correct \laura{for} output aberrations when imaging planar specular objects through a scattering medium.}

In this paper, we \laura{develop the distortion matrix approach for acoustic imaging. In view of medical ultrasound applications, this
requires a method that can go beyond imaging specular reflectors in order to tackle the more challenging case of random scattering media. Ultrasonic wave propagation in soft tissues gives rise to a speckle regime in which scattering is often due to a random distribution of unresolved scatterers. Apart from specular reflections at interfaces of tissues and organs, the reflectivity of the medium can be considered to be continuous and random. In this paper, we demonstrate: (\textit{i}) how projecting the reflection matrix into the far-field allows the suppression of 
specular reflections and multiple reverberations (clutter noise), enabling access to a purely random speckle regime, (\textit{ii}) how, in this regime, the far-field correlations of $\mathbf{D}$  enables discrimination between and correction for input and output aberrations over each isoplanatic patch, (\textit{iii}) how a position-dependent distortion matrix enables non-invasive access to the transmission matrix $\mathbf{T}$ between the plane wave basis and the entire set of image voxels, and (\textit{iv}) how a minimization of the entropy $\mathcal{H}$ enables a quantitative measurement of the wave velocity (or refractive index) in the field-of-view.} 

Throughout the paper, our theoretical developments are supported by an ultrasonic experiment \laura{using a tissue-mimicking phantom, and further applied to \textit{in vivo} ultrasound imaging of the human body}. Due to its experimental flexibility, ultrasound imaging is an ideal modality for our proof-of-concept. Nevertheless, the distortion matrix approach is by no means limited to one particular type of wave, but can be extended to any situation in which the amplitude and phase of the medium response can be recorded between multiple inputs and outputs. This study thus opens important perspectives in various domains of wave physics such as acoustics, optics, radar and seismology.

\section*{Results}

\subsection*{Confocal imaging with the reflection matrix}

\begin{figure*}[htbp]
	\centering
	\includegraphics[width=17cm]{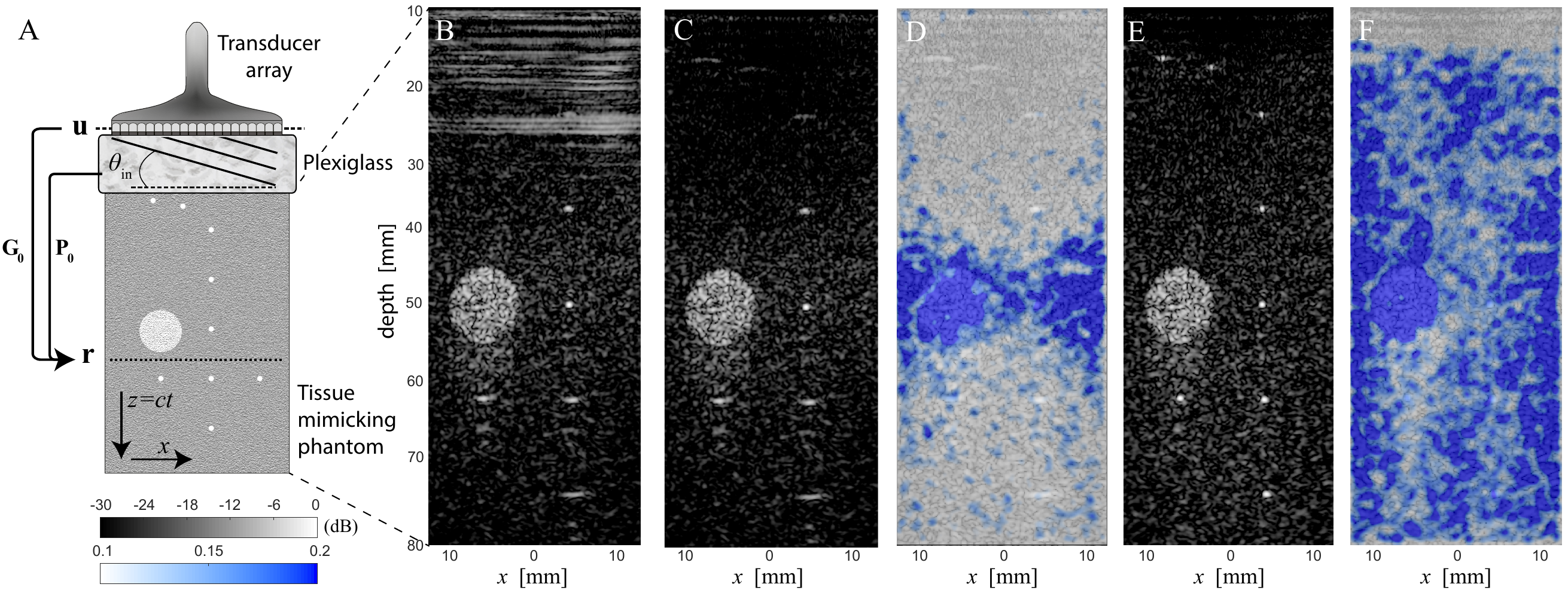}
	\caption{\textbf{Matrix imaging}. (\textbf{A}) Sketch of the experimental acquisition of $\Rut(t)$. An ultrasonic transducer array is placed in contact with a plexiglass layer, which is on top of a human tissue-mimicking phantom. For each plane wave illumination $\tin$, the backscattered wavefront is acquired as a function of array element $\uout$ and time $t$. (\textbf{B}) Original ultrasound confocal image (Eq.\ \ref{imcalc}) with $c=1800$~m/s. (\textbf{C}) The ultrasound image after the removal of multiple reflections is shown, along with (\textbf{D}) the corresponding map of the Strehl ratio $\mathcal{S}$.(\textbf{E}) The ultrasound image after matrix aberration correction is shown, with (\textbf{F}) the corresponding map of the Strehl ratio $\mathcal{S_F}$. The ultrasound images and Strehl ratio maps are displayed with the same dB- (B\&W) and linear (color) scales, respectively.}
	\label{acqfig}
\end{figure*}

The sample under study is a tissue-mimicking phantom with a speed of sound $c\p=1542 \pm 10$ m/s. It is composed of a random distribution of unresolved scatterers which generate an ultrasonic speckle characteristic of human tissue \laura{(gray background in Fig.\ \ref{acqfig}A)}. The phantom also contains \alex{eight  
subwavelength nylon monofilaments of diameter $0.1$~mm placed perpendicularly to the probe (white point-like targets). The bright circular target located at depth $z=50$~mm on the image (Fig.\ \ref{acqfig}B) is a section of a hyperechoic cylinder composed of a higher density of unresolved scatterers.}
A 15 mm-thick layer of plexiglass [$c_a\sim2750$~m/s~\cite{Carlson2003}] is placed on top of the phantom to \laura{create both strong aberrations and multiple reflections} (Fig.~\ref{acqfig}A). 

Our matrix approach begins with the experimental acquisition of the reflection matrix $\mathbf{R}$ using an ultrasonic transducer array placed in direct contact with the plexiglass layer (Fig.~\ref{acqfig}A). The reflection matrix is built by plane wave beamforming in emission, and reception by each individual element~\cite{Montaldo2009}. Acquired in this way, the reflection matrix is denoted $\Rut(t)\equiv R(\uout,\tin,t)$, where $\mathbf{u}$ defines the spatial positions of the transducers and $t$ is the time-of-flight. Details of the experimental acquisition are given in the \textit{Methods}. 
A conventional ultrasound image consists of a map of the local reflectivity of the medium. This information can be obtained from $\Rut$ by applying appropriate time delays to perform focusing in post-processing, both in emission and reception~\cite{Montaldo2009}. This focusing can also be easily performed in the frequency domain, where matrix products allow the mathematical projection of $\mathbf{R}$ between different mathematical bases~\cite{Lambert2019}. The bases implicated in this work are sketched in Fig.\ \ref{acqfig}A. They are (\textit{i}) the recording basis, which here corresponds to the transducer array elements located at $\mathbf{u}$, (\textit{ii}) the illumination basis which is composed of the incident plane waves with angle $\theta$, (\textit{iii}) the spatial Fourier basis, mapped by the transverse wave number ${k_x}$, from which the aberration and multiple reflection issues will be addressed, and (\textit{iv}) the focused basis in which the ultrasound image is built, here composed of points $\mathbf{r}=x\hat{x}+z\hat{z}$ inside the medium. In the following, we use this matrix formalism to  present our techniques for local aberration correction and clutter noise removal. This is the ideal formalism in which to develop our approach which requires that we be able to move flexibly from one basis to the other, in either input or output.

\begin{figure}[htbp]
	\centering
	\includegraphics[width=8.5cm]{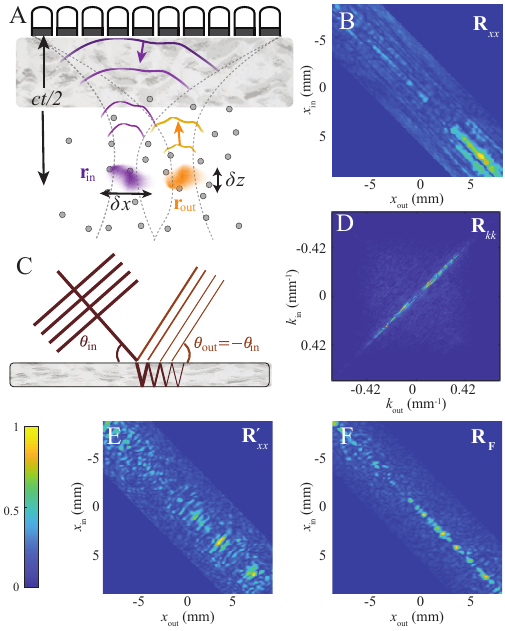}
	\caption{\textbf{Removing multiple reflections with the matrix approach.} (\textbf{A}) Focused beamforming applied to $\Rut(\omega)$ (Eqs.~\ref{projRrr}-\ref{Rrr_broadband}) yields the focused reflection matrix $\Rrr$ that contains the set of impulse responses between virtual transducers $\rin$ and $\rout$ at each depth $z$. (\textbf{B}) Modulus of 	\laura{matrix} $\mathbf{R_{xx}}$ at depth $z=25$ mm. (\textbf{C}) Sketch of multiple reflections between parallel surfaces. (\textbf{D}) Modulus of \laura{matrix} $\mathbf{R}_kk$ (Eq.~\ref{Rtt}) deduced from (\textbf{A}). (\textbf{E}) Modulus of matrix $\mathbf{R}^{\prime}_{xx}$ (Eq.~\ref{Rtt_rec}) after cancellation of the main antidiagonal ($\kin+\kout=0$) in $\mathbf{R}_{kk}$ (see \textit{Methods}).  (\textbf{F}) Modulus of filtered matrix $\mathbf{R_{F}}$ built from the estimator $\mathbf{\hat{T}}$ (Eq.~\ref{correction}). The matrices displayed in panels (\textbf{B}), (\textbf{D}),  (\textbf{E}) and  (\textbf{F}) have been normalized by their maximum value
	.}
	\label{fig_reverb}
\end{figure}

\laura{We} first apply a temporal Fourier transform to the experimentally acquired reflection matrix to obtain $\Rut(\omega)$, where $\omega=2\pi f$ is the angular frequency of the waves. To project $\Rut(\omega)$ between different bases, we then define free-space transmission matrices, $\mathbf{P}_0(\omega)$ and $\mathbf{G}_0(\omega)$, which describe the propagation of waves between the bases of interest for our experimental configuration. Their elements correspond to the 2D Green's functions which originate in the plane wave basis~\cite{Goodman1996} or at the transducer array~\cite{Watanabe} to any focal point $\mathbf{r}$ in a supposed homogeneous medium:
\begin{subequations}
	\label{propmats}
\begin{align}
\label{Gzp_eq}
P_0\left(\mathbf{r},\theta,\omega\right) 
&=\exp{\left[i k \left(z \cos\theta + x \sin \theta \right)\right]}  \\
\label{Gz_eq}
G_0\left(\mathbf{r},\mathbf{u},\omega\right)  &= - \frac{i}{4} \mathcal{H}_0^{(1)}\left (k|\mathbf{r}-\mathbf{u}| \right ),
\end{align}
\end{subequations}
where $\mathcal{H}_0^{(1)} $ is the Hankel function of the first kind. $k=\omega/c$ is the wave number. $x$ and $z$ describe the coordinates of the image pixel positions $\mathbf{r}$ in the lateral and axial directions, respectively (Fig.\ \ref{acqfig}A). $\Rut(\omega)$ can now be projected both in emission and reception to the focused basis via the matrix product~\cite{Goodman1996}
\begin{equation}
\label{projRrr}
\mathbf{R}_{\mathbf{rr}}(\omega)=\mathbf{{G}_0^*}\left(\omega\right) \times \Rut(\omega)\times \mathbf{P}_0^{\dagger}\left(\omega \right), \end{equation}
where the symbols $*$, $\dagger$ and $\times$ stands for phase conjugate, transpose conjugate and matrix product, respectively. 
Equation (\ref{projRrr}) simulates focused beamforming in post-processing in both emission and reception. For broadband signals, ballistic time gating can be performed to select only the echoes arriving at the ballistic time ($t=0$) in the focused basis. This procedure is described in more detail in Ref.~\cite{Lambert2019}. It involves considering only pairs of virtual transducers, $\rin=(\xin,z)$ and $\rout=(\xout,z)$, which are located at the same depth $z$ (Fig.\ \ref{fig_reverb}A); we denote this subspace of the focused reflection matrix as $\Rxx(z,\omega)=\left[R(\xout,\xin,z,\omega)\right]$. A coherent sum is then performed over the frequency bandwidth $\delta\omega$ to obtain the broadband focused reflection matrix
\begin{equation}
\label{Rrr_broadband}
\Rxx (z) = \int^{\omega_{+}}_{\omega_{-}} d\omega \Rxx (z,\omega),
\end{equation} 
where $\omega_{\pm}=\omega_0 \pm \delta \omega/2$ and $\omega_0$ is the central frequency. Each element of $\Rxx(z)$ contains the signal that would be detected by a virtual transducer located at $\rout=(\xout,z)$ just after a virtual source at $\rin=(\xin,z)$ emits a brief pulse of length $\Delta t= \delta \omega^{-1}$ at the central frequency $\omega_0$. Importantly, the broadband focused reflection matrix creates virtual transducers which have a greatly reduced axial dimension compared to the monochromatic focusing of $\Rxx(z,\omega)$ (Eq.\ \ref{projRrr}). This significantly improves the accuracy and spatial resolution of the subsequent analysis. 

Note that this matrix could have been directly formed in the time domain from the recorded matrix $\mathbf{R_{u\theta}}(t)$. A coherent sum of the recorded echoes coming from each focal point $\mathbf{r}$ could be performed to synthesize virtual detectors inside the medium. In practice, this would be done by applying appropriate time delays to the recorded signals~\cite{Montaldo2009}. The back-propagated wavefields obtained for each incident plane wave would then be summed coherently to generate \textit{a posteriori} a synthetic focus (i.e. a virtual source) at each focal point. Finally, the time gating step described by Eq.\ref{Rrr_broadband} consists, in the time domain, in only keeping the echoes arriving at the expected ballistic time. \\

In a recent work, the broadband focused reflection matrix $\Rxx(z)$ was shown to be a valuable observable to locally assess the quality of focusing and to quantify the multiple scattering level in the ultrasonic data~\cite{Lambert2019}. In this paper, $\Rxx(z)$ is used as a basic unit from which: (\textit{i}) a confocal image of the sample reflectivity can be built; (\textit{ii}) all of the subsequent aberration correction processes will begin.

Figure~\ref{fig_reverb}B shows $\Rxx(z)$ at depth $z=25$ mm. The brightness of its diagonal coefficients is characteristic of singly-scattered echoes~\cite{Badon2016,Blondel2018,Lambert2019}. In fact, the diagonal of $\Rxx(z)$ corresponds to the line at depth $z$ of a confocal or compounded~\cite{Montaldo2009} ultrasound image $\mathcal{I}\left(\mathbf{r} \right)$ : 
\begin{equation}
\label{imcalc}
\mathcal{I}\left(\mathbf{r}\right)\equiv \left| R\left(\xout=\xin,\xin,z \right) \right|^2.
\end{equation} 
Figure\ \ref{acqfig}B displays the resulting image $\mathcal{I}(\mathbf{r})$ of the phantom and plexiglass system. This image was created under an assumption of a homogeneous medium, with a speed of sound of $c=1800$ m/s used to calculate \laura{$ \mathbf{P}_0$} and $\mathbf{G}_0$ (Eqs.~\ref{Gzp_eq} and \ref{Gz_eq}). This value of $c$ was not chosen based on any \textit{a priori} knowledge of the medium, but rather as the value which gives the least aberrated image by eye -- a trial and error method typically used by medical practitioners and technicians. 
However, even with this optimal value for $c$, the image in Fig.\ \ref{acqfig}B remains strongly degraded by the plexiglass layer for two reasons: (1) multiple reverberations between the plexiglass walls and the probe have induced strong horizontal specular echoes, and (2) the input and output focal spots are strongly distorted (Fig.\ \ref{focusfig}C-D) 
because of the mismatch between the homogeneous propagation model and the heterogeneous reality. In the following, we show that a matrix approach to wave imaging is particularly appropriate to correct for these two issues.  \alex{A flow chart summarizing all the mathematical operations involved in this  
process is provided in \textit{SI Appendix} (Fig.~\alex{S1})}.

\subsection*{Removing multiple reverberations with the far-field reflection matrix}

\laura{Reverberation signals are a common problem in medical ultrasound imaging, often originating from multiple reflections at tissue interfaces or between bones in the human body. Here, we observe strong horizontal artifacts at shallow depths of the image (Fig.\ \ref{acqfig}B), which are due to waves which have undergone multiple reflections -- often called reverberations in the literature -- between the parallel walls of the plexiglass layer. In the following, we show that these signals can be isolated and suppressed using the reflection matrix.} 

\laura{To project the reflection matrix into the far-field,} we define a free-space transmission matrix, $\mathbf{T}_0$, which corresponds to the Fourier transform operator. Its elements link any transverse wave number $k_x$ in the Fourier space to the transverse coordinate $x$ of any point $\mathbf{r}$ in a supposed homogeneous medium:
\begin{equation}
\label{T0}
T_0\left(k_x,x \right) =\exp{\left(i {k_x}x\right)}.
\end{equation}
Each matrix $\mathbf{R}_{xx}(z)$ can now be projected in the far field via the matrix product
\begin{equation}
\label{Rtt}
    \mathbf{R}_{kk}(z)=\mathbf{T}_0 \times \mathbf{R_{xx}}(z) \times \mathbf{T}_0^{\top},
\end{equation}
where the symbol $\top$ stands for matrix transpose. The resulting matrix $\mathbf{R}_{kk}(z)=[R(\kout,\kin,z)]$ contains the reflection coefficients of the sample at depth $z$ between input and output wave numbers $\kin$ and $\kout$. An example of far-field reflection matrix $\mathbf{R}_{kk}(z)$ is displayed at depth $z=25$ mm in Fig.\ \ref{fig_reverb}D. Surprisingly, this matrix is dominated by a strongly enhanced reflected energy along its main antidiagonal ($\kout+\kin=0$). To understand this phenomenon, the reflection matrix $\mathbf{R}_{kk}(z)$ can be expressed as follows:
\begin{equation}
\label{Rtt2}
   \mathbf{R}_{kk}(z)=\mathbf{T}(z) \times \mathbf{\Gamma}(z) \times  \mathbf{T}^\top(z),
\end{equation}
where \laura{$\mathbf{\Gamma}(z) =[\Gamma(x,x^\prime,z)]$}
describes the scattering processes inside the medium. In the single scattering regime, $\mathbf{\Gamma}(z)$ is diagonal and its elements correspond to the medium reflectivity $\gamma(x,z)$ at depth $z$. $\mathbf{T}(z)$ is the true transmission matrix between the Fourier basis and the focal plane at depth $z$. Each column of this matrix corresponds to the wavefront that would be recorded in the far-field due to emission from a point source located at $\mathbf{r}=(x,z)$ inside the sample. 

In \textit{SI Appendix} (Section S1), a theoretical expression of $\mathbf{R}_{kk}$ is derived in the single scattering regime under an isoplanatic hypothesis. Interestingly, the norm-square of its coefficients $ R(\kout,\kin,z) $ is shown to be independent of aberrations. It directly yields the spatial frequency spectrum of the scattering medium at depth $z$:
\begin{equation*}
 \left |R(\kout,\kin,z) \right |^2= \left | \tilde{\gamma}(\kout+\kin,z) \right |^2 ,
\end{equation*}
where $\tilde{\gamma}(k_x,z)=\int dx \gamma(x,z) \exp (- i k_{x} x )$ is the 1D Fourier transform of the sample reflectivity $\gamma(x,z)$. \laura{In the single scattering regime, the matrix $\mathbf{R}_{kk}$ displays} a deterministic coherence along its antidiagonals~\cite{Aubry2009,Kang2015} that can be seen as a manifestation of the memory effect in reflection~\cite{Shajahan2014}. Each antidiagonal  ($\kin + \kout=$ constant) encodes one spatial frequency of the sample reflectivity. 
\laura{For the system under study here, reflections occurring between the parallel surfaces of the plexiglass obey $\kin + \kout= k_0\sin \theta_{0}$, where $k_0=\omega_0/c$ is the wave number at the central frequency and $\theta_0$ is the angle between the top face of the plexiglass and the transducer array (Fig.\ \ref{fig_reverb}C). Hence, signatures of such reflections should} arise along the main antidiagonal ($\kin +\kout=0$) of $\mathbf{R}_{kk}$.

We can take advantage of this sparse feature in $\mathbf{R}_{kk}$ to filter out signals from reverberation\laura{, independently of aberrations induced by the plexiglass} (see \textit{Methods}). Then, the inverse operation of Eq.~\ref{Rtt} can be applied to the filtered matrix $\mathbf{R}^\prime_{kk}$ to obtain a filtered focused reflection matrix: 
\begin{equation}
  \label{Rtt_rec}
  \mathbf{R}^\prime_{xx}(z)=  \mathbf{T_0^{\dag}}(z) \times \mathbf{R}^\prime_{kk}(z) \times \mathbf{T^{*}_0}(z) .
    \end{equation}
Fig.\ \ref{fig_reverb}E shows an example of $\mathbf{R}^\prime_{xx}$. Comparison with the original matrix in Fig.\ \ref{fig_reverb}B shows that the low spatial frequency components of the reflected wavefield have been removed from the diagonal of $\mathbf{R}_{xx}$. The resulting $\mathbf{R}^\prime_{xx}$ now exhibits solely random diagonal coefficients -- a characteristic of ultrasonic speckle. Finally, Fig.\ \ref{acqfig}C shows the full images calculated from $\mathbf{R}^\prime_{xx}$ (Eq.\ \ref{imcalc}). The removal of multiple reflections has enabled the discovery of previously hidden bright targets at shallow depths. However, the confocal image still suffers from aberrations, especially at small and large depths (Fig.~\ref{acqfig}C). 

\subsection*{Distortion matrix in the speckle regime}
\laura{In Ref.\ \cite{Badon2019}, the distortion matrix concept was introduced for optical imaging of extended specular reflectors in a strong aberration regime. Here, we show how to this distortion matrix approach can be extended to the speckle regime.}

\subsubsection*{Manifestation of aberrations}
In Fig.\ \ref{fig_reverb}E, a significant spreading of energy over off-diagonal coefficients of $\mathbf{R}^\prime_{xx}$ ~\cite{Blondel2018,Lambert2019} can be seen. This effect is a direct manifestation of the aberrations sketched in Fig.~\ref{fig_reverb}A, which can be understood by re-writing $\mathbf{R}^{\prime}_{xx}$ using Eqs.\ \ref{Rtt2} and \ref{Rtt_rec}:
\begin{equation}
\label{RrrMatrix}
\mathbf{R}^{\prime}_{xx}  (z)  = \mathbf{H}(z)  \times \mathbf{\Gamma}(z) \times \mathbf{H}^{\top} (z) ,
\end{equation}
where $\mathbf{H}(z)=\mathbf{T_0^{\dagger}} \times\mathbf{T}(z)$. We refer to $\mathbf{H}(z)$ as the \textit{focusing matrix}\laura{~\cite{Badon2019}}, because each line of $\mathbf{H}(z)$ corresponds to the spatial amplitude distribution of the input or output focal spots (Fig.~\ref{focusfig}C-D). 
\laura{Equation\ \ref{RrrMatrix} tells us that the off-diagonal energy spreading in $\mathbf{R}^{\prime}_{xx} (z)$ (Fig.\ \ref{fig_reverb}E) occurs when the focusing matrix $\mathbf{H}(z)$ is not diagonal, i.e. when the free-space transmission matrix $ \mathbf{T}_0$ is a poor estimator of the true transmission matrix $\mathbf{T}(z)$. This occurs when we do not have enough information about the medium to properly construct $ \mathbf{T}_0$ -- in particular, when the speed of sound distribution is unknown. This is the {cause} of sample-induced aberrations, which manifest in the off-diagonal energy spreading in $\mathbf{R}^{\prime}_{xx} (z)$, and finally, in the poor resolution in some parts of the confocal image (Fig.\ \ref{acqfig}C). }

\subsubsection*{The memory effect}
To isolate and correct for these aberration effects, we build upon a physical phenomenon often referred to as the memory effect~\cite{Freund1988,Feng1988,Osnabrugge2017} or isoplanatism~\cite{Roddier1999,Mertz2015} in wave physics. Usually, this phenomenon is considered in a plane wave basis. When an incident plane wave is rotated by an angle $\theta$, the far-field speckle image is shifted by the same angle $\theta$~\cite{Freund1988,Feng1988} (or $-\theta$ if the measurement is carried out in reflection~\cite{Katz2014,Shajahan2014}). Interestingly, this class of field-field correlations also exists in real space: Waves produced by nearby points inside a complex medium can generate highly correlated, but tilted, random speckle patterns in the far-field \laura{~\cite{ODonnell1988,Mallart1994,Walker1997,Varslot2004,Robert2008}}. In the focused basis, this corresponds to a spatially invariant point spread function (or focal spot) over an area called the isoplanatic patch. For \laura{aberration correction}, our strategy is the following: (\textit{i}) highlight these spatial correlations by building a dual-basis matrix (the distortion matrix) that connects any input focal point in the medium with the distortion exhibited by the corresponding reflected wavefront in the far-field\laura{~\cite{Badon2019}}, and (\textit{ii}) take advantage of these correlations to accurately estimate the transmission matrix $\mathbf{T}(z)$ in the same dual basis.

\subsubsection*{Revealing hidden correlations}

\begin{figure*}[t]
	\centering
	\includegraphics[width=\textwidth]{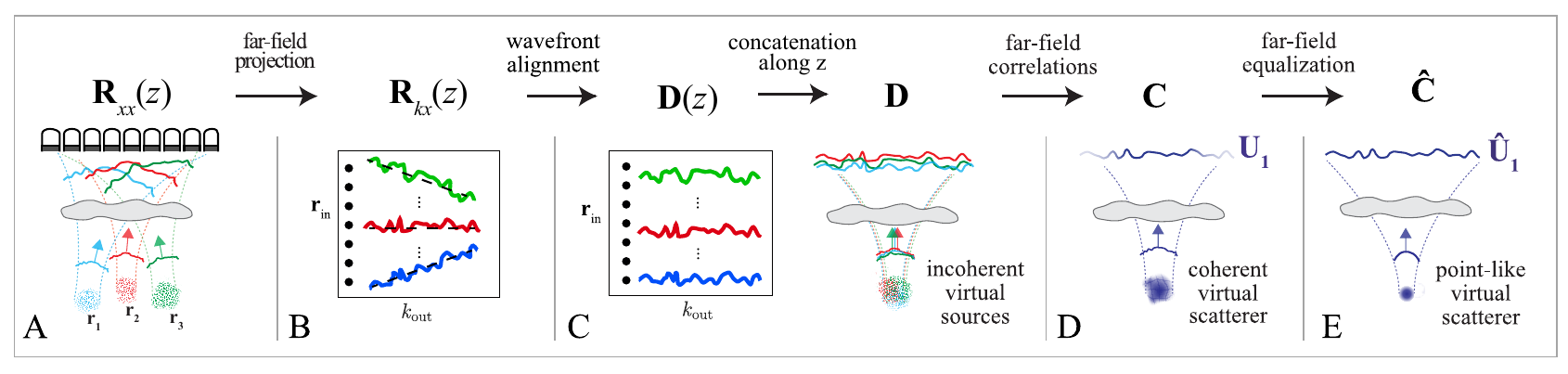} 
	\caption{\textbf{Time reversal analysis of the distortion matrix}. \laura{($\textbf{A}$) Each input focused illumination gives rise to a tilted reflected wavefront. ($\textbf{B}$) After far-field projection (Eq.~\ref{Rtrcalc}), each wavefield is stored along a line of $\mathbf{R}^T_{kx}(z)$. ($\textbf{C}$) By removing the geometrical tilt (dashed black line in B) of each reflected wavefront (Eq.~\ref{D_eq}), \alex{a set of distortion matrices $\mathbf{D}(z)$ is obtained at each depths $z$. These matrices are concatenated to yield a full-field distortion matrix $\mathbf{D}$}. $\mathbf{D}$ is equivalent to a reflection matrix but with input focal spots $H(x-\xin)$ virtually shifted at the same location (Eq.~\ref{Dth}). ($\textbf{D}$) Considering its correlation matrix $\mathbf{C}$ smooths out the sample reflectivity (Eq.~\ref{rhoaveeq}) and mimics the time reversal operator associated with a virtual specular reflector of scattering distribution $|H(x)|^2$ ($\textbf{E}$). The normalized correlation matrix $\mathbf{\hat{C}}$ makes the virtual reflector point-like (see \textit{SI Appendix}, Section S4). Its eigenvalue decomposition (Eq.~\ref{svd}) then yields the transmission matrix over each isoplanatic patch contained in the \alex{FOI} (Eq.~\ref{transiso}).}  }
	\label{focusing_speckle_img}
\end{figure*}

To isolate the effects of aberration in the reflection matrix, $\mathbf{R}^{\prime}_{xx} (z)$ is first projected into the Fourier basis in reception using the free-space transmission matrix $ \mathbf{T}_0$:
\begin{equation}
\label{Rtrcalc}
\mathbf{R}_{kx}(z)=\mathbf{T}_0 \times  \mathbf{R}^{\prime}_{xx} (z).
\end{equation}
\laura{Since the aberrating layer under consideration is laterally invariant, we might expect the memory effect to cause long-range correlation in $\mathbf{R}_{k x}(z)$, i.e. repeating patterns amongst the rows and/or columns of $\mathbf{R}_{k x}(z)$. However, this is not the case: as sketched in Fig.~\ref{focusing_speckle_img}C, input focusing points at different locations result in wavefronts with different angles in the far-field (see \textit{SI Appendix}, Section S1 and Fig.~S2 for further details). This geometric effect hides the correlations which could allow discrimination between isoplanatic patches.}

To reveal correlations in $\mathbf{R}_{k x}(z)$, the reflected wavefront can be decomposed into two contributions \alex{(see \textit{SI Appendix}, Fig.~S2)}: (1) a geometric component which would be obtained for a perfectly homogeneous medium (\laura{represented by the} black dashed line in  Fig.~\ref{focusing_speckle_img}\laura{B}) and which can be directly extracted from the reference matrix $ \mathbf{T}_0$, and (2) a distorted component due to the mismatch between the propagation model and reality (Fig.~\ref{focusing_speckle_img}\laura{C, left}). 
The key idea of this paper is to isolate the latter contribution by subtracting, from the experimentally measured wavefront, its ideal counterpart. Mathematically, this operation can be expressed as a Hadamard product between the normalized reflection matrix $\mathbf{\hat{R}}_{k x}(z)=[R(\kout,\xin,z)/|R(\kout,\xin,z)|]$ and $\mathbf{T}_0^{*}$,
\begin{equation}
\mathbf{D}(z)  =  \mathbf{\hat{R}}_{k x}(z) \circ  \mathbf{T}_0^{*} ,
\label{D_eq}
\end{equation}   
which, in terms of matrix coefficients, yields
\begin{equation}
D(\kout,\xin,z)  =  \hat{R}(\kout,\xin,z)  T_0^{*}(\xin,\kout)
\label{D_eq_coefficients} .
\end{equation}   
The matrix $\mathbf{D}=[D(\kout,\rin)]$ is the distortion matrix \alex{defined over the field-of-illumination (FOI) mapped by the set of input focusing points $\rin$}. It connects any input focal point $\rin$ to the distorted component of the reflected wavefield in the far-field. \laura{Note that, unlike conventional adaptive focusing techniques, no bright scatterer is used as a guide star in our matrix approach. In fact, this is why a normalized reflection matrix $\mathbf{\hat{R}}_{k x}$ is considered in Eq.\ \ref{D_eq}. All input focusing points $\rin$ have the same weight in $\mathbf{D}$, regardless of their reflectivity. Hence, the eight bright targets contained in the phantom (Fig.~\ref{acqfig}A) do not play the role of guide stars.}

Compared to $\mathbf{R}_{k x}$ \laura{(Fig.~\ref{focusing_speckle_img}B)}, $\mathbf{D}$ exhibits long-range correlations (see \textit{SI Appendix}, Fig.~S2): While the original reflected wavefronts display a different tilt for each focal point $\rin$, their distorted component displays an almost invariant aberration phase law over all $\rin$\laura{(Fig.~\ref{focusing_speckle_img}C)}.
To support our identification of spatial correlations in $\mathbf{D}$ with isoplanatic patches, $\mathbf{D}$ is now expressed mathematically. We begin with the simplest case of an isoplanatic aberration which implies, by definition,  a spatially-invariant input focal spot: $H(x,\xin,z)=H(x-\xin)$. Under this hypothesis, the injection of Eqs.~\ref{RrrMatrix} and \ref{Rtrcalc} into Eq.\ref{D_eq} gives the following expression for $\mathbf{D}$ (see  \textit{SI Appendix}, Section S3):
\begin{equation}
\label{Dth}
\mathbf{D}(z) =  \mathbf{T} \times \mathbf{S}(z) ,
\end{equation}
where the matrix $\mathbf{S}$ is the set of \alex{incoherent}  virtual sources re-centered at the origin such that
\begin{equation}
\label{Dthel}
S(x',\xin,z) =  \gamma(x'+\xin,z)H(x').
\end{equation}
${x'}= x-\xin$ represents a new coordinate system centered around the input focusing point. 
\laura{These virtual sources are spatially incoherent due to the random reflectivity of the medium, and their size is governed by the spatial extension of the input focal spot.} The physical meaning of Eqs.~\ref{Dth} and \ref{Dthel} is the following: Removing the geometrical component of the reflected wavefield in the far-field as done in Eq.\ref{D_eq} is equivalent to shifting each virtual source to the central point $\xin={0}$ of the imaging plane. $\mathbf{D}$ is still a type of reflection matrix, but one which contains different realizations of virtual sources all located at the origin (Fig.~\ref{focusing_speckle_img}\laura{C, right}). 
\laura{This superposition of the input focal spots will enable the unscrambling of the propagation and scattering components in the reflected wavefield}. 

\subsubsection*{\label{aberration_sec} Time reversal analysis}

The next step is to extract and exploit the correlations of $\mathbf{D}$ for imaging. \alex{In the specular scattering regime, $\mathbf{D}$ is dominated by spatial correlations in the input focal plane~\cite{Badon2019}. This is due to the the long-range coherence of the sample reflectivity for specular reflectors. Conversely, in the speckle scattering regime, the sample reflectivity $\gamma(\mathbf{r})$ is random:  $\langle\gamma(\mathbf{r})\gamma^*(\mathbf{r^\prime})\rangle= \langle\left|\gamma\right|^2\rangle\delta(\mathbf{r}-\mathbf{r^\prime})$, where $\delta$ is the Dirac distribution and the symbol $\langle \cdots \rangle$ denotes an ensemble average. In this case, correlations in the Fourier plane dominate. To extract them, the correlation matrix $\mathbf{C}=N^{-1}\mathbf{D} \mathbf{D}^{\dag}$ is an excellent tool. The coefficients of $\mathbf{C}$ are obtained by averaging the angular correlations of the distorted wave-field $D(\kout,\rin)$ over the $N$ input focusing points $\rin=(x_\textrm{in},z)$:}
\begin{equation}
\label{C0}
\alex{ C(k_x,k'_x) = N^{-1}\sum_{\rin} D(k_x,\rin) D^*(k'_x,\rin).}
 \end{equation}
$\mathbf{C}$ can be decomposed as the sum of a covariance matrix $\left \langle \mathbf{C}\right \rangle $ and a perturbation term $\delta  \mathbf{C} $:
\begin{equation}
\label{C}
   \mathbf{C}= \left \langle \mathbf{C}\right \rangle+\delta  \mathbf{C} .
\end{equation}
\alex{$\mathbf{C}$ will converge towards $\left \langle \mathbf{C}\right \rangle $ if the incoherent source term $\mathbf{S}$ of Eq.~\ref{Dth} is averaged over enough independent realizations of disorder, i.e. if the perturbation term $\delta\mathbf{C}$ tends towards zero. In fact, the intensity of $\delta\mathbf{C}$ scales} as the inverse number $M$ of resolution cells in the FOV~\cite{Robert_thesis}. In the present case, $M = L_x L_z/(\delta x \delta z)\sim 10000$, where $(L_x,L_z)$ is the spatial extent of the overall FOV and $(\delta x,\delta z)$ is the spatial extent of each resolution cell (see Fig.~\ref{fig_reverb}A). \laura{In the following, we will thus assume a convergence of $\mathbf{C}$ towards its covariance matrix $\left \langle \mathbf{C}\right \rangle$ due to disorder self-averaging.} \\

Let us now express the covariance matrix $\left \langle \mathbf{C}\right \rangle$ theoretically.  This allows $\left \langle \mathbf{C}\right \rangle$ to be written as (see  \textit{SI Appendix}, Section S4)
\begin{equation}
\label{rhoaveeq}
\left \langle \mathbf{C} \right \rangle =  \mathbf{T} \times  \mathbf{\Gamma}_H \times \mathbf{T}^{\dag} ,
\end{equation}
where $\mathbf{\Gamma}_H$ is diagonal and its coefficients are directly proportional to $|H(x)|^2$. $\mathbf{\Gamma_H}$ is equivalent to a scattering matrix associated with a virtual \alex{coherent} reflector whose scattering distribution corresponds to the input focal spot intensity $|H(x)|^2$ (Fig.~\ref{focusing_speckle_img}\laura{D}). Expressed in the form of Eq.\ \ref{rhoaveeq}, \alex{$\langle\mathbf{C}\rangle$ is analogous to a reflection matrix associated with a single scatterer of reflectivity $|H(x)|^2$.}
For such an experimental configuration, it has been shown that an iterative time reversal process converges towards a wavefront that focuses perfectly through the heterogeneous medium onto this scatterer~\cite{Prada1994,Prada1996}. Interestingly, this time-reversal invariant can also be deduced from the eigenvalue decomposition of the time-reversal operator $\mathbf{R}\mathbf{R}^{\dag}$~\cite{Prada1994,Prada1996,Prada2003}. 
The same decomposition could thus be applied to $\mathbf{C}$ in order to retrieve the wavefront that would perfectly compensate for aberrations and optimally focus on the virtual reflector\laura{. This effect is illustrat\alex{ed} in Fig.~\ref{focusing_speckle_img}\laura{D}. It is important to emphasize, however, \alex{that the induced focal spot} is enlarged compared to the diffraction limit~\cite{Aubry2006,Robert2009}. \alex{For the \laura{goal} 
of diffraction-limited imaging, the size of this focal spot should be reduced.} In the following, we express this situation mathematically, and show how to resolve it.}  

\alex{By 
\laura{the van Cittert-Zernike} theorem~\cite{Mallart1994}, the correlation coefficients $C(k'_x,k_x)$ are directly proportional to the Fourier transform of the scattering distribution $|H(x)|^2$ (see \textit{SI Appendix}, Section S4, for details). To reduce the size of the virtual reflector, one can equalize the Fourier spectrum of its scattering distribution. Interestingly,} this can be done by normalizing the correlation matrix coefficients as follows  
\begin{equation}
\label{hat}
    \hat{C}(k^{\prime}_x,k_x)={{{C}}(k^{\prime}_x,k_x)}/{{|{C}}(k^{\prime}_x,k_x)|}.
\end{equation}
\laura{This operation is illustrated by Fig.~\ref{focusing_speckle_img}\laura{E}.} The normalized correlation matrix $\mathbf{\hat{C}}=\left [\hat{C}(k^{\prime}_x,k_x) \right]$ can be expressed as: 
\begin{equation}
\label{rhoaveeq_norm}
 \mathbf{\hat{C}} =  \mathbf{T} \times\mathbf{{\Gamma}}_{\delta} \times  \mathbf{T}^{\dag}\laura{.}
\end{equation}
\laura{In contrast to the operator $\mathbf{{\Gamma}}_{{H}}$ of Eq.\ \ref{rhoaveeq}, $\mathbf{{\Gamma}}_{{\delta}}$ is a scattering matrix associated with a point-like (diffraction-limited) reflector at the origin \laura{(Fig\ \ref{focusing_speckle_img}E)}}. A reflection matrix associated with such a point-like reflector is of rank 1~\cite{Prada1994,Prada1996}; this property should also hold for the normalized correlation matrix $\mathbf{\hat{C}}$ in the case of spatially-invariant aberrations. \alex{As we will see, the first eigenvector of $\mathbf{\hat{C}}$} yields the distorted component of the wavefront, and its phase conjugation enables compensation for aberration, resulting in optimal focusing within the corresponding isoplanatic patch.

Beyond the isoplanatic case, the eigenvalue decomposition of $\mathbf{\hat{C}} $ can be written as follows: 
\begin{equation}
\label{svd}
\mathbf{\hat{C}} =\mathbf{U \Sigma  U^{\dag}} .
\end{equation}
$\mathbf{\Sigma}$ is a diagonal matrix containing the eigenvalues $\sigma_i$ in descending order: $\sigma_1>\sigma_2>..>\sigma_N$. $\mathbf{U}$ is a unitary matrix that contains the orthonormal set of eigenvectors $\mathbf{U}_i$. In a conventional iterative time reversal experiment~\cite{Prada1994,Prada1996}, there is a one-to-one association between each eigenstate of the reflection matrix and each point-like target in the medium. The corresponding eigenvalue $\sigma_i$ is related to the scatterer reflectivity and the eigenvector $\mathbf{U}_i$ yields the transmitted wavefront that focuses on the corresponding reflector. In this work, iterative time reversal is applied to $\mathbf{\hat{C}}$. Each isoplanatic patch in the \alex{FOI} gives rise to a virtual reflector at the origin associated with a different aberration phase law. We thus expect a one-to-one association between each isoplanatic patch $p$ and each eigenstate of $\mathbf{\hat{C}}$: for each isoplanatic patch, the eigenvector $\mathbf{U\p}=[U\p(k_x)]$ should yield the \alex{corresponding distorted wavefront} in Fourier space, and the 
\laura{eigenvalue} $\sigma\p$ should provide an indicator of the focusing quality in that patch.
\begin{figure*}[t]
	\centering
	\includegraphics[width=\textwidth]{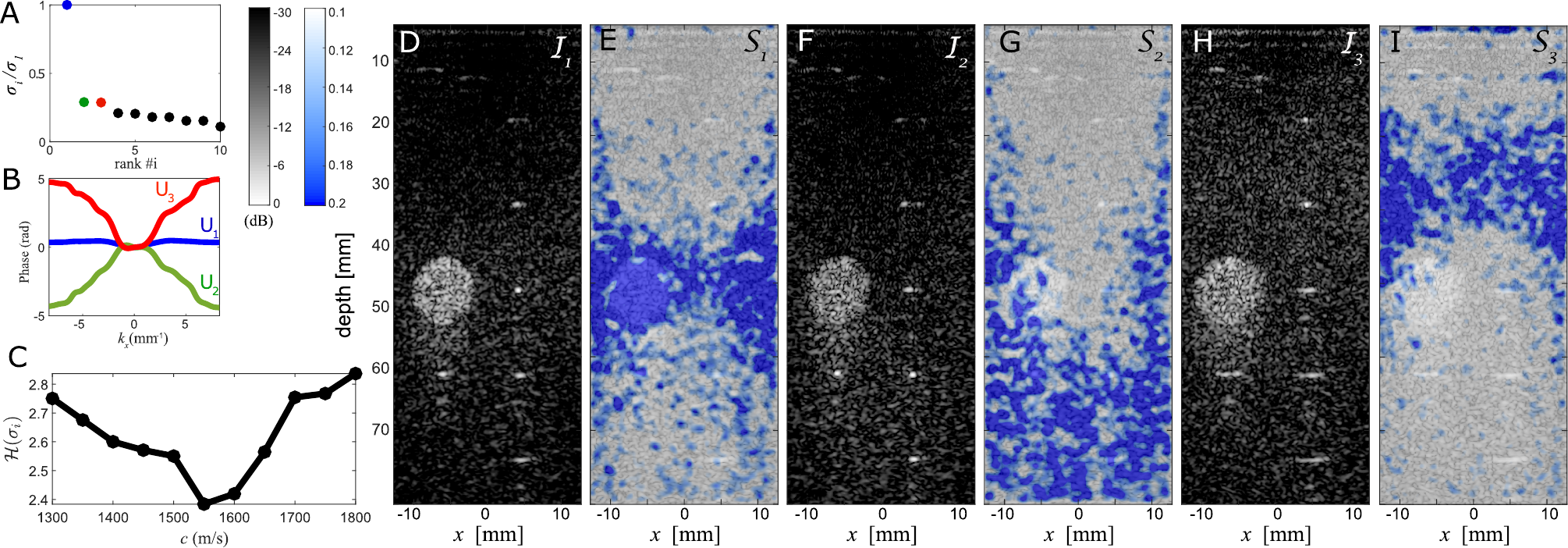} 
	\caption{\textbf{Retrieving the transmission matrix $\mathbf{T}$ from the correlation matrix $\mathbf{\hat{C}}$.} Results of the SVD of $\mathbf{\hat{C}}$ are shown: (\textbf{A}) Normalized eigenvalues $\tilde{\sigma}_i$, and (\textbf{B}) the phase of the three first eigenvectors, $\mathbf{U}_i$. (\textbf{C}) Entropy \alex{$\mathcal{H}$} (Eq.~\ref{entropy}) of the eigenvalues \alex{$\sigma_i$} is plotted versus the model speed of sound $c$. (\textbf{D}-\textbf{H}) Confocal images $\mathcal{I}\p(\mathbf{r})$ are shown with their corresponding Strehl ratio maps $\mathcal{S}\p(\mathbf{r})$, deduced from the three first transmission matrices $\mathbf{\bar{T}}\p$ (\laura{Eq.~\ref{S}}). \laura{The ultrasound images and Strehl ratio maps are displayed with the same dynamic (B\&W) and linear (color) scales, respectively.}}
	\label{resultsfig} 
\end{figure*}

\subsection*{\alex{Isoplanatic patches and Shannon Entropy}}

\subsubsection*{\alex{Field-of-view decomposition into isoplanatic patches}}

We now apply our theoretical predictions to the experimental ultrasound imaging data. Figure~\ref{resultsfig}A displays the normalized eigenvalues $\hat{\sigma}_i=\sigma_i/\sum_{j=1}^N\sigma_j$ of the correlation matrix $\mathbf{\hat{C}}$. If the convergence towards the covariance matrix were complete, the rank of $\mathbf{\hat{C}}$ should yield the number of isoplanatic patches in the ultrasound image. In Fig.\ \ref{resultsfig}A, a few eigenvalues seem to dominate, but it is not clear how many are significantly above the noise background. To solve this problem, we consider the Shannon entropy $\mathcal{H}$ of the eigenvalues $\hat{\sigma}_i$~\cite{Campbell1960,Roberts1999}:
\begin{equation}
  \mathcal{H}(\hat{\sigma}_i)= -\sum_{i=1}^{N} \hat{\sigma}_i \log_2 \left ( \hat{\sigma}_i \right ).
   \label{entropy}
\end{equation}
Shannon entropy yields the least biased  estimate possible for the information available, i.e. the data set with the least artifact for a given signal-to-noise ratio. Thus, it can be used here as an indicator of how many eigenstates are required to create an adequate ultrasound image without being affected by the perturbation term in Eq.\ref{C}\laura{~\cite{Badon2019}}. The eigenvalues of Fig.~\ref{resultsfig}A \laura{(calculated using the model wave velocity $c=1800$~m/s}) have an entropy of $\mathcal{H}\simeq 2.85$ (\laura{Fig.\ \ref{resultsfig}C}). Hence, only the three first eigenstates should be required to construct an unaberrated image of the medium.  Figure~\ref{resultsfig}B shows the phase of the three first eigenvectors $\mathbf{U}\p$. $\mathbf{U}_1$ is almost flat and exhibits \alex{a phase standard deviation of $0.28$~radians}, indicating that no correction for aberration (or a very minimal one) is required for optimal focusing in the isoplanatic patch associated with that vector. $\mathbf{U}_2$ and $\mathbf{U}_3$, however, display \alex{a phase standard deviation of 1.36 and 1.62 radians}, respectively. They are probably associated with the most aberrated parts of the image. Now that the important eigenstates are known, an estimator $\mathbf{\bar{T}\p}$ of the transmission matrix can now be calculated by combining the free-space $ \mathbf{T}_0$ matrix and the normalized eigenvector $\mathbf{\hat{U}\p}=[U\p(k_x)/|U\p(k_x)|]$:
\begin{equation}
\label{transiso}
   \mathbf{\bar{T}\p}=\mathbf{\hat{U}\p} \circ  \mathbf{T}_0 .
\end{equation}
\alex{The normalization of $\mathbf{U_p}$ ensures an equal contribution of each spatial frequency in $\mathbf{\bar{T}\p}$}. Then, the transmission matrices $\mathbf{\bar{T}\p}$ can be used to project the reflection matrix into the focused basis: 
\begin{equation}
\label{Rttcalc}
\mathbf{R\p}  =\mathbf{\bar{T}\p}^{\dag} \times \mathbf{R}^{\prime}_{kk} \times\mathbf{\bar{T}\p}^{*} .
\end{equation}
Figures \ref{fig_reverb}E and F illustrate the benefit of our matrix approach at depth $z=25$ mm. While the original matrix $\mathbf{R}^{\prime}_{xx}$ exhibits a significant spreading of the backscattered energy over its off-diagonal elements (Fig.~\ref{fig_reverb}E), the \alex{corrected} reflection matrix $\mathbf{R}_{3}$ (Eq.~\ref{Rttcalc}) is almost diagonal  (Fig.~\ref{fig_reverb}F). This feature demonstrates that the input and output focal spots are now close to be diffraction-limited and that aberrations have been almost fully corrected by the transmission matrix $\mathbf{T_3}$ at depth $z=25$ mm. 

The resulting ultrasound images, calculated from the diagonal elements of $\mathbf{R\p}$, are displayed in Figs~\ref{resultsfig}D, F and H: each estimator $\mathbf{\bar{T}\p}$ of the transmission matrix reveals a well-resolved and contrasted image of the phantom over distinct isoplanatic patches. $\mathbf{U_1}$ is associated with an isoplanatic patch at mid-depth ($z\simeq 45-55$~mm). As previously anticipated, correction by $\mathbf{U_1}$ leaves the image almost unchanged (compare Fig~\ref{acqfig}C to Fig~\ref{resultsfig}D).  This isoplanatic patch does not require aberration correction  because the model wave velocity $c=1800$~m/s is already close to the integrated speed of sound value at mid-depth. However, the phases of $\mathbf{U_2}$ and $\mathbf{U_3}$ exhibit curved shapes which indicate an incorrect model for the speed of sound $c$ (Fig~\ref{resultsfig}B). While the convex shape of $\mathbf{U}_2$ suggests an underestimation of $c$, the concave shape of $\mathbf{U}_3$ indicates overestimation. Correction with each eigenvector compensates for the associated distortion effect: the confocal images show an optimized contrast and resolution at large depths ($z>70$~mm) for $\mathbf{\bar{T}_2}$ (see Fig.~\ref{resultsfig}F) and shallow depths ($25 < z < 40$~mm) for $\mathbf{\bar{T}_3}$ (see Fig.~\ref{resultsfig}H). 

The gain in image quality can quantified by the Strehl ratio, $\mathcal{S}$~\cite{mahajan1982strehl}. Initially introduced in the context of optics, $\mathcal{S}$ is defined as the ratio of the peak intensity of the imaging system point spread function with aberration to that without. Equivalently, it can also be defined in the far-field as the squared magnitude of the mean aberration phase factor. $\mathcal{S}$ is directly proportional to the focusing parameter introduced by Mallart and Fink in the context of ultrasound imaging~\cite{Mallart1994}. Here, we can calculate a spatially-resolved Strehl ratio using the distortion matrices $\mathbf{{D}\p}= \mathbf{R\p}\circ \mathbf{T_0}^*$ computed after aberration correction:
\begin{equation}
\label{S}
  \mathcal{S}\p(\rin)= \left | \left \langle  {D}\p( \kout,\rin) \right \rangle_{\kout} \right |^2 ,
\end{equation}
where the symbol $\langle \cdots \rangle$ denotes an average over the variable in the subscript (which here is the output transverse wave number $\kout$). The Strehl ratio ranges from 0 for a completely degraded focal spot to 1 for a perfect focusing.  The maps of Strehl ratio corresponding to the confocal images $\mathcal{I}\p$ are shown in Figs.~\ref{resultsfig}E, G and I. These maps enable direct visualization of the isoplanatic area in which each different aberration correction is effective, allowing quantitative confirmation of our previous qualitative analysis of confocal images. Moreover, $\mathcal{S}\p$ enables an estimation of the focus quality at each point of the image. Compared to the initial value $\mathcal{S}_1$ displayed in Fig.~\ref{resultsfig}E, $\mathcal{S}_2$ and $\mathcal{S}_3$ show an improvement of the focusing quality by a factor 3 at large and shallow depths, respectively (Figs.~\ref{resultsfig}G--I). 

\alex{The results displayed in Fig.\ \ref{resultsfig} show that the decomposition of the imaging problem into isoplanatic patches, originally demonstrated with $\mathbf{D}$ for large specular reflectors in optics~\cite{Badon2019}, also holds in a random speckle regime if we consider, this time, the normalized correlation matrix $\mathbf{\hat{C}}$. \laura{In fact, \laura{the process actually performs even better in speckle than for specular reflectors,} 
since it is possible to discriminate between aberrations in input and output, and hence to correct for each independently.} The drawback here lies in the fact that corrections over each isoplanatic patch are difficult to combine.} 
\alex{To \laura{address this issue,} 
a first option is to refine the propagation model (i.e the speed of sound distribution).} 
\laura{To that end, the Shannon entropy  $\mathcal{H}(\sigma_i)$ is a valuable tool.}

\subsubsection*{\alex{Shannon entropy minimization}}

The first path towards full-field imaging is based on a minimization of the distortion entropy $\mathcal{H}(\sigma_i)$ (Eq.~\ref{entropy}). The logic is as follows: 
\laura{(a) In the speckle regime, there is a 
\alex{direct relation} 
between the \alex{Shannon entropy $\mathcal{H}(\sigma_i)$} 
of $\mathbf{\hat{C}}$ and the number \alex{$N_p$} of isoplanatic patches supported by the \alex{FOI} (shown in the previous section). }
\alex{(b) When the propagation model inside the FOI gets closer to reality, the number $N_p$ of isoplanatic patches within the FOI decreases \laura{(see \textit{SI Appendix}, Fig.~S\alex{3})}. (c) \laura{Thus, $\mathcal{H}(\sigma_i)$ is}  
minimal when the propagation model matches the speed of sound distribution in the \alex{FOI}.
\laura{This rationale is only valid if the convergence of $\mathbf{\hat{C}}$ towards $\langle \mathbf{\hat{C}} \rangle$ is achieved. To ensure this convergence, we note that, as shown in \textit{SI Appendix} (Section S5), the standard deviation of the coefficients of $\mathbf{\hat{C}}$ is proportional to the width of the aberrated PSF. Thus, for a more precise measure of entropy, we calculate $\mathcal{H}(\sigma_i)$ from the corrected distortion matrix $\mathbf{D_1}$.}} 


Figure~\ref{resultsfig}C provides a first proof-of-concept of this idea. It shows the entropy $\mathcal{H}(\sigma_i)$ as a function of the speed of sound $c$ used to model the propagation of ultrasonic waves \laura{in the \alex{FOI considered} (here, the phantom down to $z=80$ mm)}. 
\laura{$\mathcal{H}(\sigma_i)$} exhibits a minimum around $c=1550$~m/s, which is close to the speed of sound $c_p$ in the phantom. \alex{Using this optimized wave velocity in the propagation model,}  a single adaptive focusing correction then enables compensation for aberrations over the whole field of view (see \textit{SI Appendix}, Fig.~\alex{S3}).

Note that \laura{while} the entropy $\mathcal{H}_1(\sigma_i)$ displays a minimum, it does not reach the ideal value of 1. A first reason \laura{for this is the perturbation term in Eq.~\ref{C}: experimental noise and an insufficient number of input focal points can hinder perfect smoothing of the fluctuations caused by the random sample reflectivity.}
\alex{Another potential reason is that 
\laura{imperfections in the probe or plexiglass layer could} induce lateral variations of the aberrations upstream of the FOV.}

\subsubsection*{\alex{Discussion}}

\alex{To obtain a spatial map of the speed of sound and a full-field image of \laura{a heterogeneous medium}, 
one would need to repeat the same entropy minimization process \laura{described above,} but over a 
\laura{finite} and moving \alex{FOI}. The value of \laura{$c$ which minimizes} 
the entropy would be the speed of sound averaged over this \alex{FOI}. \laura{However, a} compromise 
\laura{must} be made between the spatial resolution (\alex{FOI size}) and the precision of the speed of sound measurement (see \textit{SI Appendix}, Section S5). Moreover, note that for highly-resolved mapping, this approach may prove prohibitively computationally expensive.}

\subsection*{\alex{Transmission matrix imaging}}

\subsubsection*{\alex{Phantom imaging: depth-dependent aberrations}}

The second route towards full-field imaging is more general, and goes far beyond the case of spatially-invariant aberrations. It consists in locally estimating each coefficient of the transmission matrix $\mathbf{T}$ that links the far-field and focused bases. The idea is to consider a sub-distortion matrix $\mathbf{D^\prime}(\mathbf{r\p})$ centered on each pixel $\mathbf{r\p}$ of the image over a limited \alex{FOI}: 
\begin{equation}
    D^\prime(\kout,\rin,\mathbf{r\p})=D(\kout,\rin)W(\rin-\mathbf{r\p}) ,
\end{equation}
where $W(\mathbf{r})$ is the spatial window function
\begin{equation*}
\label{window}
W(\mathbf{r})=\left\{
\begin{array}{ll}
1 & \, \mbox{for }|x|<\Delta x \mbox{ and }|z|<\Delta z  \\
0 & \,  \mbox{otherwise}.
\end{array} 
\right.
\end{equation*} 
The extent $(\Delta x, \Delta z)$ of this FOI should be subject to the following compromise: It should be large enough to average the fluctuations linked to disorder, but sufficiently small to enable a local measurement of aberrations \alex{(see \textit{SI Appendix}, Section S5)}. 
\laura{\alex{Here,} the dimensions of this window have been \alex{empirically} set to \alex{$\Delta z=5$~mm, with $\Delta x=25.6$~mm} (the lateral extent of the image)}. For each image pixel $\mathbf{r\p}$, a normalized correlation matrix $\mathbf{\hat{C}'}(\mathbf{r\p})$ can be deduced from $\mathbf{{D}'}(\mathbf{r\p})$. The first eigenvector $\mathbf{U_1}(\mathbf{r\p})$ yields a local aberration phase law for each pixel of the image. It can then be used to build an estimator $\mathbf{\bar{T}}$ of the global transmission matrix $\mathbf{T}$: 
\begin{equation}
\bar{T}(k_x,\mathbf{r\p})={U}_1(k_x,\mathbf{r\p})T_0(k_x,\mathbf{r\p}) .
\end{equation}
$\mathbf{\bar{T}}$ is then used to compensate for all of the phase distortions undergone by the incident and reflected wavefronts. Mathematically, this is accomplished by applying the phase conjugate of $\mathbf{\bar{T}}$ to both sides of the far-field reflection matrix $\mathbf{R}^{\prime}_{k k}$:
\begin{equation}
\label{correction}
    \mathbf{R_F}=\mathbf{\bar{T}}^{\dag} \times \mathbf{R}^{\prime}_{k k}(z) \times \mathbf{\bar{T}}^{*} .
\end{equation}
The diagonal elements of the full-field reflection matrix $\mathbf{R_F}$ yield the confocal image $\mathcal{I}_F(\mathbf{r})$ displayed in Fig.~\ref{acqfig}E. The corresponding Strehl ratio map $\mathcal{S}_F$ is shown in Fig.~\ref{acqfig}F. The clarity of the ultrasound image compared to the initial (Fig.~\ref{acqfig}B) and intermediate (Fig.~\ref{acqfig}C) ones, and the marked improvement in $\mathcal{S}_F$ compared to that of Fig.~\ref{acqfig}D demonstrate the effectiveness of this transmission matrix approach. 
A satisfying Strehl ratio $\mathcal{S}_F \sim 0.4$ is reached over the entire field of view, and a factor of five improvement is observed at shallow and large depths where the impact of the aberrating layer is the strongest. Such an improvement of the focusing quality is far from being negligible as it translates to a gain of 14 dB in image contrast.

\laura{This proof of concept experiment opens a number of additional questions. First,} despite our best efforts, the measured Strehl ratio $\mathcal{S}_F$ does not approach the ideal value of 1. Several reasons \alex{can} account for this: 
\laura{\alex{(\textit{i})} a part of the reflected wavefield has been lost at shallow depth when specular reflections and clutter noise have been removed; \alex{(\textit{ii})}  experimental noise and multiple scattering events taking place upstream of the focal plane could hamper our measure of the Strehl ratio, especially at large depths; \alex{(\textit{iii})} the same correction applies to the whole frequency bandwidth while the aberrations are likely to be dispersive (although for the phantom/plexiglass system considered here, dispersion should not be very strong).} \alex{Second, this experiment only involves depth-dependent aberrations. \laura{While such a configuration is of interest for, for example, imaging the liver through fat or muscle layers, or the brain through the skull, it lacks generality.} 
In the next section, both \alex{lateral and depth variations} of aberrations are addressed by means of an in vivo imaging experiment.}

\subsubsection*{\alex{In vivo ultrasound imaging: 
\laura{spatially-distributed} aberrations}}
\label{invivo_section}
\begin{figure*}[t]
	\centering
	\includegraphics[width=17 cm]{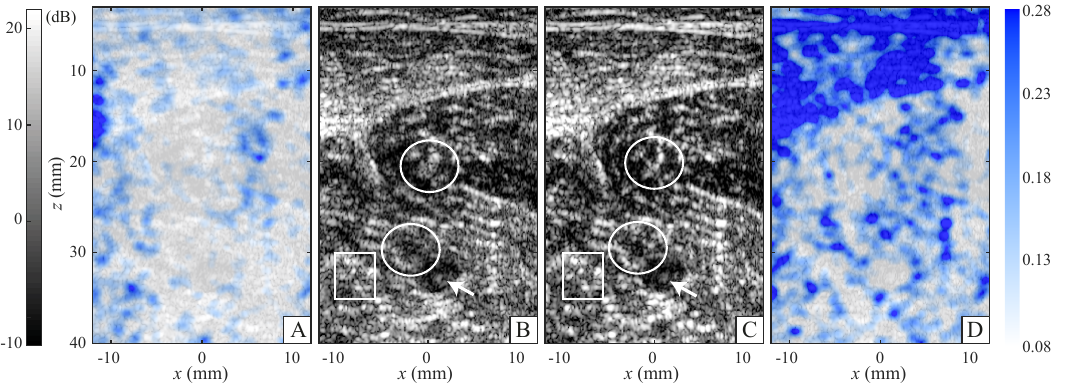}
	\caption{\alex{\textbf{Aberration correction of in vivo imaging.} Before correction, (\textbf{A}) the Strehl ratio map and (\textbf{B}) the confocal image exhibit the consequences of sample-induced distortion. (\textbf{C}) The corrected confocal image has improved contrast and resolution, as evidenced by the corresponding Strehl ratio map in (\textbf{D}). \laura{\alex{Rectangular and} circular areas in \textbf{B}, \textbf{C} highlight areas for comparison of images before and after correction.}}}
	\label{resultsfig_invivo}
\end{figure*}

\laura{We now apply the aberration correction technique to a dataset acquired in vivo from a human calf \alex{(see Methods)}. The uncorrected image is shown in Fig.\ \ref{resultsfig_invivo}B. Larger structures can be clearly identified, such as the vein (white arrow) near $(x,z)\approx(1,33)$~mm. Some smaller structures are visible, such as muscle fibers running perpendicular to the field of view (bright spots), but blurring of many of these structures is visible by eye \alex{(see \laura{e.g.} 
the \alex{highlighted areas} in Fig.\ \ref{resultsfig_invivo}B)}. This observation is confirmed by the accompanying Strehl ratio map in Fig.\ \ref{resultsfig_invivo}A, which shows values \alex{inferior to} 0.1 over most areas of the image. In the previous section, Strehl ratio maps of the phantom/plexiglass systems showed values \alex{smaller than} 0.1 for areas which were the most strongly affected by aberration.
These results suggest that the image in Fig.\ \ref{resultsfig_invivo}B is significantly aberrated over the entire spatial area.} 

\laura{To correct for aberration, we apply the technique described in previous sections. Due to the heterogeneity of the tissues examined, it can be expected that there are multiple isoplanatic patches, which should not be assumed to be laterally invariant or of the same spatial extent. For full-field imaging, we thus extend the \alex{
\laura{FOI} scanning method} to an iterative approach that consists in gradually decreasing the \alex{spatial extent of the FOI}. Specifically, this entails correcting as in Eq.\ \ref{correction}, recalculating a new $\mathbf{D}$, and performing a new correction with a smaller window size. The process is iterated until optimal focusing is achieved -- maximization of the Strehl ratio for each focal point. Four window sizes were used: $W(\Delta x,\Delta z)=[(10,20), (7.5,15), (5,10), (3,7.5)]$~mm.} \alex{In \textit{SI Appendix} (Fig. S4), the spatial distribution of aberrations is \laura{exhibited in the}  
evolution of the phase of $\mathbf{\hat{U}_1}$ across the field-of-view. Aberrations are shown to be strongly position-dependent and \laura{to} display high spatial frequencies. Such a configuration would be particularly complicated for conventional adaptive focusing techniques. 
\laura{In contrast,} the transmission matrix is the ideal tool to overcome such complex aberrations.} 

\laura{After correction \alex{(Fig.\ \ref{resultsfig_invivo}C)}, the ultrasound image is \alex{indeed} sharper, with better contrast, and smaller structures can be more easily discerned 
\laura{(see \alex{highlighted areas} in Figs.\ \ref{resultsfig_invivo}B and C)}. The Strehl ratio map shows that the image resolution has been improved over most regions of the image (Fig.\ \ref{resultsfig_invivo}D), with the most significant improvements being at muscle fibers, e.g. at $(x,z)\approx(8,28)$~mm or boundaries between different tissue types, e.g. at $(x,z)\approx(5,12)$~mm. The values of $\mathcal{S}_F\sim 0.2-0.4$ are encouraging, as this is the range of values observed when diffraction-limited resolution was achieved for the phantom  
\laura{system}. \alex{Again, experimental noise and multiple scattering are potential 
\laura{causes for the less than ideal values of $\mathcal{S}_F< 1$}.}}

\subsubsection*{\laura{Discussion}}

\laura{Unlike} \alex{conventional} adaptive focusing whose efficiency range is limited to a single isoplanatic patch 
\laura{(Fig.\ \ref{focusfig})},
a full-field image of the medium under investigation is obtained with diffraction-limited resolution. \laura{Note that other recent approaches for acoustic imaging have also proposed analogous spatial sectioning to correct for spatially-distributed aberration~\cite{Jaeger2015,PFM2019}}. \alex{\laura{In this respect, a} 
key parameter is the choice of the FOI at each iteration. As shown in \textit{SI Appendix} (Section S5), the SVD of $\mathbf{\hat{C}}$ will succeed in \alex{properly} extracting the \alex{local aberration transmittance} if the number $N$ of input focusing points in each FOI is at least four times larger than the number of transverse resolution cells mapping the aberrated focal spot, $M_\delta=\delta x /\delta x_0$.} \laura{This is why the FOI can be reduced after each iteration of the aberration correction process. As the imaging PSF narrows, $M_\delta$ becomes smaller and the required number $N$ of input focusing points decreases.} 


\section*{{Discussion and Conclusion}}

\laura{The distortion matrix approach} provides a powerful tool for imaging inside a heterogeneous medium with \textit{a priori} unknown characteristics. Aberrations can be corrected without any guide stars or prior knowledge of the speed of sound distribution in the medium. While our method is inspired by previous works in ultrasound imaging~\cite{ODonnell1988,Mallart1994,Varslot2004,Robert2008,Montaldo2011}, and is  \laura{built on the recent introduction of the distortion matrix in optics~\cite{Badon2019}}, it  features several distinct and important \laura{advances}. 

The first is its primary building block: the broadband focused reflection matrix that precisely selects the echoes originating from a single scattering event at each depth. 
This operation 
is decisive in terms of signal-to-noise ratio since it drastically reduces the detrimental contribution of out-of-focus and multiply-scattered echoes. Equally importantly, \laura{this matrix} captures all of the input-output spatial correlations of these singly-scattered echoes. 

\laura{T}he approach presented \laura{here} 
also introduces the projection of the reflection matrix in the far-field. \laura{This enables the elimination of artifacts from multiple reflections between parallel surfaces, revealing previously hidden parts of the image. Here, we have only examined reflections from surfaces which are parallel to the ultrasound array, which is more relevant for imaging layered materials than it is for imaging human tissue. While signatures of other flat surfaces should be identifiable as correlations in off-antidiagonal lines of \alex{$\mathbf{R}_{kk}$} or in other mathematical bases~\cite{RodriguezMolares2017}, reverberations from uneven or curved surfaces can not, at present, be addressed with this method.}

\laura{For aberration correction, projection of the reflection matrix into a dual basis allows the isolation of the distorted component. Then,}   
all of the input focal spots can be superimposed onto the same (virtual) location. The normalized correlation of these distorted wavefields, and an average over disorder, then enables the synthesis of a virtual reflector. Unlike related works \laura{in acoustics~\cite{Varslot2004,Robert2008,Montaldo2011,PFM2019},} this virtual scatterer is point-like\laura{, i.e. not limited by the size of the aberrated focal spot}. Moreover, \laura{this approach constitutes a significant advance \alex{over recent works which were limited to aberration correction at either input~\cite{PFM2019} or output~\cite{Badon2019}. Here, we demonstrate} how the randomness of a scattering medium can be leveraged to 
\laura{identify} and correct for aberrations at both input and output. By retrieving the transmission matrix between the elements of the probe and each focal point in the medium, \alex{spatially distributed aberrations can be overcome.} A full-field and diffraction-limited image is recovered. Our approach is thus straightforward, not requiring a tedious iterative \alex{focusing process} to be repeated over each isoplanatic patch.}

\laura{It is important to note that, although the first experimental proof-of-concept involved a relatively simple multi-layered wave velocity distribution, our approach is not at all limited to laterally-invariant aberrations. As shown by the in vivo imaging experiment, the distortion matrix approach also corrects for complex position-dependent aberrations caused by an unknown speed of sound distribution in the medium.}

\alex{Last but not least, we furthermore exploit concepts from information theory. In particular,} \laura{we introduce the idea that, by minimizing the Shannon entropy of the correlation matrix $\hat{\mathbf{C}}$, the \alex{local} acoustic velocity $c$ can be estimated 
\laura{for a chosen field of \alex{illumination}}. Further work will focus on refining this technique for detailed mapping of $c$ \alex{even through strong inhomogeneities}. The wave velocity is actually a quantitative marker for structural health monitoring or biomedical diagnosis. A particularly pertinent example is the measurement of the speed of sound in the human liver, which is decisive for the early detection of non-alcoholic fatty liver diseases~\cite{Imbault2017}, but which must be measured through aberrating layers of fat and muscle. }

\laura{Despite all these exciting perspectives, our matrix approach still suffers from several drawbacks that should be tackled in the near future, One limitation is its restriction to a speckle scattering regime. Theoretically, for specular reflection, the SVD of $\mathbf{D}$ should be examined rather than of $\mathbf{\hat{C}}$~\cite{Badon2019}. Future theoretical developments will examine the exact relation between the singular states of $\mathbf{\hat{C}}$ and $\mathbf{D}$ in a regime which combines speckle and specular scattering.} 
\laura{A second limit lies in our broadband analysis; for a dispersive medium, time reversal of wave distortions -- rather than a simple phase conjugation -- will be required. Thirdly, we are limited by the  size of isoplanatic patch that can be resolved; treatment of high order aberrations requires that sufficiently small patches be resolvable. 
This issue, however, can be partially overcome by gradually reducing the FOI for the distortion matrix. Finally, on a related note, the contribution of multiple scattering has not been thoroughly treated. Although the distortion matrix approach can eliminate most of the multiple scattering background using an SVD process~\cite{Badon2019}, it could in the future take advantage of multiple scattering to using the medium as a scattering lens and improve the resolution beyond the diffraction limit~\cite{Derode1995}. }

\alex{To conclude, the distortion matrix \laura{concept} can be applied} to any field of wave physics for which multi-element technology is available. A reflection matrix approach to wave imaging has already been initiated both in optical microscopy~\cite{Kang2015,Badon2016,Kang2017,Badon2019, Kim2019}, MIMO radar imaging~\cite{Roberts2010} and seismology~\cite{Blondel2018}. 
\laura{The ability to apply the distortion matrix to random media (not just specular reflectors) should be valuable for optical deep imaging in biological tissues~\cite{badon2017multiple}}. At the other end of the \laura{spatial} scale, volcanoes and fault zones are particularly heterogeneous areas~\cite{Blondel2018} in which the distortion matrix concept could be fruitful for a bulk seismic imaging of the Earth's crust beyond a few kilometers in depth. The reflection/distortion matrix concept is thus universal. The potential range of applications of this approach is wide and highly promising, whether it be for a direct imaging of the medium reflectivity, or a quantitative and local characterization of the wave speed~\cite{Imbault2017,Lambert2019},
absorption~\cite{Aubry2011}, and scattering~\cite{Aubry2008,Mohanty2017} parameters.

\section*{Material and Methods}

\subsection*{\alex{Tissue mimicking phantom experiment}}
The experimental setup consisted in an \laura{1D} ultrasound phased-array probe (SuperLinear$^{\text{TM}}$ SL15-4) connected to an ultrafast scanner (Aixplorer\textregistered, SuperSonic Imagine, Aix-en-provence, France). \laura{The array contains 256 elements with pitch} $p=0.2$~mm. The acquisition sequence consisted of emission of plane waves at 49 incident angles $\tin$ spanning $-24^o$ to $24^o$. The emitted signal was a {sinusoidal burst} of central frequency $f_0=7.5$~MHz \laura{and bandwidth of} $2.5$ to $10$~MHz. In reception, all elements were used to record the reflected wavefield over a time length $\Delta t= 124$~$\mu$s at a sampling frequency of $30$~MHz. 

\alex{\subsection*{\textit{In vivo} imaging experiment}
The in vivo ultrasound dataset was collected by the SuperSonic Imagine company on a healthy volunteer from which informed consent had been obtained. Before being put at our disposal, this dataset
was previously fully anonymized following standard practice defined by Commission nationale de l'information et des libert\'{e}s (CNIL). The experimental setup consisted in a \laura{1D} $5-18$~MHz linear transducer array (SL18-5, Supersonic Imagine) connected to an ultrafast scanner (Aixplorer Mach-30, Supersonic Imagine, Aix-en-Provence, France). \laura{The array contains 192 elements with pitch} 
$p=0.2$~mm. The probe was placed in direct contact with the calf of the healthy volunteer, orthogonally to the muscular fibers. An informed consent was obtained from the healthy volunteer. The ultrasound sequence consisted in transmitting $101$ \laura{plane waves} \laura{at} incident angles spanning $-25^{o}$ to $25^{o}$, \laura{calculated using a speed of sound hypothesis of $c_0=1580$~m/s. The pulse repetition rate was $1000$~Hz. The emitted signal was a sinusoidal burst of three half periods of central frequency $f_c=7.5$~MHz. For each excitation, all elements recorded the backscattered signal over a time length $\Delta t = 80 \mu$s at a sampling frequency of $40$~MHz.} This ultrasound emission sequence meets the FDA Track 3 Recommendations.}

\subsection*{Multiple reflection filter}
The multiple reflection filter consists in applying an adaptive Gaussian filter to remove the specular contribution that lies along the main antidiagonal of $\mathbf{R}_{kk}$, such that
\begin{equation}
\laura{ R(\kout,\kin) =  R(\kout,\kin) \left [ 1 - \alpha e^{-|\kout+\kin|^2/\delta k^2} \right]. } 
\end{equation}
The width $\delta k$ of the Gaussian filter scales as the inverse of the transverse dimension $\Delta x$ of the field of view: $\delta k=\Delta x^{-1}$. The parameter $\alpha$ defines the strength of the filter:
    \begin{equation}
\alpha= \frac{\langle |  R(\kout,\kin) |  \rangle_{\Delta k > \delta k }}{\langle |  R(\kout,\kin) |  \rangle_{\Delta k < \delta k }}-1.
\end{equation}
where the symbol $\langle \cdots \rangle$ denotes an average over the couples $(\kout,\kin)$ separated by a distance $\Delta k=|\kout-\kin|$ smaller or larger than $\delta k$. When the specular component dominates, the parameter $\alpha$ tends to 1 and the Gaussian filter is fully applied: The main antidiagonal of $\mathbf{R}_{kk}$ is then set to zero (see Fig.\ref{fig_reverb}f). When there is no peculiar specular contribution, the parameter $\alpha$ tends to 0 and the Gaussian filter is not applied: The main antidiagonal of $\mathbf{R}_{kk}$ remains unchanged.

\subsection*{Data availability}
Data used in this manuscript have been deposited at figshare
(\href{https://figshare.com/projects/Distortion_matrix_approach_for_full-field_imaging_of_random_scattering_media_2020/78141}{https://figshare.com/projects/Distortion\_matrix\_approach\_for\_full-field\_imaging\_of\_random\_scattering\_media\_2020/78141}).

\section*{ACKNOWLEDGMENTS}

The authors wish to thank Victor Barolle, Amaury Badon and Thibaud Blondel whose own research works in optics and seismology inspired this study. The authors are grateful for the funding provided by Labex WIFI (Laboratory of Excellence within the French Program Investments for the Future) (ANR-10-LABX-24 and ANR-10-IDEX-0001-02 PSL*). W.L. acknowledges financial support from the Supersonic Imagine company. L.C. acknowledges financial support from the European Union's Horizon 2020 research and innovation programme under the Marie Sk\l{}odowska-Curie grant agreement No. 744840.  This project has received funding from the European Research Council (ERC) under the European Union's Horizon 2020 research and innovation programme (grant agreement No. 819261, REMINISCENCE). 

\vspace{\baselineskip}

\clearpage 

\clearpage

\renewcommand{\thetable}{S\arabic{table}}
\renewcommand{\thefigure}{S\arabic{figure}}
\renewcommand{\theequation}{S\arabic{equation}}
\renewcommand{\thesection}{S\arabic{section}}

\setcounter{equation}{0}
\setcounter{figure}{0}

\begin{center}
\Large{\bf{Supplementary Information}}
\end{center}
\normalsize
This Supplementary Information is dedicated to the theoretical derivation of the reflection, distortion and corresponding correlation matrices in the isoplanatic limit.

\section{Reflection matrix in the far-field basis}

As stated in the accompanying paper (Eq. 6), the far-field reflection matrix $\mathbf{R}_{kk}$ can be expressed as follows:
\begin{equation}
\label{Rtt2b}
   \mathbf{R}_{kk}(z)=\mathbf{T} \times \mathbf{\Gamma}(z) \times  \mathbf{T}^{\top} .
\end{equation}
\alex{where $\mathbf{\Gamma}(z)=[\Gamma(x,x^\prime,z)]$ describes the scattering processes inside the medium.} In the isoplanatic limit, the aberrations can be modelled by a far-field phase screen of transmittance $\mathbf{\tilde{H}}=[\tilde{H}(k_x)]$, where $\tilde{H}(k_x)=\int dx H(x) \exp (- i k_{x} x )$ is the 1D Fourier transform of the input or output point spread function\laura{,} $H(x)$\laura{, which is} defined in Eq.~9 of the accompanying paper. The transmission matrix $\mathbf{T}$ can then be expressed as an Hadamard product between $\mathbf{\tilde{H}}$ and $\mathbf{T_0}$, the free-space transmission matrix,
\begin{equation}
\label{isoplanatic}
\mathbf{T}= \mathbf{\tilde{H}} \circ \mathbf{T_0}.
\end{equation}
The injection of this last equation into Eq.~\ref{Rtt2b} yields the following expression for the far-field reflection matrix coefficients 
\begin{equation}
\label{isoplanatic3}
R(\kout,\kin,z)=\tilde{H}(\kin)\tilde{\gamma}(\kin+\kout,z)  \tilde{H}(\kout) ,
\end{equation}
where $\tilde{\gamma}(k_x,z)=\int dx\, \gamma(x,z) \exp (- i k_{x} x )$ is the 1D Fourier transform of the sample reflectivity $\gamma(x,z)$. Assuming the aberration as a phase screen ($|\tilde{H}(k_x)|=1$), the norm-square of the coefficients of $\mathbf{R}_{kk}(z)$ (Eq.~\ref{isoplanatic3}) are given by 
\begin{equation}
 |   R(\kout,\kin,z) |^2= \left  | \tilde{\gamma}(\kout+\kin,z) \right |^2 .
\end{equation}
Each antidiagonal of $ \mathbf{R}_{kk}$ (\laura{where} $\kin+ \kout=$ constant) encodes one spatial frequency of the sample reflectivity.

\section{Reflection matrix in the dual basis}

The dual reflection matrix $\mathbf{R}_{k \mathbf{r}}=    \mathbf{R}_{k x}(z)$ is obtained by projecting $\mathbf{R}_{kk}$ into the focused basis in emission:
\begin{equation}
    \mathbf{R}_{k x}(z)=\mathbf{R}_{k k}(z)\times \mathbf{T}_0^* .
\end{equation}
Injecting Eq.~\ref{Rtt2b} into this last equation leads to the following expression for $\mathbf{R}_{k x}(z)$:
\begin{equation}
\label{Rtr}
    \mathbf{R}_{k x}(z)=\mathbf{T}  \times \mathbf{\Gamma}(z) \times  \mathbf{H}^{\top}  ,
\end{equation}
where $\mathbf{H}=\mathbf{T}_0^{\dagger} \mathbf{T} $ is the focusing matrix whose columns corresponds to the input focal spots $H(x,\xin)$. In the isoplanatic limit, $H(x,\xin)=H(x-\xin)$. The elements of $\mathbf{R}_{k \mathbf{r}}$ can then be expressed as
\laura{\begin{equation}
\label{Rkr}
R(\kout,\mathbf{r_{in}})= \int dx\: T (\kout,x) \gamma(x,z) H(x-\xin) .
\end{equation}}

To investigate the far-field correlations of the reflected wavefield, the spatial correlation matrix 
\laura{$\mathbf{B}=\mathbf{R}_{k \mathbf{r}} \mathbf{R}_{k \mathbf{r}}^{\dag}$} should be considered.
$\mathbf{B}$ can be decomposed as the sum of a covariance matrix $\left \langle \mathbf{B}\right \rangle $ and a perturbation term $\delta  \mathbf{B} $:
\begin{equation}
\label{B}
   \mathbf{B}= \left \langle \mathbf{B}\right \rangle+\delta  \mathbf{B} ,
\end{equation}
where the symbol $\langle \cdots \rangle$ denotes an ensemble average.


In the speckle regime, the random nature of the sample reflectivity $\gamma(\mathbf{r})$ means that  $\langle\gamma(\mathbf{r})\gamma^*(\mathbf{r^\prime})\rangle= \langle\left|\gamma\right|^2\rangle\delta(\mathbf{r}-\mathbf{r^\prime})$, where $\delta$ is the Dirac distribution. The correlation matrix should converge towards the covariance matrix $\left \langle \mathbf{B}\right \rangle$ for a large number of independent realizations. More precisely, the intensity of the perturbation term in Eq.\ref{B}, 
\laura{$|\delta B(k,k')|$}, should scale as the inverse of $M$, the number of independent resolution cells contained in the field of view~\cite{Robert_thesis,Priestley1988,Goodman2000}.

Assuming the convergence of $\mathbf{B}$ towards $\langle \mathbf{B} \rangle $ in the speckle regime, the correlation coefficients $B(k_\textrm{out},k^\prime_\mathrm{out})$ can be expressed as follows 
\begin{equation}
\label{B2}
 B (\kout,k'_\mathrm{out}) = \langle \left | \gamma \right |^2 \rangle  \int x\: T(\kout,x) \gamma_R(x) T^*(k'_\mathrm{out},x)\\
\end{equation}
where 
\begin{equation}
\label{GammaR}
  \gamma_R(x)=\int d \mathbf{r_{in}}\: \Omega (\xin,z) \left | H(x-\xin) \right |^2 . 
  \end{equation}
The function $\Omega (\mathbf{r})$ denotes the spatial domain occupied by the field of view:
\begin{equation*}
\label{window2}
\Omega(\mathbf{r})=\left\{
\begin{array}{ll}
1 & \, \mbox{for }\mathbf{r}\mbox{ inside the field of view}  \\
0 & \,  \mbox{otherwise}.
\end{array} 
\right.
\end{equation*} 
Equation~\ref{B2} can be rewritten as the following matrix product
\begin{eqnarray}
\label{B3}
\mathbf{B} \propto \mathbf{T} \times \mathbf{\Gamma_R}  \times \mathbf{T^{\dag}} ,
\end{eqnarray}
where $\mathbf{\Gamma_R}$ is a diagonal scattering matrix associated with a virtual object. Its coefficients $\gamma_R(x)$ correspond to the convolution of the input focal spot intensity $|H(x)|^2$ with the whole field of view $\Omega(x)$. Its spatial extent thus spans the entire field of view 
\laura{(Fig.\ \ref{fdort})}.

Expressed in the form of Eq.\ \ref{B3}, $\mathbf{B}$ is analogous to the time-reversal operator obtained for a single scatterer of reflectivity $\gamma_R(x)$~\cite{Varslot2004,Robert2008}. If this virtual scatterer were point-like, $\mathbf{B}$
would be of rank 1 and its eigenvector would correspond to the wavefront that focuses perfectly through the heterogeneous medium onto the virtual scatterer, even in presence of strong aberrations~\cite{Prada1994,Prada1996}. Here, however, this is far from being the case. The eigenvalue decomposition of $\mathbf{B}$ yields a set of eigenmodes which focus 
on different parts of the same virtual scatterer over restricted angular domains~\cite{Aubry2006,Robert2009}. One solution to this degeneracy is to limit the field of view in order to reduce the size of the virtual reflector and increase the angular aperture of the eigenwavefronts. However, a restricted field of view means a lack of averaging over disorder. The perturbation term in Eq.~\ref{B} would no longer be negligible and iterative time reversal would not converge towards the optimal aberration phase law. As we will see in the next section, the distortion matrix concept avoids this impossible compromise.

\section{Distortion matrix in the dual basis}

The distortion matrix $\mathbf{{D}}(z)$ is defined as the Hadamard product between the reflection matrix $\mathbf{{R}}_{kx}(z)$ and the reference transmission matrix $\mathbf{T}_0^{*}$. In terms of matrix coefficients, this can be written  
\begin{equation}
D(\kout,\rin)  =  {R}(\kout,\rin)  T_0^{*}(\kout,\rin)
\label{D_eq_coefficients2} .
\end{equation}   
Injecting Eqs.~\ref{isoplanatic} and \ref{Rkr} into the last equation yields the following expression for $D(\kout,\rin)$: 
\begin{equation}
\label{Dkr}
D(\kout,\mathbf{r_{in}})= \hat{H}(\kout)\int dx\: T_0(\kout,x-\xin) \gamma(x,z) H(x-\xin) .
\end{equation}
To investigate the far-field correlations of the distorted wavefield, the spatial correlation matrix $\mathbf{C} = \mathbf{D} \mathbf{D}^{\dag}$ is investigated.
As previously observed with $\mathbf{B}$, $\mathbf{C}$ should converge towards the covariance matrix $\left \langle \mathbf{C}\right \rangle$ for a large number of independent realizations, i.e. a large number $N$ of input focusing points. Assuming this condition is fulfilled in the speckle regime, the correlation coefficients $C(\kout,k'_\mathrm{out})$ can be expressed as follows 
\begin{equation}
\label{C2}
C (\kout,k^\prime_\mathrm{out}) = \langle \left | \gamma \right |^2 \rangle  \int dx^\prime T(\kout,x^\prime) \gamma_D(x^\prime) T^*(k^\prime_\mathrm{out},x^\prime) ,
\end{equation}
where 
\begin{equation}
\label{GammaD}
  \gamma_D(x^\prime)= \left | H(x^\prime) \right |^2 .
\end{equation}
Equation~\ref{C2} can be rewritten as the following matrix product
\begin{eqnarray}
\label{C3}
\mathbf{C} \propto \mathbf{T} \times \mathbf{\Gamma_D}  \times \mathbf{T^{\dag}} ,
\end{eqnarray}
where $\mathbf{\Gamma_D}$ is a diagonal scattering matrix associated with a virtual object centered at the origin. Its coefficients $\gamma_D(x)$ correspond to the input focal spot intensity $|H(x)|^2$\laura{.} 
Expressed in the form of Eq.\ \ref{C3}, $\mathbf{C}$ is analogous to the time-reversal operator obtained for a single scatterer of reflectivity $\gamma_D(x)$~\cite{Varslot2004,Robert2008}.
The unscrambling of input focal spots which is made possible by $\mathbf{D}$ allows the size of the virtual reflector to be reduced to $\delta x$, the dimension of the aberrated focal spot \alex{(see Fig.~4D of the accompanying paper)}. This is an important improvement over what is offered by $\mathbf{B}$, the correlation matrix constructed from reflection matrix $\mathbf{R}$.

\section{Normalized correlation matrix}

As we will see now, this virtual reflector can even be made point-like by considering a normalized correlation matrix (Eq.~18 of the accompanying paper). To demonstrate this assertion, Eq.\ref{C3} is rewritten with the help of Eq.~\ref{isoplanatic}:
\begin{equation}
\label{C4}
C (\kout,k'_\mathrm{out}) \propto \langle \left | \gamma \right |^2 \rangle \hat{H}(\kout) \hat{H}^*(k'_\mathrm{out}) \left [\hat{H} \ast \hat{H} \right ](\kout-k'_\mathrm{out}),
\end{equation}
where the symbol $\ast$ stands for a correlation product. This correlation term in Eq.~\ref{C4} results from the Fourier transform of the input focal spot $|H(x)|^2$ in Eq.\ref{C3}. This formulation is reminiscent of the Van Cittert Zernicke theorem for an aberrating layer, which links the spatial correlation of a wavefield to the Fourier transform of the intensity distribution from an incoherent source (here the input focal spots)~ \cite{Mallart1994}. In other words, the support of the  coherence function $\left [\hat{H} \ast \hat{H} \right ](\kout-k'_\mathrm{out})$ scales as the inverse of the input focal spot size $\delta x$.

The approach for reducing the size of this virtual scatterer is to render the autocorrelation term flat. Since $|H(k_x)|=1$, this can be done by considering the normalized correlation matrix $\mathbf{\hat{C}}$ (Eq. 17) whose coefficients are given by 
\begin{equation}
\label{Ch}
\hat{C} (\kout,k'_\mathrm{out}) \propto \langle \left | \gamma \right |^2 \rangle \hat{H}(\kout) \hat{H}^*(k'_\mathrm{out}). 
\end{equation}
This last equation is valid if the convergence of $\mathbf{C}$ towards the covariance matrix $\langle \mathbf{C} \rangle$ is achieved, i.e. if a large enough number $N$ of input focusing points is considered. Equation~\ref{Ch} can be rewritten as the following matrix product
\begin{eqnarray}
\label{Ch2}
\mathbf{C} \propto \mathbf{T} \times \mathbf{\Gamma_\delta}  \times \mathbf{T^{\dag}} ,
\end{eqnarray}
where $\mathbf{\Gamma_\delta}$ is a diagonal scattering matrix associated with a point-like scatterer centered at the origin, such that $\gamma_\delta(x)=\delta(x)$ (see Fig.~4E of the accompanying paper). 

Expressed in the form of Eq.\ \ref{Ch2}, $\mathbf{\hat{C}}$ is analogous to the time-reversal operator obtained for a point-like scatterer. Equation~\ref{Ch} confirms that $\mathbf{\hat{C}}$ is of rank 1 and the corresponding eigenvector $\mathbf{U_1}$ directly provides the aberration phase law:
\begin{equation}
    \mathbf{U_1} \equiv [\hat{H}(\kout)] .
\end{equation}
An estimator $\mathbf{\bar{T}}$ of the transmission matrix can then be deduced (Eq.22 of the accompanying paper).

\section{Convergence of the matrix approach}
\alex{
Until now, for sake of simplicity, we have assumed that $\mathbf{\hat{C}}$ \laura{converges} 
towards the covariance matrix $\langle \mathbf{\hat{C}}\rangle$. However, this is not always the case for restricted fields-of-illumination. Even if this convergence is not \laura{achieved}
, the covariance matrix $\langle \mathbf{\hat{C}}\rangle$ can still be retrieved by means of the eigenvalue decomposition of $\mathbf{\hat{C}}$. In the isoplanatic limit, the covariance matrix $\langle \mathbf{\hat{C}} \rangle$ is indeed of rank 1 (Eq.~\ref{Ch}). We expect the aberration transmittance $\hat{H}(\kout)$ to be \laura{contained in} 
first eigenvector $\mathbf{\hat{U}_1}$ of $\mathbf{\hat{C}}$. To derive a necessary condition for this, the matrix $\mathbf{\hat{C}}$ should be written as a sum of the covariance matrix $\left \langle \mathbf{\hat{C}} \right \rangle  $ and a perturbation term $ \delta \mathbf{\hat{C}}   $:
\begin{equation}
\label{eqhatC}
 \mathbf{\hat{C}}=\left \langle \mathbf{\hat{C}} \right \rangle +  \delta \mathbf{\hat{C}} \laura{.}
\end{equation}
The matrix $\langle \mathbf{\hat{C}} \rangle$ is of rank 1 and associated 
\laura{with} a single eigenvalue $\hat{\sigma}_1$. The matrix $\delta \mathbf{\hat{C}} $ can be considered as a correlated random matrix. Its rank is equal to the number $M_\delta$ of independent speckle grains in the far-field. In first approximation, the eigenvalues of $\delta \mathbf{\hat{C}} $ can be assumed to follow the eigenvalue distribution of a Hermitian random matrix of size $M_{\delta}$. $\langle \mathbf{\hat{C}} \rangle$ will emerge along the first eigenstate of  $\mathbf{\hat{C}}$ if~\cite{Aubry2010}
\begin{equation}
\label{cond}
    \frac{\hat{\sigma}_1^2}{\left \langle \sum_{i=1}^{M_\delta} \sigma_i^2 \right \rangle } >   \frac{4}{{M_\delta}} .
\end{equation}
The factor 4 comes from the superior bound of the Marcenko-Pastur law~\cite{marcenko}\laura { -- }
the distribution that the normalized squared eigenvalues of $\delta \mathbf{\hat{C}} $ are supposed to follow. To make this last inequality more explicit, we 
\laura{express} the first eigenvalue $\hat{\sigma}_1$ of the covariance matrix and the mean sum of the squared eigenvalues of $\mathbf{C}$, $\left \langle  \sum_{i=1}^{M_\delta} \sigma_i^2 \right \rangle $. On 
\laura{one} hand, because $\langle \mathbf{\hat{C}}\rangle $ is of rank 1, the square of its eigenvalue $\hat{\sigma}_1^2$ is equal to the trace of $\langle \mathbf{\hat{C}}\rangle \langle \mathbf{\hat{C}}\rangle ^{\dag} $:
\begin{equation}
\label{sigma1}
    \hat{\sigma}_1^2= \sum_{\kout} \sum_{\kpout} |\langle \hat{C} \rangle (\kout,\kpout)  |^2 = N_k^2 \laura{,}
\end{equation}
where $N_k$ is the dimension of the matrix $\mathbf{C}$. On the other hand, the mean sum of the squared eigenvalues $\sigma_i^2$ is equal to the trace of \alex{$\mathbf{C}\mathbf{C}^\dag$}:
\begin{equation}
\left  \langle   \sum_{i=1}^{M_\delta} \sigma_i^2 \right \rangle  =  \left  \langle \sum_{\kout} \sum_{\kpout} |\hat{C}(\kout,k'_\mathrm{out})|^2  \right \rangle \laura{.}
\end{equation}
Injecting Eq.~\ref{eqhatC} into the last equation yields
\begin{eqnarray}
\left  \langle   \sum_{i=1}^{M_\delta} \sigma_i^2 \right \rangle & = & \sum_{\kout} \sum_{k'_\mathrm{out}}  | \langle \hat{C} \rangle (\kout,k'_\mathrm{out})  |^2 + \sum_{\kout} \sum_{k'_\mathrm{out}}  \langle | \delta \hat{C} (\kout,k'_\mathrm{out})  |^2 \rangle \nonumber \\
&= & N_k^2 + \sum_{\kout} \sum_{k'_\mathrm{out}}  \langle | \delta \hat{C} (\kout,k'_\mathrm{out})  |^2 \rangle \laura{.}
\label{eq0}
\end{eqnarray}
Since the coefficients $\hat{C}(\kout,k'_\textrm{out})$ are of modulus 1, their variance is directly given by their phase fluctuations:
\begin{equation}
\label{eq1}
  \left \langle \left | \delta \hat{C}(\kout,k'_\textrm{out}) \right |^2 \right \rangle    = \left \langle  \left |  \mbox{arg} \left \lbrace \hat{C}(\kout,k'_\mathrm{out}) \right \rbrace \right |^2 \laura{.} \right\rangle 
\end{equation}
For a large number 
\laura{$N$} of input focusing points, the variance $\left \langle \left | \delta \hat{C}(\kout,k'_\mathrm{out}) \right |^2 \right \rangle$ of the normalized correlation matrix coefficients can be expressed as follows~\cite{Priestley1988,Robert_thesis}
\begin{equation}
\label{eq2}
\left \langle  \left |  \mbox{arg} \left \lbrace \hat{C}(\kout,k'_\mathrm{out}) \right \rbrace \right |^2 \right\rangle  \simeq  N^{-1} \left( |C( \kout,k'_\mathrm{out})|^{-2} -1 \right) \laura{.}
\end{equation}
Injecting Eqs.~\ref{eq1} and \ref{eq2} into Eq.~\ref{eq0} leads to
\begin{equation}
\left  \langle   \sum_{i=1}^{M_{\delta}} \sigma_i^2 \right \rangle 
\simeq N^{-1} \sum_{\kout} \sum_{k'_\mathrm{out}}    |C( \kout,k'_\mathrm{out})|^{-2}  \laura{.}
\label{eq0b}
\end{equation}
For analytical tractability, we will replace $|C( \kout,k'_\mathrm{out})|$ by its average over all 
\laura{pairs} $(\kout,k'_\mathrm{out})$. Interestingly, this mean correlation value 
\laura{scales directly} as the inverse of $M_{\delta}$, \laura{where $M_{\delta}$ is} the number of independent speckle grains in the far-field~\cite{Badon2019}\laura{.} The previous equation can then be simplified as follows:
\begin{equation}
\label{eq4}
\left  \langle   \sum_{i=1}^{M_{\delta}} \sigma_i^2 \right \rangle  \sim   N_k^2 M_{\delta}^2 /N  \laura{.}
\end{equation}
Injecting Eqs.~\ref{sigma1} and \ref{eq4} into Eq.~\ref{cond} yields our final expression of the success condition:
\begin{equation}
\label{eq5}
 N > 4 M_{\delta} \laura{.}
\end{equation}
The number $N$ of input focusing points in the field-of-illumination should be large compared to the number $M_{\delta}$ of independent speckle grains in the far-field. 
The latter quantity is equal to the ratio between the support $\Delta k$ of 
\laura{the distorted wavefield} in the spatial frequency domain\laura{,} and its correlation width $\delta k$: 
\begin{equation}
    M_\delta=\Delta k/\delta k  \laura{.}
\end{equation}
Each distorted 
\laura{wavefield} is produced by a virtual incoherent source in the focal plane whose size is given by the extension $\delta x$ of the input focal spot. \laura{By} 
the van Cittert Zernike theorem, the correlation width $\delta k$ 
\laura{scales} as the inverse of $\delta x$. Reciprocally, the spatial frequency support $\Delta k$ scales as the inverse of the coherence length of the 
\laura{wavefield} in the focal plane, i.e the resolution cell $\delta x_0$. $M_\delta$ is thus equal to the number of resolution cells $\delta x_0$ mapping the aberrated focal spot:
\begin{equation}
    M_{\delta}={\delta x}/{\delta x_0} \laura{.}
\end{equation}
The condition of Eq.~\ref{eq5} can thus be translated as follows
\laura{: t}he number $N$ of input focusing points forming the field-of-illumination should be one order of magnitude larger than the number of resolution cells mapping the aberrated focal spot. This result is fundamental since it governs our strategy for full-field imaging. An iterative procedure is employed and consists in progressively correcting aberrations over smaller and smaller fields-of-illumination.}

\begin{figure*}[htbp]
	\centering
	\includegraphics[width=17cm]{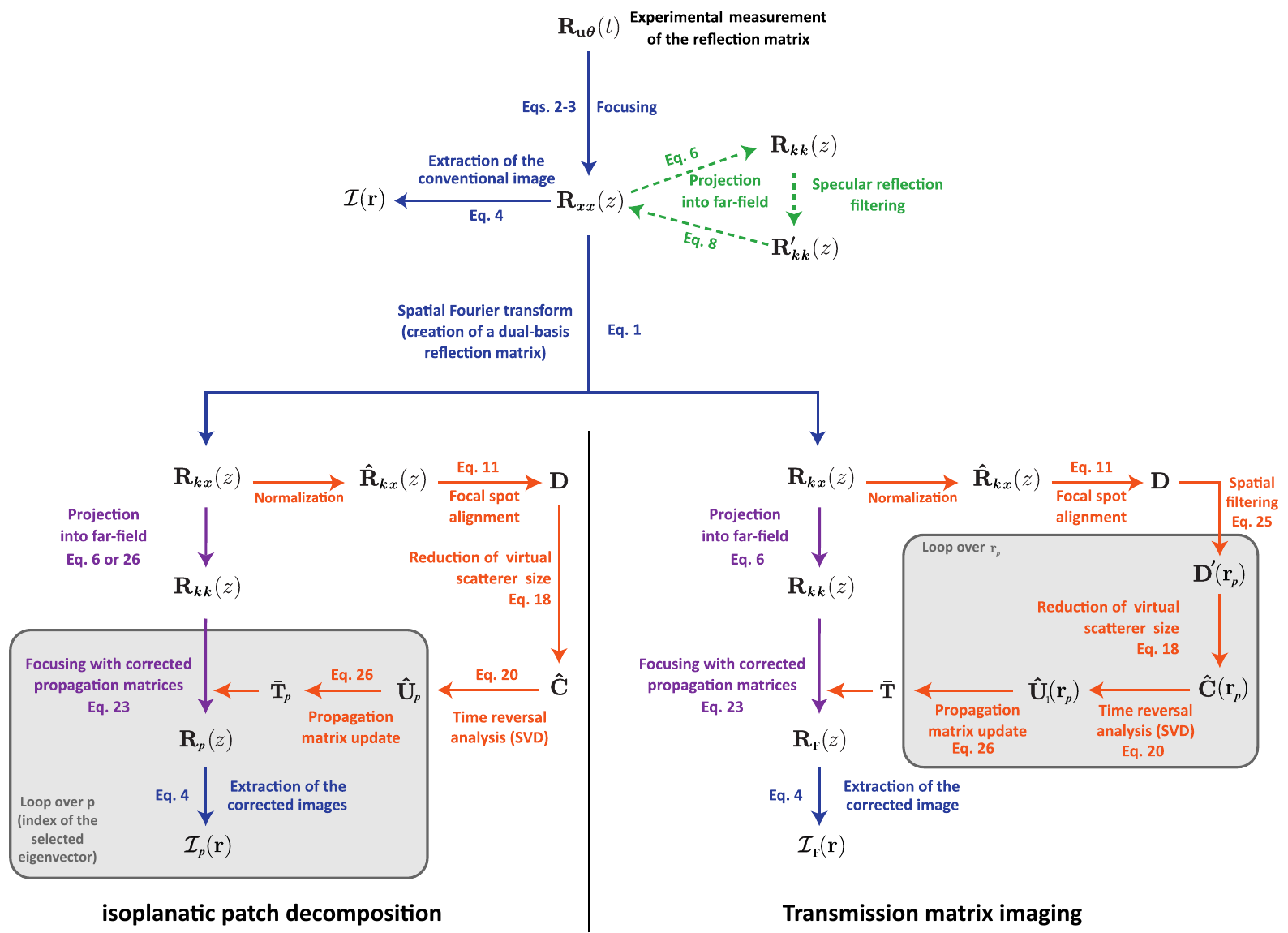}
	\caption{\laura{\textbf{Schematic of mathematical operations for the matrix approach to ultrasound imaging}: along the green path (top dashed lines), reverberation suppression is performed. Along the orange path (bottom 
	\laura{loops}), the distortion matrix approach is applied to extract aberration laws, and to use them to update the estimation of the propagation matrix $\mathbf{\bar{T}_i}$. \laura{Two different options are possible: \textit{(i)} decomposition of the field-of-view into multiple isoplanatic patches, i.e. correction with multiple eigenvectors (bottom left loop), or \textit{(ii)} transmission matrix imaging, in which the reflection matrix and image are corrected for a finite, moving spatial window (bottom right loop).} Along the purple path\laura{s} (vertical arrows), the experimental data is corrected for aberration. Equation numbers correspond to those in the main text.}}
\label{fig_schema}
\end{figure*}

\begin{figure*}[t]
	\centering
	\includegraphics[width=10cm]{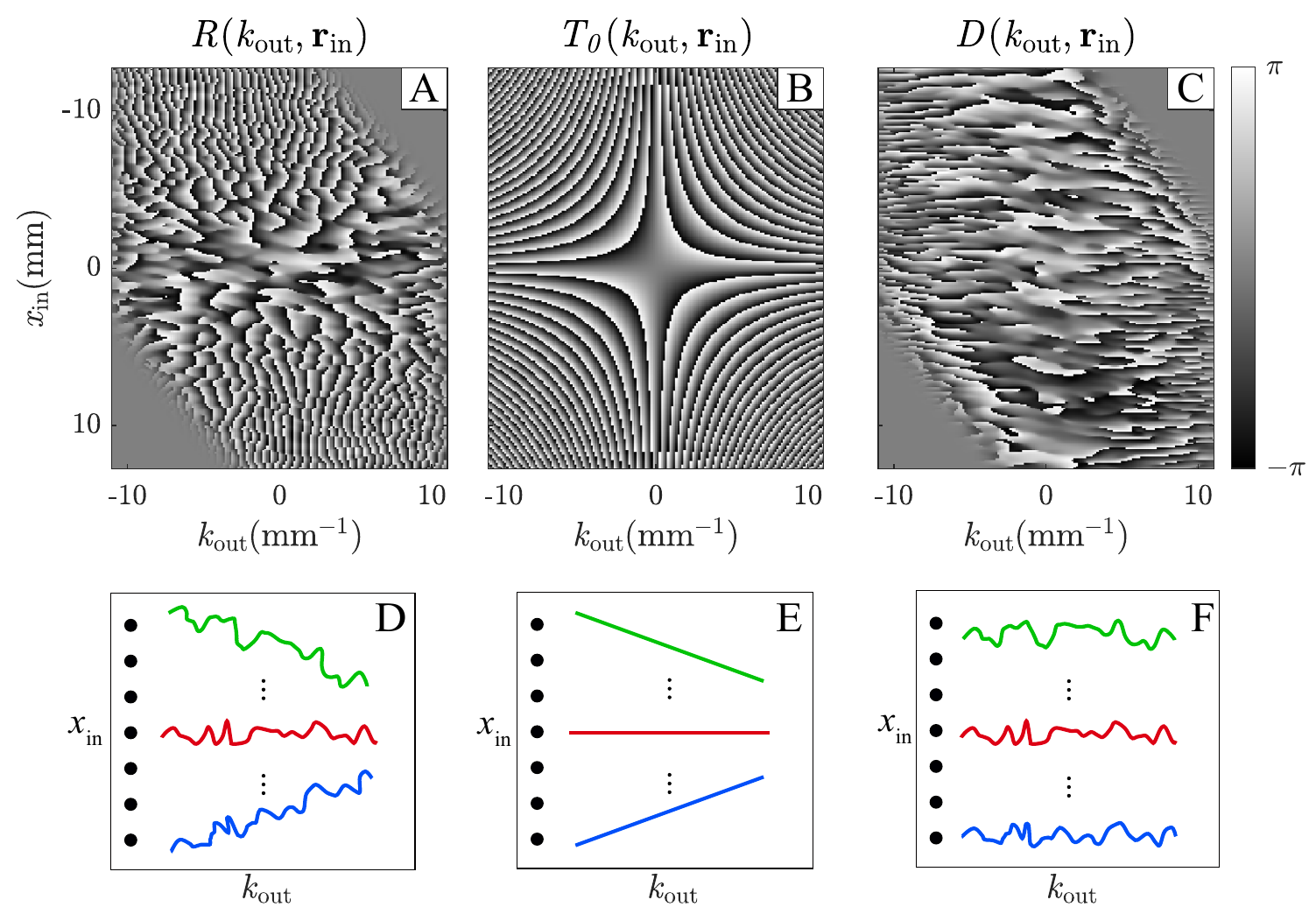}
	\caption{\textbf{Revealing the correlations of the reflected wavefield.} (\textbf{A}) Phase of the reflection matrix $\mathbf{R}_{kx}(z)$ at depth $z=30$ mm. (\textbf{B}) Phase of the free-space transmission matrix $\mathbf{T}_0$ at the same depth ($c=1800$ m/s). (\textbf{C}) Phase of the distortion matrix $\mathbf{D}$ deduced from $\mathbf{R}_{kx}(z)$ and $\mathbf{T}_0$. (\textbf{D},\textbf{E}, \textbf{F}) Sketches of the wavefronts contained in the matrices displayed in (\textbf{A},\textbf{B},\textbf{C})}.
	\label{Dsketches}
\end{figure*}

\begin{figure*}
	\centering
	\includegraphics[width=17 cm]{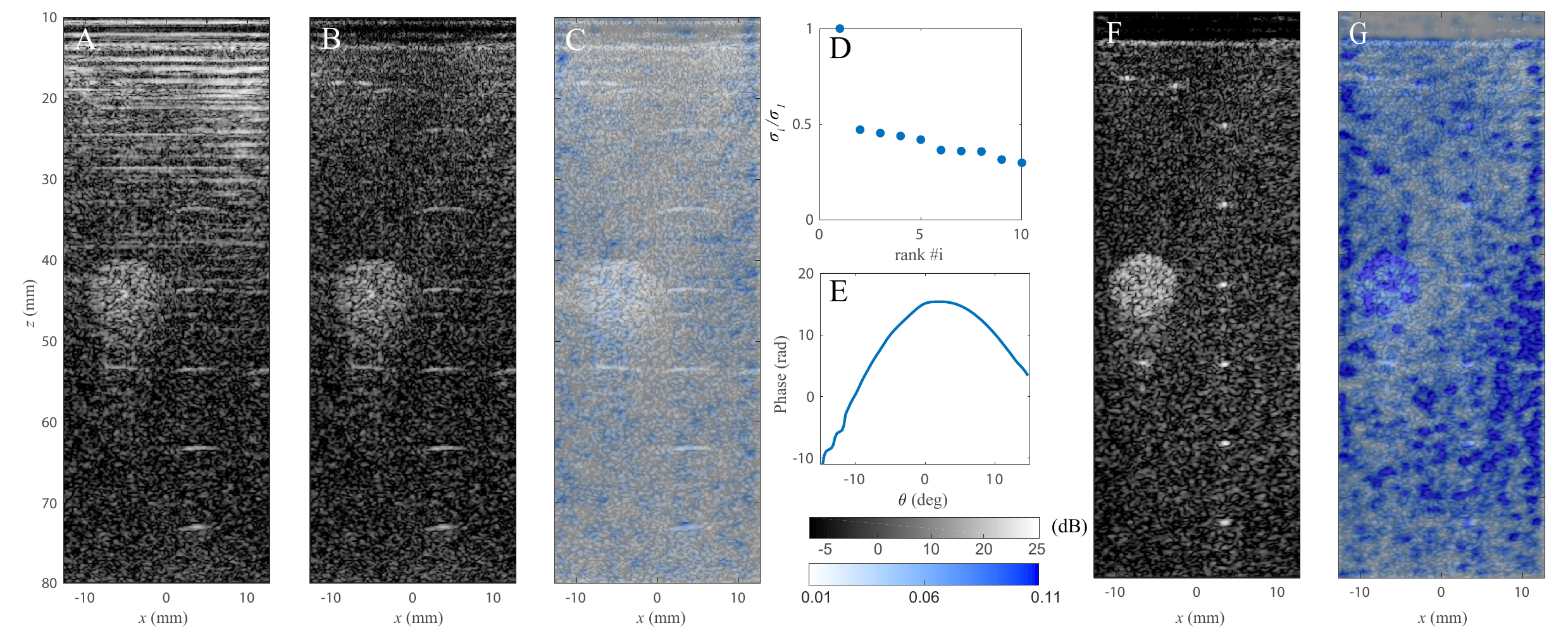}
	\caption{\textbf{Matrix imaging based with a correct phantom wave velocity model.}  (\textbf{A}) Original ultrasound confocal image (Eq.~4 of the accompanying paper) with $c=1542$~m/s. (\textbf{B}) The ultrasound image after the removal of multiple reflections is shown, along with (\textbf{C}) the corresponding map of the Strehl ratio $\mathcal{S}$.
(\textbf{D}) The normalized correlation matrix displays an eigenvalue spectrum dominated by one eigenstate. (\textbf{E}) The phase of the corresponding eigenvector is used to correct for aberrations both at input and output.
	(\textbf{F}) The ultrasound image after matrix aberration correction is shown, with (\textbf{G}) the corresponding map of the Strehl ratio $\mathcal{S_F}$.  The ultrasound images and Strehl ratio maps are displayed with the same dB- (B\&W) and linear (color) scales, respectively. Because the wave velocity model is correct in the phantom, the field of view is contained in a single isoplanatic patch. However, the Strehl ratio reached after correction ($\mathcal{S}_F\sim \alex{0.2}$) is lower than with a wave velocity model c=1800 m/s ($\mathcal{S}_F\sim \laura{0.4}$, see Fig.~5 of the accompanying paper). Aberrations are indeed stronger in the current case because of a larger contrast with the aberrating layer ($c \sim 2750$ m/s). The signal-to-noise ratio is thus weaker and the aberration correction process less efficient. }
	\label{resultsfig2} 
\end{figure*}

\begin{figure*}
	\centering
	\includegraphics[width=15 cm]{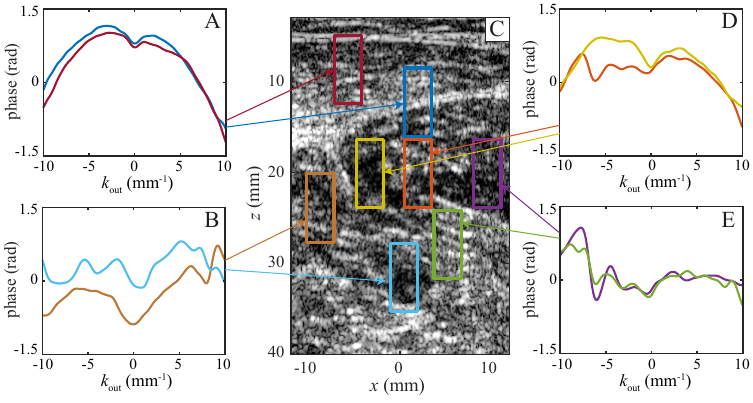}
	\caption{\alex{\textbf{Spatial distribution of the aberration transmittance in the in vivo imaging experiment.} The phase of the first singular vector $\mathbf{\hat{U}_1}$ is displayed for the different rectangular areas of identical color superimposed to the ultrasound image displayed in C. 
	\laura{In each panel A, B, D, and E, two aberration phase laws are shown: each of these pairs (and their corresponding rectangular areas on the ultrasound image) belong to the same isoplanatic patch.}}}
	\label{resultsfig3} 
\end{figure*}

\begin{figure}
	\centering
	\includegraphics[width=8.5 cm]{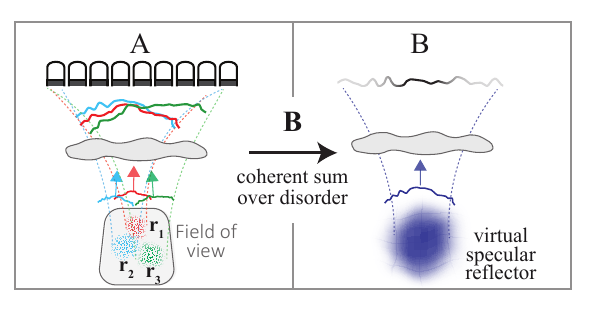}
	\caption{\textbf{Time reversal analysis of the reflection matrix}. ($\textbf{A}$) Each row of $\mathbf{R}_{kx}(z)$ corresponds to the reflected wavefield induced by each input focal point $\rin$. ($\textbf{B}$)  
	\laura{When $\mathbf{B}$ (the correlation matrix of $\mathbf{R}_{kx}(z)$) is calculated, the sample reflectivity is smoothed out (spatially averaged). $\mathbf{B}$ thus mimics the time reversal operator associated with a virtual specular reflector of scattering distribution $\gamma_R(x)$ that scales as the field of view $\Omega(x)$ (Eq.~\ref{GammaR}).}}
	\label{fdort}
\end{figure}

\clearpage


\begin{thebibliography}{75}%
\makeatletter
\providecommand \@ifxundefined [1]{%
 \@ifx{#1\undefined}
}%
\providecommand \@ifnum [1]{%
 \ifnum #1\expandafter \@firstoftwo
 \else \expandafter \@secondoftwo
 \fi
}%
\providecommand \@ifx [1]{%
 \ifx #1\expandafter \@firstoftwo
 \else \expandafter \@secondoftwo
 \fi
}%
\providecommand \natexlab [1]{#1}%
\providecommand \enquote  [1]{``#1''}%
\providecommand \bibnamefont  [1]{#1}%
\providecommand \bibfnamefont [1]{#1}%
\providecommand \citenamefont [1]{#1}%
\providecommand \href@noop [0]{\@secondoftwo}%
\providecommand \href [0]{\begingroup \@sanitize@url \@href}%
\providecommand \@href[1]{\@@startlink{#1}\@@href}%
\providecommand \@@href[1]{\endgroup#1\@@endlink}%
\providecommand \@sanitize@url [0]{\catcode `\\12\catcode `\$12\catcode
  `\&12\catcode `\#12\catcode `\^12\catcode `\_12\catcode `\%12\relax}%
\providecommand \@@startlink[1]{}%
\providecommand \@@endlink[0]{}%
\providecommand \url  [0]{\begingroup\@sanitize@url \@url }%
\providecommand \@url [1]{\endgroup\@href {#1}{\urlprefix }}%
\providecommand \urlprefix  [0]{URL }%
\providecommand \Eprint [0]{\href }%
\providecommand \doibase [0]{http://dx.doi.org/}%
\providecommand \selectlanguage [0]{\@gobble}%
\providecommand \bibinfo  [0]{\@secondoftwo}%
\providecommand \bibfield  [0]{\@secondoftwo}%
\providecommand \translation [1]{[#1]}%
\providecommand \BibitemOpen [0]{}%
\providecommand \bibitemStop [0]{}%
\providecommand \bibitemNoStop [0]{.\EOS\space}%
\providecommand \EOS [0]{\spacefactor3000\relax}%
\providecommand \BibitemShut  [1]{\csname bibitem#1\endcsname}%
\let\auto@bib@innerbib\@empty
\bibitem [{\citenamefont {Roddier}(1999)}]{Roddier1999}%
  \BibitemOpen
  \bibfield  {author} {\bibinfo {author} {\bibfnamefont {F.}~\bibnamefont
  {Roddier}},\ }\href@noop {} {\emph {\bibinfo {title} {Adaptive optics in
  astronomy}}}\ (\bibinfo  {publisher} {Cambridge University Press,
  Cambridge},\ \bibinfo {year} {1999})\BibitemShut {NoStop}%
\bibitem [{\citenamefont {Angelsen}(2000)}]{Angelsen2000}%
  \BibitemOpen
  \bibfield  {author} {\bibinfo {author} {\bibfnamefont {B.~A.}\ \bibnamefont
  {Angelsen}},\ }\href@noop {} {\emph {\bibinfo {title} {Ultrasound imaging:
  waves, signals, and signal processing.}}}\ (\bibinfo  {publisher} {Emantec},\
  \bibinfo {year} {2000})\BibitemShut {NoStop}%
\bibitem [{\citenamefont {Booth}(2007)}]{Booth2007}%
  \BibitemOpen
  \bibfield  {author} {\bibinfo {author} {\bibfnamefont {M.~J.}\ \bibnamefont
  {Booth}},\ }\href@noop {} {\bibfield  {journal} {\bibinfo  {journal} {Philos.
  Trans. R. Soc. A}\ }\textbf {\bibinfo {volume} {365}},\ \bibinfo {pages}
  {2829 } (\bibinfo {year} {2007})}\BibitemShut {NoStop}%
\bibitem [{\citenamefont {Labeyrie}(1970)}]{Labeyrie1970}%
  \BibitemOpen
  \bibfield  {author} {\bibinfo {author} {\bibfnamefont {A.}~\bibnamefont
  {Labeyrie}},\ }\href@noop {} {\bibfield  {journal} {\bibinfo  {journal}
  {Astron. Astrophys.}\ }\textbf {\bibinfo {volume} {6}},\ \bibinfo {pages}
  {85} (\bibinfo {year} {1970})}\BibitemShut {NoStop}%
\bibitem [{\citenamefont {O'Donnell}\ and\ \citenamefont
  {Flax}(1988)}]{ODonnell1988}%
  \BibitemOpen
  \bibfield  {author} {\bibinfo {author} {\bibfnamefont {M.}~\bibnamefont
  {O'Donnell}}\ and\ \bibinfo {author} {\bibfnamefont {S.}~\bibnamefont
  {Flax}},\ }\href@noop {} {\bibfield  {journal} {\bibinfo  {journal} {IEEE
  Trans. Ultrason., Ferroelectr., Freq. Control}\ }\textbf {\bibinfo {volume}
  {35}},\ \bibinfo {pages} {768} (\bibinfo {year} {1988})}\BibitemShut
  {NoStop}%
\bibitem [{\citenamefont {Mallart}\ and\ \citenamefont
  {Fink}(1994)}]{Mallart1994}%
  \BibitemOpen
  \bibfield  {author} {\bibinfo {author} {\bibfnamefont {R.}~\bibnamefont
  {Mallart}}\ and\ \bibinfo {author} {\bibfnamefont {M.}~\bibnamefont {Fink}},\
  }\href@noop {} {\bibfield  {journal} {\bibinfo  {journal} {J. Acoust. Soc.
  Am.}\ }\textbf {\bibinfo {volume} {96}},\ \bibinfo {pages} {2718} (\bibinfo
  {year} {1994})}\BibitemShut {NoStop}%
\bibitem [{\citenamefont {M{\aa}s{\o}y}\ \emph {et~al.}(2005)\citenamefont
  {M{\aa}s{\o}y}, \citenamefont {Varslot},\ and\ \citenamefont
  {Angelsen}}]{Masoy2005}%
  \BibitemOpen
  \bibfield  {author} {\bibinfo {author} {\bibfnamefont {S.-E.}\ \bibnamefont
  {M{\aa}s{\o}y}}, \bibinfo {author} {\bibfnamefont {T.}~\bibnamefont
  {Varslot}}, \ and\ \bibinfo {author} {\bibfnamefont {B.}~\bibnamefont
  {Angelsen}},\ }\href@noop {} {\bibfield  {journal} {\bibinfo  {journal} {J.
  Acoust. Soc. Am.}\ }\textbf {\bibinfo {volume} {117}},\ \bibinfo {pages}
  {450} (\bibinfo {year} {2005})}\BibitemShut {NoStop}%
\bibitem [{\citenamefont {Montaldo}\ \emph {et~al.}(2011)\citenamefont
  {Montaldo}, \citenamefont {Tanter},\ and\ \citenamefont
  {Fink}}]{Montaldo2011}%
  \BibitemOpen
  \bibfield  {author} {\bibinfo {author} {\bibfnamefont {G.}~\bibnamefont
  {Montaldo}}, \bibinfo {author} {\bibfnamefont {M.}~\bibnamefont {Tanter}}, \
  and\ \bibinfo {author} {\bibfnamefont {M.}~\bibnamefont {Fink}},\ }\href@noop
  {} {\bibfield  {journal} {\bibinfo  {journal} {Phys. Rev. Lett.}\ }\textbf
  {\bibinfo {volume} {106}},\ \bibinfo {pages} {054301} (\bibinfo {year}
  {2011})}\BibitemShut {NoStop}%
\bibitem [{\citenamefont {Kang}\ \emph {et~al.}(2017)\citenamefont {Kang},
  \citenamefont {Kang}, \citenamefont {Jeong}, \citenamefont {Kwon},
  \citenamefont {Yang}, \citenamefont {Hong}, \citenamefont {Kim},
  \citenamefont {Song}, \citenamefont {Park}, \citenamefont {Lee},
  \citenamefont {Kim}, \citenamefont {Kim},\ and\ \citenamefont
  {Choi}}]{Kang2017}%
  \BibitemOpen
  \bibfield  {author} {\bibinfo {author} {\bibfnamefont {S.}~\bibnamefont
  {Kang}}, \bibinfo {author} {\bibfnamefont {P.}~\bibnamefont {Kang}}, \bibinfo
  {author} {\bibfnamefont {S.}~\bibnamefont {Jeong}}, \bibinfo {author}
  {\bibfnamefont {Y.}~\bibnamefont {Kwon}}, \bibinfo {author} {\bibfnamefont
  {T.~D.}\ \bibnamefont {Yang}}, \bibinfo {author} {\bibfnamefont {J.~H.}\
  \bibnamefont {Hong}}, \bibinfo {author} {\bibfnamefont {M.}~\bibnamefont
  {Kim}}, \bibinfo {author} {\bibfnamefont {K.-D.}\ \bibnamefont {Song}},
  \bibinfo {author} {\bibfnamefont {J.~H.}\ \bibnamefont {Park}}, \bibinfo
  {author} {\bibfnamefont {J.~H.}\ \bibnamefont {Lee}}, \bibinfo {author}
  {\bibfnamefont {M.~J.}\ \bibnamefont {Kim}}, \bibinfo {author} {\bibfnamefont
  {K.~H.}\ \bibnamefont {Kim}}, \ and\ \bibinfo {author} {\bibfnamefont
  {W.}~\bibnamefont {Choi}},\ }\href@noop {} {\bibfield  {journal} {\bibinfo
  {journal} {Nat. Commun.}\ }\textbf {\bibinfo {volume} {8}},\ \bibinfo {pages}
  {2157} (\bibinfo {year} {2017})}\BibitemShut {NoStop}%
\bibitem [{\citenamefont {Kim}\ \emph {et~al.}(2019)\citenamefont {Kim},
  \citenamefont {Jo}, \citenamefont {Hong}, \citenamefont {Kim}, \citenamefont
  {Yoon}, \citenamefont {Song}, \citenamefont {Kang}, \citenamefont {Lee},
  \citenamefont {Kim}, \citenamefont {Park},\ and\ \citenamefont
  {Choi}}]{Kim2019}%
  \BibitemOpen
  \bibfield  {author} {\bibinfo {author} {\bibfnamefont {M.}~\bibnamefont
  {Kim}}, \bibinfo {author} {\bibfnamefont {Y.}~\bibnamefont {Jo}}, \bibinfo
  {author} {\bibfnamefont {J.~H.}\ \bibnamefont {Hong}}, \bibinfo {author}
  {\bibfnamefont {S.}~\bibnamefont {Kim}}, \bibinfo {author} {\bibfnamefont
  {S.}~\bibnamefont {Yoon}}, \bibinfo {author} {\bibfnamefont {K.-D.}\
  \bibnamefont {Song}}, \bibinfo {author} {\bibfnamefont {S.}~\bibnamefont
  {Kang}}, \bibinfo {author} {\bibfnamefont {B.}~\bibnamefont {Lee}}, \bibinfo
  {author} {\bibfnamefont {G.~H.}\ \bibnamefont {Kim}}, \bibinfo {author}
  {\bibfnamefont {H.-C.}\ \bibnamefont {Park}}, \ and\ \bibinfo {author}
  {\bibfnamefont {W.}~\bibnamefont {Choi}},\ }\href@noop {} {\bibfield
  {journal} {\bibinfo  {journal} {Nat. Commun.}\ }\textbf {\bibinfo {volume}
  {10}},\ \bibinfo {pages} {3152} (\bibinfo {year} {2019})}\BibitemShut
  {NoStop}%
\bibitem [{\citenamefont {Muller}\ and\ \citenamefont
  {Buffington}(1974)}]{Muller1974}%
  \BibitemOpen
  \bibfield  {author} {\bibinfo {author} {\bibfnamefont {R.~A.}\ \bibnamefont
  {Muller}}\ and\ \bibinfo {author} {\bibfnamefont {A.}~\bibnamefont
  {Buffington}},\ }\href@noop {} {\bibfield  {journal} {\bibinfo  {journal} {J.
  Opt. Soc. Am.}\ }\textbf {\bibinfo {volume} {64}},\ \bibinfo {pages} {1200}
  (\bibinfo {year} {1974})}\BibitemShut {NoStop}%
\bibitem [{\citenamefont {Booth}\ \emph {et~al.}(2002)\citenamefont {Booth},
  \citenamefont {Neil}, \citenamefont {Ju{\v s}kaitis},\ and\ \citenamefont
  {Wilson}}]{Booth5788}%
  \BibitemOpen
  \bibfield  {author} {\bibinfo {author} {\bibfnamefont {M.~J.}\ \bibnamefont
  {Booth}}, \bibinfo {author} {\bibfnamefont {M.~A.~A.}\ \bibnamefont {Neil}},
  \bibinfo {author} {\bibfnamefont {R.}~\bibnamefont {Ju{\v s}kaitis}}, \ and\
  \bibinfo {author} {\bibfnamefont {T.}~\bibnamefont {Wilson}},\ }\href@noop {}
  {\bibfield  {journal} {\bibinfo  {journal} {Proc. Natl. Acad. Sci.}\ }\textbf
  {\bibinfo {volume} {99}},\ \bibinfo {pages} {5788} (\bibinfo {year}
  {2002})}\BibitemShut {NoStop}%
\bibitem [{\citenamefont {Nock}\ \emph {et~al.}(1989)\citenamefont {Nock},
  \citenamefont {Trahey},\ and\ \citenamefont {Smith}}]{Nock1989}%
  \BibitemOpen
  \bibfield  {author} {\bibinfo {author} {\bibfnamefont {L.}~\bibnamefont
  {Nock}}, \bibinfo {author} {\bibfnamefont {G.~E.}\ \bibnamefont {Trahey}}, \
  and\ \bibinfo {author} {\bibfnamefont {S.~W.}\ \bibnamefont {Smith}},\
  }\href@noop {} {\bibfield  {journal} {\bibinfo  {journal} {J. Acoust. Soc.
  Am.}\ }\textbf {\bibinfo {volume} {85}},\ \bibinfo {pages} {1819} (\bibinfo
  {year} {1989})}\BibitemShut {NoStop}%
\bibitem [{\citenamefont {D{\'e}barre}\ \emph {et~al.}(2009)\citenamefont
  {D{\'e}barre}, \citenamefont {Botcherby}, \citenamefont {Watanabe},
  \citenamefont {Srinivas}, \citenamefont {Booth},\ and\ \citenamefont
  {Wilson}}]{Debarre2009}%
  \BibitemOpen
  \bibfield  {author} {\bibinfo {author} {\bibfnamefont {D.}~\bibnamefont
  {D{\'e}barre}}, \bibinfo {author} {\bibfnamefont {E.~J.}\ \bibnamefont
  {Botcherby}}, \bibinfo {author} {\bibfnamefont {T.}~\bibnamefont {Watanabe}},
  \bibinfo {author} {\bibfnamefont {S.}~\bibnamefont {Srinivas}}, \bibinfo
  {author} {\bibfnamefont {M.~J.}\ \bibnamefont {Booth}}, \ and\ \bibinfo
  {author} {\bibfnamefont {T.}~\bibnamefont {Wilson}},\ }\href@noop {}
  {\bibfield  {journal} {\bibinfo  {journal} {Opt. Lett.}\ }\textbf {\bibinfo
  {volume} {34}},\ \bibinfo {pages} {2495} (\bibinfo {year}
  {2009})}\BibitemShut {NoStop}%
\bibitem [{\citenamefont {Ji}\ \emph {et~al.}(2010)\citenamefont {Ji},
  \citenamefont {Milkie},\ and\ \citenamefont {Betzig}}]{Ji2010}%
  \BibitemOpen
  \bibfield  {author} {\bibinfo {author} {\bibfnamefont {N.}~\bibnamefont
  {Ji}}, \bibinfo {author} {\bibfnamefont {D.~E.}\ \bibnamefont {Milkie}}, \
  and\ \bibinfo {author} {\bibfnamefont {E.}~\bibnamefont {Betzig}},\
  }\href@noop {} {\bibfield  {journal} {\bibinfo  {journal} {Nature Methods}\
  }\textbf {\bibinfo {volume} {7}},\ \bibinfo {pages} {141} (\bibinfo {year}
  {2010})}\BibitemShut {NoStop}%
\bibitem [{\citenamefont {Ji}\ \emph {et~al.}(2012)\citenamefont {Ji},
  \citenamefont {Sato},\ and\ \citenamefont {Betzig}}]{Ji22}%
  \BibitemOpen
  \bibfield  {author} {\bibinfo {author} {\bibfnamefont {N.}~\bibnamefont
  {Ji}}, \bibinfo {author} {\bibfnamefont {T.~R.}\ \bibnamefont {Sato}}, \ and\
  \bibinfo {author} {\bibfnamefont {E.}~\bibnamefont {Betzig}},\ }\href@noop {}
  {\bibfield  {journal} {\bibinfo  {journal} {Proc. Natl. Acad. Sci.}\ }\textbf
  {\bibinfo {volume} {109}},\ \bibinfo {pages} {22} (\bibinfo {year}
  {2012})}\BibitemShut {NoStop}%
\bibitem [{\citenamefont {Adie}\ \emph {et~al.}(2012)\citenamefont {Adie},
  \citenamefont {Graf}, \citenamefont {Ahmad}, \citenamefont {Carney},\ and\
  \citenamefont {Boppart}}]{Adie7175}%
  \BibitemOpen
  \bibfield  {author} {\bibinfo {author} {\bibfnamefont {S.~G.}\ \bibnamefont
  {Adie}}, \bibinfo {author} {\bibfnamefont {B.~W.}\ \bibnamefont {Graf}},
  \bibinfo {author} {\bibfnamefont {A.}~\bibnamefont {Ahmad}}, \bibinfo
  {author} {\bibfnamefont {P.~S.}\ \bibnamefont {Carney}}, \ and\ \bibinfo
  {author} {\bibfnamefont {S.~A.}\ \bibnamefont {Boppart}},\ }\href@noop {}
  {\bibfield  {journal} {\bibinfo  {journal} {Proc. Natl. Acad. Sci.}\ }\textbf
  {\bibinfo {volume} {109}},\ \bibinfo {pages} {7175} (\bibinfo {year}
  {2012})}\BibitemShut {NoStop}%
\bibitem [{\citenamefont {Dahl}\ \emph {et~al.}(2005)\citenamefont {Dahl},
  \citenamefont {Soo},\ and\ \citenamefont {Trahey}}]{Dahl2005}%
  \BibitemOpen
  \bibfield  {author} {\bibinfo {author} {\bibfnamefont {J.~J.}\ \bibnamefont
  {Dahl}}, \bibinfo {author} {\bibfnamefont {M.~S.}\ \bibnamefont {Soo}}, \
  and\ \bibinfo {author} {\bibfnamefont {G.~E.}\ \bibnamefont {Trahey}},\
  }\href@noop {} {\bibfield  {journal} {\bibinfo  {journal} {IEEE Trans.
  Ultrason. Ferroelectr. Freq. Control}\ }\textbf {\bibinfo {volume} {52}},\
  \bibinfo {pages} {1504} (\bibinfo {year} {2005})}\BibitemShut {NoStop}%
\bibitem [{\citenamefont {Judkewitz}\ \emph {et~al.}(2015)\citenamefont
  {Judkewitz}, \citenamefont {Horstmeyer}, \citenamefont {Vellekoop},
  \citenamefont {Papadopoulos},\ and\ \citenamefont {Yang}}]{Judkewitz2015}%
  \BibitemOpen
  \bibfield  {author} {\bibinfo {author} {\bibfnamefont {B.}~\bibnamefont
  {Judkewitz}}, \bibinfo {author} {\bibfnamefont {R.}~\bibnamefont
  {Horstmeyer}}, \bibinfo {author} {\bibfnamefont {I.~M.}\ \bibnamefont
  {Vellekoop}}, \bibinfo {author} {\bibfnamefont {I.~N.}\ \bibnamefont
  {Papadopoulos}}, \ and\ \bibinfo {author} {\bibfnamefont {C.}~\bibnamefont
  {Yang}},\ }\href@noop {} {\bibfield  {journal} {\bibinfo  {journal} {Nat.
  Phys.}\ }\textbf {\bibinfo {volume} {11}},\ \bibinfo {pages} {684} (\bibinfo
  {year} {2015})}\BibitemShut {NoStop}%
\bibitem [{\citenamefont {Ali}\ and\ \citenamefont {Dahl}(2018)}]{Ali2018}%
  \BibitemOpen
  \bibfield  {author} {\bibinfo {author} {\bibfnamefont {R.}~\bibnamefont
  {Ali}}\ and\ \bibinfo {author} {\bibfnamefont {J.}~\bibnamefont {Dahl}},\
  }in\ \href@noop {} {\emph {\bibinfo {booktitle} {IEEE Int. Ultrason.
  Symp.}}}\ (\bibinfo {year} {2018})\ pp.\ \bibinfo {pages} {1--4}\BibitemShut
  {NoStop}%
\bibitem [{\citenamefont {Rau}\ \emph {et~al.}(2019)\citenamefont {Rau},
  \citenamefont {Schweizer}, \citenamefont {Vishnevskiy},\ and\ \citenamefont
  {Goksel}}]{Rau2019}%
  \BibitemOpen
  \bibfield  {author} {\bibinfo {author} {\bibfnamefont {R.}~\bibnamefont
  {Rau}}, \bibinfo {author} {\bibfnamefont {D.}~\bibnamefont {Schweizer}},
  \bibinfo {author} {\bibfnamefont {V.}~\bibnamefont {Vishnevskiy}}, \ and\
  \bibinfo {author} {\bibfnamefont {O.}~\bibnamefont {Goksel}},\ }in\
  \href@noop {} {\emph {\bibinfo {booktitle} {IEEE Int. Ultrason. Symp.}}}\
  (\bibinfo  {publisher} {IEEE},\ \bibinfo {address} {Glasgow},\ \bibinfo
  {year} {2019})\ pp.\ \bibinfo {pages} {2003--2006}\BibitemShut {NoStop}%
\bibitem [{\citenamefont {Jaeger}\ \emph
  {et~al.}(2015{\natexlab{a}})\citenamefont {Jaeger}, \citenamefont {Robinson},
  \citenamefont {G\"{u}nhan~Akar\c{c}ay},\ and\ \citenamefont
  {Frenz}}]{Jaeger2015SosMap}%
  \BibitemOpen
  \bibfield  {author} {\bibinfo {author} {\bibfnamefont {M.}~\bibnamefont
  {Jaeger}}, \bibinfo {author} {\bibfnamefont {E.}~\bibnamefont {Robinson}},
  \bibinfo {author} {\bibfnamefont {H.}~\bibnamefont {G\"{u}nhan~Akar\c{c}ay}},
  \ and\ \bibinfo {author} {\bibfnamefont {M.}~\bibnamefont {Frenz}},\
  }\href@noop {} {\bibfield  {journal} {\bibinfo  {journal} {Phys. Med. Biol.}\
  }\textbf {\bibinfo {volume} {60}},\ \bibinfo {pages} {4497} (\bibinfo {year}
  {2015}{\natexlab{a}})}\BibitemShut {NoStop}%
\bibitem [{\citenamefont {Chau}\ \emph {et~al.}(2019)\citenamefont {Chau},
  \citenamefont {Jakovljevic}, \citenamefont {Lavarello},\ and\ \citenamefont
  {Dahl}}]{Chau2019}%
  \BibitemOpen
  \bibfield  {author} {\bibinfo {author} {\bibfnamefont {G.}~\bibnamefont
  {Chau}}, \bibinfo {author} {\bibfnamefont {M.}~\bibnamefont {Jakovljevic}},
  \bibinfo {author} {\bibfnamefont {R.}~\bibnamefont {Lavarello}}, \ and\
  \bibinfo {author} {\bibfnamefont {J.}~\bibnamefont {Dahl}},\ }\href@noop {}
  {\bibfield  {journal} {\bibinfo  {journal} {Ultrason. Imag.}\ }\textbf
  {\bibinfo {volume} {41}},\ \bibinfo {pages} {3} (\bibinfo {year}
  {2019})}\BibitemShut {NoStop}%
\bibitem [{\citenamefont {Bendjador}\ \emph {et~al.}(2020)\citenamefont
  {Bendjador}, \citenamefont {Deffieux},\ and\ \citenamefont
  {Tanter}}]{PFM2019}%
  \BibitemOpen
  \bibfield  {author} {\bibinfo {author} {\bibfnamefont {H.}~\bibnamefont
  {Bendjador}}, \bibinfo {author} {\bibfnamefont {T.}~\bibnamefont {Deffieux}},
  \ and\ \bibinfo {author} {\bibfnamefont {M.}~\bibnamefont {Tanter}},\ }\href
  {\doibase 10.1109/tmi.2020.2986830} {\bibfield  {journal} {\bibinfo
  {journal} {{IEEE} Trans. Med. Imag.}} (\bibinfo
  {year} {2020})},\ \bibinfo {note} {in press}\BibitemShut {NoStop}%
\bibitem [{\citenamefont {Prada}\ and\ \citenamefont {Fink}(1994)}]{Prada1994}%
  \BibitemOpen
  \bibfield  {author} {\bibinfo {author} {\bibfnamefont {C.}~\bibnamefont
  {Prada}}\ and\ \bibinfo {author} {\bibfnamefont {M.}~\bibnamefont {Fink}},\
  }\href@noop {} {\bibfield  {journal} {\bibinfo  {journal} {Wave Motion}\
  }\textbf {\bibinfo {volume} {20}},\ \bibinfo {pages} {151} (\bibinfo {year}
  {1994})}\BibitemShut {NoStop}%
\bibitem [{\citenamefont {Tanter}\ \emph {et~al.}(2000)\citenamefont {Tanter},
  \citenamefont {Thomas},\ and\ \citenamefont {Fink}}]{Tanter2000}%
  \BibitemOpen
  \bibfield  {author} {\bibinfo {author} {\bibfnamefont {M.}~\bibnamefont
  {Tanter}}, \bibinfo {author} {\bibfnamefont {J.-L.}\ \bibnamefont {Thomas}},
  \ and\ \bibinfo {author} {\bibfnamefont {M.}~\bibnamefont {Fink}},\
  }\href@noop {} {\bibfield  {journal} {\bibinfo  {journal} {J. Acoust. Soc.
  Am.}\ }\textbf {\bibinfo {volume} {108}},\ \bibinfo {pages} {223} (\bibinfo
  {year} {2000})}\BibitemShut {NoStop}%
\bibitem [{\citenamefont {Derode}\ \emph {et~al.}(2003)\citenamefont {Derode},
  \citenamefont {Tourin}, \citenamefont {de~Rosny}, \citenamefont {Tanter},
  \citenamefont {Yon},\ and\ \citenamefont {Fink}}]{Derode2003}%
  \BibitemOpen
  \bibfield  {author} {\bibinfo {author} {\bibfnamefont {A.}~\bibnamefont
  {Derode}}, \bibinfo {author} {\bibfnamefont {A.}~\bibnamefont {Tourin}},
  \bibinfo {author} {\bibfnamefont {J.}~\bibnamefont {de~Rosny}}, \bibinfo
  {author} {\bibfnamefont {M.}~\bibnamefont {Tanter}}, \bibinfo {author}
  {\bibfnamefont {S.}~\bibnamefont {Yon}}, \ and\ \bibinfo {author}
  {\bibfnamefont {M.}~\bibnamefont {Fink}},\ }\href@noop {} {\bibfield
  {journal} {\bibinfo  {journal} {Phys. Rev. Lett.}\ }\textbf {\bibinfo
  {volume} {90}},\ \bibinfo {pages} {014301} (\bibinfo {year}
  {2003})}\BibitemShut {NoStop}%
\bibitem [{\citenamefont {Popoff}\ \emph
  {et~al.}(2010{\natexlab{a}})\citenamefont {Popoff}, \citenamefont {Lerosey},
  \citenamefont {Carminati}, \citenamefont {Fink}, \citenamefont {Boccara},\
  and\ \citenamefont {Gigan}}]{Popoff2010}%
  \BibitemOpen
  \bibfield  {author} {\bibinfo {author} {\bibfnamefont {S.~M.}\ \bibnamefont
  {Popoff}}, \bibinfo {author} {\bibfnamefont {G.}~\bibnamefont {Lerosey}},
  \bibinfo {author} {\bibfnamefont {R.}~\bibnamefont {Carminati}}, \bibinfo
  {author} {\bibfnamefont {M.}~\bibnamefont {Fink}}, \bibinfo {author}
  {\bibfnamefont {A.~C.}\ \bibnamefont {Boccara}}, \ and\ \bibinfo {author}
  {\bibfnamefont {S.}~\bibnamefont {Gigan}},\ }\href@noop {} {\bibfield
  {journal} {\bibinfo  {journal} {Phys. Rev. Lett.}\ }\textbf {\bibinfo
  {volume} {104}},\ \bibinfo {pages} {100601} (\bibinfo {year}
  {2010}{\natexlab{a}})}\BibitemShut {NoStop}%
\bibitem [{\citenamefont {Kim}\ \emph {et~al.}(2012)\citenamefont {Kim},
  \citenamefont {Choi}, \citenamefont {Yoon}, \citenamefont {Choi},
  \citenamefont {Kim}, \citenamefont {Park},\ and\ \citenamefont
  {Choi}}]{Kim2012}%
  \BibitemOpen
  \bibfield  {author} {\bibinfo {author} {\bibfnamefont {M.}~\bibnamefont
  {Kim}}, \bibinfo {author} {\bibfnamefont {Y.}~\bibnamefont {Choi}}, \bibinfo
  {author} {\bibfnamefont {C.}~\bibnamefont {Yoon}}, \bibinfo {author}
  {\bibfnamefont {W.}~\bibnamefont {Choi}}, \bibinfo {author} {\bibfnamefont
  {J.}~\bibnamefont {Kim}}, \bibinfo {author} {\bibfnamefont {Q.-H.}\
  \bibnamefont {Park}}, \ and\ \bibinfo {author} {\bibfnamefont
  {W.}~\bibnamefont {Choi}},\ }\href@noop {} {\bibfield  {journal} {\bibinfo
  {journal} {Nat. Photonics}\ }\textbf {\bibinfo {volume} {6}},\ \bibinfo
  {pages} {583} (\bibinfo {year} {2012})}\BibitemShut {NoStop}%
\bibitem [{\citenamefont {Popoff}\ \emph
  {et~al.}(2010{\natexlab{b}})\citenamefont {Popoff}, \citenamefont {Lerosey},
  \citenamefont {Fink}, \citenamefont {Boccara},\ and\ \citenamefont
  {Gigan}}]{Popoff2010b}%
  \BibitemOpen
  \bibfield  {author} {\bibinfo {author} {\bibfnamefont {S.~M.}\ \bibnamefont
  {Popoff}}, \bibinfo {author} {\bibfnamefont {G.}~\bibnamefont {Lerosey}},
  \bibinfo {author} {\bibfnamefont {M.}~\bibnamefont {Fink}}, \bibinfo {author}
  {\bibfnamefont {A.~C.}\ \bibnamefont {Boccara}}, \ and\ \bibinfo {author}
  {\bibfnamefont {S.}~\bibnamefont {Gigan}},\ }\href@noop {} {\bibfield
  {journal} {\bibinfo  {journal} {Nat. Commun.}\ }\textbf {\bibinfo {volume}
  {1}},\ \bibinfo {pages} {81} (\bibinfo {year}
  {2010}{\natexlab{b}})}\BibitemShut {NoStop}%
\bibitem [{\citenamefont {Cizmar}\ and\ \citenamefont
  {Dholakia}(2012)}]{Cizmar2012}%
  \BibitemOpen
  \bibfield  {author} {\bibinfo {author} {\bibfnamefont {T.}~\bibnamefont
  {Cizmar}}\ and\ \bibinfo {author} {\bibfnamefont {K.}~\bibnamefont
  {Dholakia}},\ }\href@noop {} {\bibfield  {journal} {\bibinfo  {journal} {Nat.
  Commun.}\ }\textbf {\bibinfo {volume} {3}},\ \bibinfo {pages} {1027}
  (\bibinfo {year} {2012})}\BibitemShut {NoStop}%
\bibitem [{\citenamefont {Borcea}\ \emph {et~al.}(2002)\citenamefont {Borcea},
  \citenamefont {Papanicolaou}, \citenamefont {Tsogka},\ and\ \citenamefont
  {Berryman}}]{Borcea2002}%
  \BibitemOpen
  \bibfield  {author} {\bibinfo {author} {\bibfnamefont {L.}~\bibnamefont
  {Borcea}}, \bibinfo {author} {\bibfnamefont {G.}~\bibnamefont
  {Papanicolaou}}, \bibinfo {author} {\bibfnamefont {C.}~\bibnamefont
  {Tsogka}}, \ and\ \bibinfo {author} {\bibfnamefont {J.}~\bibnamefont
  {Berryman}},\ }\href@noop {} {\bibfield  {journal} {\bibinfo  {journal}
  {Inverse Problems}\ }\textbf {\bibinfo {volume} {18}},\ \bibinfo {pages}
  {1247} (\bibinfo {year} {2002})}\BibitemShut {NoStop}%
\bibitem [{\citenamefont {Popoff}\ \emph {et~al.}(2011)\citenamefont {Popoff},
  \citenamefont {Aubry}, \citenamefont {Lerosey}, \citenamefont {Fink},
  \citenamefont {Boccara},\ and\ \citenamefont {Gigan}}]{Popoff2011}%
  \BibitemOpen
  \bibfield  {author} {\bibinfo {author} {\bibfnamefont {S.~M.}\ \bibnamefont
  {Popoff}}, \bibinfo {author} {\bibfnamefont {A.}~\bibnamefont {Aubry}},
  \bibinfo {author} {\bibfnamefont {G.}~\bibnamefont {Lerosey}}, \bibinfo
  {author} {\bibfnamefont {M.}~\bibnamefont {Fink}}, \bibinfo {author}
  {\bibfnamefont {A.~C.}\ \bibnamefont {Boccara}}, \ and\ \bibinfo {author}
  {\bibfnamefont {S.}~\bibnamefont {Gigan}},\ }\href@noop {} {\bibfield
  {journal} {\bibinfo  {journal} {Phys. Rev. Lett.}\ }\textbf {\bibinfo
  {volume} {107}},\ \bibinfo {pages} {263901} (\bibinfo {year}
  {2011})}\BibitemShut {NoStop}%
\bibitem [{\citenamefont {Shajahan}\ \emph {et~al.}(2014)\citenamefont
  {Shajahan}, \citenamefont {Aubry}, \citenamefont {Rupin}, \citenamefont
  {Chassignole},\ and\ \citenamefont {Derode}}]{Shajahan2014}%
  \BibitemOpen
  \bibfield  {author} {\bibinfo {author} {\bibfnamefont {S.}~\bibnamefont
  {Shajahan}}, \bibinfo {author} {\bibfnamefont {A.}~\bibnamefont {Aubry}},
  \bibinfo {author} {\bibfnamefont {F.}~\bibnamefont {Rupin}}, \bibinfo
  {author} {\bibfnamefont {B.}~\bibnamefont {Chassignole}}, \ and\ \bibinfo
  {author} {\bibfnamefont {A.}~\bibnamefont {Derode}},\ }\href@noop {}
  {\bibfield  {journal} {\bibinfo  {journal} {Appl. Phys. Lett.}\ }\textbf
  {\bibinfo {volume} {104}},\ \bibinfo {pages} {234105} (\bibinfo {year}
  {2014})}\BibitemShut {NoStop}%
\bibitem [{\citenamefont {Badon}\ \emph {et~al.}(2016)\citenamefont {Badon},
  \citenamefont {Li}, \citenamefont {Lerosey}, \citenamefont {Boccara},
  \citenamefont {Fink},\ and\ \citenamefont {Aubry}}]{Badon2016}%
  \BibitemOpen
  \bibfield  {author} {\bibinfo {author} {\bibfnamefont {A.}~\bibnamefont
  {Badon}}, \bibinfo {author} {\bibfnamefont {D.}~\bibnamefont {Li}}, \bibinfo
  {author} {\bibfnamefont {G.}~\bibnamefont {Lerosey}}, \bibinfo {author}
  {\bibfnamefont {A.~C.}\ \bibnamefont {Boccara}}, \bibinfo {author}
  {\bibfnamefont {M.}~\bibnamefont {Fink}}, \ and\ \bibinfo {author}
  {\bibfnamefont {A.}~\bibnamefont {Aubry}},\ }\href@noop {} {\bibfield
  {journal} {\bibinfo  {journal} {Sci. Adv.}\ }\textbf {\bibinfo {volume}
  {2}},\ \bibinfo {pages} {e1600370} (\bibinfo {year} {2016})}\BibitemShut
  {NoStop}%
\bibitem [{\citenamefont {Blondel}\ \emph {et~al.}(2018)\citenamefont
  {Blondel}, \citenamefont {Chaput}, \citenamefont {Derode}, \citenamefont
  {Campillo},\ and\ \citenamefont {Aubry}}]{Blondel2018}%
  \BibitemOpen
  \bibfield  {author} {\bibinfo {author} {\bibfnamefont {T.}~\bibnamefont
  {Blondel}}, \bibinfo {author} {\bibfnamefont {J.}~\bibnamefont {Chaput}},
  \bibinfo {author} {\bibfnamefont {A.}~\bibnamefont {Derode}}, \bibinfo
  {author} {\bibfnamefont {M.}~\bibnamefont {Campillo}}, \ and\ \bibinfo
  {author} {\bibfnamefont {A.}~\bibnamefont {Aubry}},\ }\href@noop {}
  {\bibfield  {journal} {\bibinfo  {journal} {J. Geophys. Res.: Solid Earth}\
  }\textbf {\bibinfo {volume} {123}},\ \bibinfo {pages} {10936} (\bibinfo
  {year} {2018})}\BibitemShut {NoStop}%
\bibitem [{\citenamefont {Choi}\ \emph {et~al.}(2013)\citenamefont {Choi},
  \citenamefont {Hillman}, \citenamefont {Choi}, \citenamefont {Lue},
  \citenamefont {Dasari}, \citenamefont {So}, \citenamefont {Choi},\ and\
  \citenamefont {Yaqoob}}]{Choi2013}%
  \BibitemOpen
  \bibfield  {author} {\bibinfo {author} {\bibfnamefont {Y.}~\bibnamefont
  {Choi}}, \bibinfo {author} {\bibfnamefont {T.~R.}\ \bibnamefont {Hillman}},
  \bibinfo {author} {\bibfnamefont {W.}~\bibnamefont {Choi}}, \bibinfo {author}
  {\bibfnamefont {N.}~\bibnamefont {Lue}}, \bibinfo {author} {\bibfnamefont
  {R.~R.}\ \bibnamefont {Dasari}}, \bibinfo {author} {\bibfnamefont {P.~T.~C.}\
  \bibnamefont {So}}, \bibinfo {author} {\bibfnamefont {W.}~\bibnamefont
  {Choi}}, \ and\ \bibinfo {author} {\bibfnamefont {Z.}~\bibnamefont
  {Yaqoob}},\ }\href@noop {} {\bibfield  {journal} {\bibinfo  {journal} {Phys.
  Rev. Lett.}\ }\textbf {\bibinfo {volume} {111}},\ \bibinfo {pages} {243901}
  (\bibinfo {year} {2013})}\BibitemShut {NoStop}%
\bibitem [{\citenamefont {Jeong}\ \emph {et~al.}(2018)\citenamefont {Jeong},
  \citenamefont {Lee}, \citenamefont {Choi}, \citenamefont {Kang},
  \citenamefont {Hong}, \citenamefont {Park}, \citenamefont {Lim},
  \citenamefont {Park},\ and\ \citenamefont {Choi}}]{Jeong2018}%
  \BibitemOpen
  \bibfield  {author} {\bibinfo {author} {\bibfnamefont {S.}~\bibnamefont
  {Jeong}}, \bibinfo {author} {\bibfnamefont {Y.-R.}\ \bibnamefont {Lee}},
  \bibinfo {author} {\bibfnamefont {W.}~\bibnamefont {Choi}}, \bibinfo {author}
  {\bibfnamefont {S.}~\bibnamefont {Kang}}, \bibinfo {author} {\bibfnamefont
  {J.~H.}\ \bibnamefont {Hong}}, \bibinfo {author} {\bibfnamefont {J.-S.}\
  \bibnamefont {Park}}, \bibinfo {author} {\bibfnamefont {Y.-S.}\ \bibnamefont
  {Lim}}, \bibinfo {author} {\bibfnamefont {H.-G.}\ \bibnamefont {Park}}, \
  and\ \bibinfo {author} {\bibfnamefont {W.}~\bibnamefont {Choi}},\ }\href@noop
  {} {\bibfield  {journal} {\bibinfo  {journal} {Nat. Photonics}\ }\textbf
  {\bibinfo {volume} {12}},\ \bibinfo {pages} {277} (\bibinfo {year}
  {2018})}\BibitemShut {NoStop}%
\bibitem [{\citenamefont {Varslot}\ \emph {et~al.}(2004)\citenamefont
  {Varslot}, \citenamefont {Krogstad}, \citenamefont {Mo},\ and\ \citenamefont
  {Angelsen}}]{Varslot2004}%
  \BibitemOpen
  \bibfield  {author} {\bibinfo {author} {\bibfnamefont {T.}~\bibnamefont
  {Varslot}}, \bibinfo {author} {\bibfnamefont {H.}~\bibnamefont {Krogstad}},
  \bibinfo {author} {\bibfnamefont {E.}~\bibnamefont {Mo}}, \ and\ \bibinfo
  {author} {\bibfnamefont {B.~A.}\ \bibnamefont {Angelsen}},\ }\href@noop {}
  {\bibfield  {journal} {\bibinfo  {journal} {J. Acoust. Soc. Am.}\ }\textbf
  {\bibinfo {volume} {115}},\ \bibinfo {pages} {3068} (\bibinfo {year}
  {2004})}\BibitemShut {NoStop}%
\bibitem [{\citenamefont {Robert}\ and\ \citenamefont
  {Fink}(2008)}]{Robert2008}%
  \BibitemOpen
  \bibfield  {author} {\bibinfo {author} {\bibfnamefont {J.-L.}\ \bibnamefont
  {Robert}}\ and\ \bibinfo {author} {\bibfnamefont {M.}~\bibnamefont {Fink}},\
  }\href@noop {} {\bibfield  {journal} {\bibinfo  {journal} {J. Acoust. Soc.
  Am.}\ }\textbf {\bibinfo {volume} {123}},\ \bibinfo {pages} {866} (\bibinfo
  {year} {2008})}\BibitemShut {NoStop}%
\bibitem [{\citenamefont {Kang}\ \emph {et~al.}(2015)\citenamefont {Kang},
  \citenamefont {Jeong}, \citenamefont {Choi}, \citenamefont {Yang},
  \citenamefont {Joo}, \citenamefont {Lee}, \citenamefont {Lim}, \citenamefont
  {Park},\ and\ \citenamefont {Choi}}]{Kang2015}%
  \BibitemOpen
  \bibfield  {author} {\bibinfo {author} {\bibfnamefont {S.}~\bibnamefont
  {Kang}}, \bibinfo {author} {\bibfnamefont {S.}~\bibnamefont {Jeong}},
  \bibinfo {author} {\bibfnamefont {H.}~\bibnamefont {Choi}, \bibfnamefont
  {W.~Ko}}, \bibinfo {author} {\bibfnamefont {T.~D.}\ \bibnamefont {Yang}},
  \bibinfo {author} {\bibfnamefont {J.~H.}\ \bibnamefont {Joo}}, \bibinfo
  {author} {\bibfnamefont {J.-S.}\ \bibnamefont {Lee}}, \bibinfo {author}
  {\bibfnamefont {Y.-S.}\ \bibnamefont {Lim}}, \bibinfo {author} {\bibfnamefont
  {Q.-H.}\ \bibnamefont {Park}}, \ and\ \bibinfo {author} {\bibfnamefont
  {W.}~\bibnamefont {Choi}},\ }\href@noop {} {\bibfield  {journal} {\bibinfo
  {journal} {Nat. Photonics}\ }\textbf {\bibinfo {volume} {9}},\ \bibinfo
  {pages} {1} (\bibinfo {year} {2015})}\BibitemShut {NoStop}%
\bibitem [{\citenamefont {Badon}\ \emph {et~al.}(2020)\citenamefont {Badon},
  \citenamefont {Barolle}, \citenamefont {Irsch}, \citenamefont {Boccara},
  \citenamefont {Fink},\ and\ \citenamefont {Aubry}}]{Badon2019}%
  \BibitemOpen
  \bibfield  {author} {\bibinfo {author} {\bibfnamefont {A.}~\bibnamefont
  {Badon}}, \bibinfo {author} {\bibfnamefont {V.}~\bibnamefont {Barolle}},
  \bibinfo {author} {\bibfnamefont {K.}~\bibnamefont {Irsch}}, \bibinfo
  {author} {\bibfnamefont {A.~C.}\ \bibnamefont {Boccara}}, \bibinfo {author}
  {\bibfnamefont {M.}~\bibnamefont {Fink}}, \ and\ \bibinfo {author}
  {\bibfnamefont {A.}~\bibnamefont {Aubry}},\ }\href@noop {} {\bibfield
  {journal} {\bibinfo  {journal} {Sci. Adv.}\ } (\bibinfo {year}
  {2020})},\ \bibinfo {note} {in press}\BibitemShut {NoStop}%
\bibitem [{\citenamefont {Prada}\ \emph {et~al.}(1996)\citenamefont {Prada},
  \citenamefont {Manneville}, \citenamefont {Spoliansky},\ and\ \citenamefont
  {Fink}}]{Prada1996}%
  \BibitemOpen
  \bibfield  {author} {\bibinfo {author} {\bibfnamefont {C.}~\bibnamefont
  {Prada}}, \bibinfo {author} {\bibfnamefont {S.}~\bibnamefont {Manneville}},
  \bibinfo {author} {\bibfnamefont {D.}~\bibnamefont {Spoliansky}}, \ and\
  \bibinfo {author} {\bibfnamefont {M.}~\bibnamefont {Fink}},\ }\href@noop {}
  {\bibfield  {journal} {\bibinfo  {journal} {J. Acoust. Soc. Am.}\ }\textbf
  {\bibinfo {volume} {99}} (\bibinfo {year} {1996})}\BibitemShut {NoStop}%
\bibitem [{\citenamefont {Aubry}\ and\ \citenamefont
  {Derode}(2009)}]{Aubry2009}%
  \BibitemOpen
  \bibfield  {author} {\bibinfo {author} {\bibfnamefont {A.}~\bibnamefont
  {Aubry}}\ and\ \bibinfo {author} {\bibfnamefont {A.}~\bibnamefont {Derode}},\
  }\href@noop {} {\bibfield  {journal} {\bibinfo  {journal} {J. Appl. Phys.}\
  }\textbf {\bibinfo {volume} {106}},\ \bibinfo {pages} {044903} (\bibinfo
  {year} {2009})}\BibitemShut {NoStop}%
\bibitem [{\citenamefont {Lambert}\ \emph {et~al.}(2020)\citenamefont
  {Lambert}, \citenamefont {Cobus}, \citenamefont {Fink},\ and\ \citenamefont
  {Aubry}}]{Lambert2019}%
  \BibitemOpen
  \bibfield  {author} {\bibinfo {author} {\bibfnamefont {W.}~\bibnamefont
  {Lambert}}, \bibinfo {author} {\bibfnamefont {L.~A.}\ \bibnamefont {Cobus}},
	\bibinfo {author} {\bibfnamefont {M.}~\bibnamefont {Couade}},
  \bibinfo {author} {\bibfnamefont {M.}~\bibnamefont {Fink}}, \ and\ \bibinfo
  {author} {\bibfnamefont {A.}~\bibnamefont {Aubry}},\ }\href@noop {}
  {\bibfield  {journal} {\bibinfo  {journal} {Phys. Rev. X}\ }\textbf {\bibinfo
  {volume} {10}},\ \bibinfo {pages} {021048} (\bibinfo {year} {2020})}\BibitemShut {NoStop}%
\bibitem [{\citenamefont {Montaldo}\ \emph {et~al.}(2009)\citenamefont
  {Montaldo}, \citenamefont {Tanter}, \citenamefont {Bercoff}, \citenamefont
  {Benech},\ and\ \citenamefont {Fink}}]{Montaldo2009}%
  \BibitemOpen
  \bibfield  {author} {\bibinfo {author} {\bibfnamefont {G.}~\bibnamefont
  {Montaldo}}, \bibinfo {author} {\bibfnamefont {M.}~\bibnamefont {Tanter}},
  \bibinfo {author} {\bibfnamefont {J.}~\bibnamefont {Bercoff}}, \bibinfo
  {author} {\bibfnamefont {N.}~\bibnamefont {Benech}}, \ and\ \bibinfo {author}
  {\bibfnamefont {M.}~\bibnamefont {Fink}},\ }\href@noop {} {\bibfield
  {journal} {\bibinfo  {journal} {IEEE Trans. Ultrason., Ferroelectr., Freq.
  Control}\ }\textbf {\bibinfo {volume} {56}},\ \bibinfo {pages} {489}
  (\bibinfo {year} {2009})}\BibitemShut {NoStop}%
\bibitem [{\citenamefont {{Carlson}}\ \emph {et~al.}(2003)\citenamefont
  {{Carlson}}, \citenamefont {{van Deventer}}, \citenamefont {{Scolan}},\ and\
  \citenamefont {{Carlander}}}]{Carlson2003}%
  \BibitemOpen
  \bibfield  {author} {\bibinfo {author} {\bibfnamefont {J.~E.}\ \bibnamefont
  {{Carlson}}}, \bibinfo {author} {\bibfnamefont {J.}~\bibnamefont {{van
  Deventer}}}, \bibinfo {author} {\bibfnamefont {A.}~\bibnamefont {{Scolan}}},
  \ and\ \bibinfo {author} {\bibfnamefont {C.}~\bibnamefont {{Carlander}}},\
  }in\ \href@noop {} {\emph {\bibinfo {booktitle} {IEEE Symposium on
  Ultrasonics, 2003}}},\ Vol.~\bibinfo {volume} {1}\ (\bibinfo {year} {2003})\
  pp.\ \bibinfo {pages} {885--888}\BibitemShut {NoStop}%
\bibitem [{\citenamefont {Goodman}(1996)}]{Goodman1996}%
  \BibitemOpen
  \bibfield  {author} {\bibinfo {author} {\bibfnamefont {J.~W.}\ \bibnamefont
  {Goodman}},\ }\href@noop {} {\emph {\bibinfo {title} {{Introduction to
  Fourier Optics}}}}\ (\bibinfo  {publisher} {McGraw-Hill, Inc.},\ \bibinfo
  {year} {1996})\ p.\ \bibinfo {pages} {491}\BibitemShut {NoStop}%
\bibitem [{\citenamefont {Watanabe}(2014)}]{Watanabe}%
  \BibitemOpen
  \bibfield  {author} {\bibinfo {author} {\bibfnamefont {K.}~\bibnamefont
  {Watanabe}},\ }\href@noop {} {\emph {\bibinfo {title} {Integral transform
  techniques for Green's functions. Chapter 2: Green's Functions for Laplace
  and Wave Equations}}}\ (\bibinfo  {publisher} {Springer, Cham, Switzerland},\
  \bibinfo {year} {2014})\BibitemShut {NoStop}%
\bibitem [{\citenamefont {Freund}\ \emph {et~al.}(1988)\citenamefont {Freund},
  \citenamefont {Rosenbluh},\ and\ \citenamefont {Feng}}]{Freund1988}%
  \BibitemOpen
  \bibfield  {author} {\bibinfo {author} {\bibfnamefont {I.}~\bibnamefont
  {Freund}}, \bibinfo {author} {\bibfnamefont {M.}~\bibnamefont {Rosenbluh}}, \
  and\ \bibinfo {author} {\bibfnamefont {S.}~\bibnamefont {Feng}},\ }\href@noop
  {} {\bibfield  {journal} {\bibinfo  {journal} {Phys. Rev. Lett.}\ }\textbf
  {\bibinfo {volume} {61}},\ \bibinfo {pages} {2328} (\bibinfo {year}
  {1988})}\BibitemShut {NoStop}%
\bibitem [{\citenamefont {Feng}\ \emph {et~al.}(1988)\citenamefont {Feng},
  \citenamefont {Kane}, \citenamefont {Lee},\ and\ \citenamefont
  {Stone}}]{Feng1988}%
  \BibitemOpen
  \bibfield  {author} {\bibinfo {author} {\bibfnamefont {S.}~\bibnamefont
  {Feng}}, \bibinfo {author} {\bibfnamefont {C.}~\bibnamefont {Kane}}, \bibinfo
  {author} {\bibfnamefont {P.~A.}\ \bibnamefont {Lee}}, \ and\ \bibinfo
  {author} {\bibfnamefont {A.~D.}\ \bibnamefont {Stone}},\ }\href@noop {}
  {\bibfield  {journal} {\bibinfo  {journal} {Phys. Rev. Lett.}\ }\textbf
  {\bibinfo {volume} {61}},\ \bibinfo {pages} {834} (\bibinfo {year}
  {1988})}\BibitemShut {NoStop}%
\bibitem [{\citenamefont {Osnabrugge}\ \emph {et~al.}(2017)\citenamefont
  {Osnabrugge}, \citenamefont {Horstmeyer}, \citenamefont {Papadopoulos},
  \citenamefont {Judkewitz},\ and\ \citenamefont {Vellekoop}}]{Osnabrugge2017}%
  \BibitemOpen
  \bibfield  {author} {\bibinfo {author} {\bibfnamefont {G.}~\bibnamefont
  {Osnabrugge}}, \bibinfo {author} {\bibfnamefont {R.}~\bibnamefont
  {Horstmeyer}}, \bibinfo {author} {\bibfnamefont {I.~N.}\ \bibnamefont
  {Papadopoulos}}, \bibinfo {author} {\bibfnamefont {B.}~\bibnamefont
  {Judkewitz}}, \ and\ \bibinfo {author} {\bibfnamefont {I.~M.}\ \bibnamefont
  {Vellekoop}},\ }\href@noop {} {\bibfield  {journal} {\bibinfo  {journal}
  {Optica}\ }\textbf {\bibinfo {volume} {4}},\ \bibinfo {pages} {886} (\bibinfo
  {year} {2017})}\BibitemShut {NoStop}%
\bibitem [{\citenamefont {Mertz}\ \emph {et~al.}(2015)\citenamefont {Mertz},
  \citenamefont {Paudel},\ and\ \citenamefont {Bifano}}]{Mertz2015}%
  \BibitemOpen
  \bibfield  {author} {\bibinfo {author} {\bibfnamefont {J.}~\bibnamefont
  {Mertz}}, \bibinfo {author} {\bibfnamefont {H.}~\bibnamefont {Paudel}}, \
  and\ \bibinfo {author} {\bibfnamefont {T.~G.}\ \bibnamefont {Bifano}},\
  }\href@noop {} {\bibfield  {journal} {\bibinfo  {journal} {Appl. Opt.}\
  }\textbf {\bibinfo {volume} {54}},\ \bibinfo {pages} {3498} (\bibinfo {year}
  {2015})}\BibitemShut {NoStop}%
\bibitem [{\citenamefont {Katz}\ \emph {et~al.}(2014)\citenamefont {Katz},
  \citenamefont {Heidmann}, \citenamefont {Fink},\ and\ \citenamefont
  {Gigan}}]{Katz2014}%
  \BibitemOpen
  \bibfield  {author} {\bibinfo {author} {\bibfnamefont {O.}~\bibnamefont
  {Katz}}, \bibinfo {author} {\bibfnamefont {P.}~\bibnamefont {Heidmann}},
  \bibinfo {author} {\bibfnamefont {M.}~\bibnamefont {Fink}}, \ and\ \bibinfo
  {author} {\bibfnamefont {S.}~\bibnamefont {Gigan}},\ }\href@noop {}
  {\bibfield  {journal} {\bibinfo  {journal} {Nat. Photonics}\ }\textbf
  {\bibinfo {volume} {8}},\ \bibinfo {pages} {784} (\bibinfo {year}
  {2014})}\BibitemShut {NoStop}%
\bibitem [{\citenamefont {Walker}\ and\ \citenamefont
  {Trahey}(1997)}]{Walker1997}%
  \BibitemOpen
  \bibfield  {author} {\bibinfo {author} {\bibfnamefont {W.~F.}\ \bibnamefont
  {Walker}}\ and\ \bibinfo {author} {\bibfnamefont {G.~E.}\ \bibnamefont
  {Trahey}},\ }\href@noop {} {\bibfield  {journal} {\bibinfo  {journal} {J.
  Acoust. Soc. Am.}\ }\textbf {\bibinfo {volume} {101}},\ \bibinfo {pages}
  {1847} (\bibinfo {year} {1997})}\BibitemShut {NoStop}%
\bibitem [{\citenamefont {Robert}(2007)}]{Robert_thesis}%
  \BibitemOpen
  \bibfield  {author} {\bibinfo {author} {\bibfnamefont {J.-L.}\ \bibnamefont
  {Robert}},\ }\emph {\bibinfo {title} {Evaluation of Green's functions in
  complex media by decomposition of the Time Reversal Operator: Application to
  Medical Imaging and aberration correction}},\ \href@noop {} {Ph.D. thesis},\
  \bibinfo  {school} {Universite Paris 7 - Denis Diderot} (\bibinfo {year}
  {2007})\BibitemShut {NoStop}%
\bibitem [{\citenamefont {Prada}\ and\ \citenamefont
  {Thomas}(2003)}]{Prada2003}%
  \BibitemOpen
  \bibfield  {author} {\bibinfo {author} {\bibfnamefont {C.}~\bibnamefont
  {Prada}}\ and\ \bibinfo {author} {\bibfnamefont {J.-L.}\ \bibnamefont
  {Thomas}},\ }\href@noop {} {\bibfield  {journal} {\bibinfo  {journal} {J.
  Acoust. Soc. Am.}\ }\textbf {\bibinfo {volume} {114}},\ \bibinfo {pages}
  {235} (\bibinfo {year} {2003})}\BibitemShut {NoStop}%
\bibitem [{\citenamefont {Aubry}\ \emph {et~al.}(2006)\citenamefont {Aubry},
  \citenamefont {de~Rosny}, \citenamefont {Minonzio}, \citenamefont {Prada},\
  and\ \citenamefont {Fink}}]{Aubry2006}%
  \BibitemOpen
  \bibfield  {author} {\bibinfo {author} {\bibfnamefont {A.}~\bibnamefont
  {Aubry}}, \bibinfo {author} {\bibfnamefont {J.}~\bibnamefont {de~Rosny}},
  \bibinfo {author} {\bibfnamefont {J.-G.}\ \bibnamefont {Minonzio}}, \bibinfo
  {author} {\bibfnamefont {C.}~\bibnamefont {Prada}}, \ and\ \bibinfo {author}
  {\bibfnamefont {M.}~\bibnamefont {Fink}},\ }\href@noop {} {\bibfield
  {journal} {\bibinfo  {journal} {J. Acoust. Soc. Am.}\ }\textbf {\bibinfo
  {volume} {120}},\ \bibinfo {pages} {2746} (\bibinfo {year}
  {2006})}\BibitemShut {NoStop}%
\bibitem [{\citenamefont {Robert}\ and\ \citenamefont
  {Fink}(2009)}]{Robert2009}%
  \BibitemOpen
  \bibfield  {author} {\bibinfo {author} {\bibfnamefont {J.-L.}\ \bibnamefont
  {Robert}}\ and\ \bibinfo {author} {\bibfnamefont {M.}~\bibnamefont {Fink}},\
  }\href@noop {} {\bibfield  {journal} {\bibinfo  {journal} {J. Acoust. Soc.
  Am.}\ }\textbf {\bibinfo {volume} {125}},\ \bibinfo {pages} {218} (\bibinfo
  {year} {2009})}\BibitemShut {NoStop}%
\bibitem [{\citenamefont {Campbell}(1960)}]{Campbell1960}%
  \BibitemOpen
  \bibfield  {author} {\bibinfo {author} {\bibfnamefont {L.~L.}\ \bibnamefont
  {Campbell}},\ }\href@noop {} {\bibfield  {journal} {\bibinfo  {journal} {Inf.
  Control}\ }\textbf {\bibinfo {volume} {3}},\ \bibinfo {pages} {360} (\bibinfo
  {year} {1960})}\BibitemShut {NoStop}%
\bibitem [{\citenamefont {Roberts}\ \emph {et~al.}(1999)\citenamefont
  {Roberts}, \citenamefont {Penny},\ and\ \citenamefont {Rezek}}]{Roberts1999}%
  \BibitemOpen
  \bibfield  {author} {\bibinfo {author} {\bibfnamefont {S.~J.}\ \bibnamefont
  {Roberts}}, \bibinfo {author} {\bibfnamefont {W.}~\bibnamefont {Penny}}, \
  and\ \bibinfo {author} {\bibfnamefont {L.}~\bibnamefont {Rezek}},\
  }\href@noop {} {\bibfield  {journal} {\bibinfo  {journal} {Med. Biol. Eng.
  Comput.}\ }\textbf {\bibinfo {volume} {37}},\ \bibinfo {pages} {93} (\bibinfo
  {year} {1999})}\BibitemShut {NoStop}%
\bibitem [{\citenamefont {Mahajan}(1982)}]{mahajan1982strehl}%
  \BibitemOpen
  \bibfield  {author} {\bibinfo {author} {\bibfnamefont {V.~N.}\ \bibnamefont
  {Mahajan}},\ }\href@noop {} {\bibfield  {journal} {\bibinfo  {journal} {J.
  Opt. Soc. Am.}\ }\textbf {\bibinfo {volume} {72}},\ \bibinfo {pages} {1258}
  (\bibinfo {year} {1982})}\BibitemShut {NoStop}%
\bibitem [{\citenamefont {Jaeger}\ \emph
  {et~al.}(2015{\natexlab{b}})\citenamefont {Jaeger}, \citenamefont {Held},
  \citenamefont {Peeters}, \citenamefont {Preisser}, \citenamefont
  {Gr{\"{u}}nig},\ and\ \citenamefont {Frenz}}]{Jaeger2015}%
  \BibitemOpen
  \bibfield  {author} {\bibinfo {author} {\bibfnamefont {M.}~\bibnamefont
  {Jaeger}}, \bibinfo {author} {\bibfnamefont {G.}~\bibnamefont {Held}},
  \bibinfo {author} {\bibfnamefont {S.}~\bibnamefont {Peeters}}, \bibinfo
  {author} {\bibfnamefont {S.}~\bibnamefont {Preisser}}, \bibinfo {author}
  {\bibfnamefont {M.}~\bibnamefont {Gr{\"{u}}nig}}, \ and\ \bibinfo {author}
  {\bibfnamefont {M.}~\bibnamefont {Frenz}},\ }\href@noop {} {\bibfield
  {journal} {\bibinfo  {journal} {Ultrasound Med. Biol.}\ }\textbf {\bibinfo
  {volume} {41}},\ \bibinfo {pages} {235} (\bibinfo {year}
  {2015}{\natexlab{b}})}\BibitemShut {NoStop}%
\bibitem [{\citenamefont {Rodriguez-Molares}\ \emph {et~al.}(2017)\citenamefont
  {Rodriguez-Molares}, \citenamefont {Fatemi}, \citenamefont {L{\o}vstakken},\
  and\ \citenamefont {Torp}}]{RodriguezMolares2017}%
  \BibitemOpen
  \bibfield  {author} {\bibinfo {author} {\bibfnamefont {A.}~\bibnamefont
  {Rodriguez-Molares}}, \bibinfo {author} {\bibfnamefont {A.}~\bibnamefont
  {Fatemi}}, \bibinfo {author} {\bibfnamefont {L.}~\bibnamefont
  {L{\o}vstakken}}, \ and\ \bibinfo {author} {\bibfnamefont {H.}~\bibnamefont
  {Torp}},\ }\href@noop {} {\bibfield  {journal} {\bibinfo  {journal} {IEEE
  Trans. Ultrason. Ferroelec. Freq. Contr.}\ }\textbf {\bibinfo {volume}
  {64}},\ \bibinfo {pages} {1285} (\bibinfo {year} {2017})}\BibitemShut
  {NoStop}%
\bibitem [{\citenamefont {Imbault}\ \emph {et~al.}(2017)\citenamefont
  {Imbault}, \citenamefont {Faccinetto}, \citenamefont {Osmanski},
  \citenamefont {Tissier}, \citenamefont {Deffieux}, \citenamefont {Gennisson},
  \citenamefont {Vilgrain},\ and\ \citenamefont {Tanter}}]{Imbault2017}%
  \BibitemOpen
  \bibfield  {author} {\bibinfo {author} {\bibfnamefont {M.}~\bibnamefont
  {Imbault}}, \bibinfo {author} {\bibfnamefont {A.}~\bibnamefont {Faccinetto}},
  \bibinfo {author} {\bibfnamefont {B.-F.}\ \bibnamefont {Osmanski}}, \bibinfo
  {author} {\bibfnamefont {A.}~\bibnamefont {Tissier}}, \bibinfo {author}
  {\bibfnamefont {T.}~\bibnamefont {Deffieux}}, \bibinfo {author}
  {\bibfnamefont {J.-L.}\ \bibnamefont {Gennisson}}, \bibinfo {author}
  {\bibfnamefont {V.}~\bibnamefont {Vilgrain}}, \ and\ \bibinfo {author}
  {\bibfnamefont {M.}~\bibnamefont {Tanter}},\ }\href@noop {} {\bibfield
  {journal} {\bibinfo  {journal} {Phys. Med. Biol.}\ }\textbf {\bibinfo
  {volume} {62}},\ \bibinfo {pages} {3582} (\bibinfo {year}
  {2017})}\BibitemShut {NoStop}%
\bibitem [{\citenamefont {Derode}\ \emph {et~al.}(1995)\citenamefont {Derode},
  \citenamefont {Roux},\ and\ \citenamefont {Fink}}]{Derode1995}%
  \BibitemOpen
  \bibfield  {author} {\bibinfo {author} {\bibfnamefont {A.}~\bibnamefont
  {Derode}}, \bibinfo {author} {\bibfnamefont {P.}~\bibnamefont {Roux}}, \ and\
  \bibinfo {author} {\bibfnamefont {M.}~\bibnamefont {Fink}},\ }\href@noop {}
  {\bibfield  {journal} {\bibinfo  {journal} {Phys. Rev. Lett.}\ }\textbf
  {\bibinfo {volume} {75}},\ \bibinfo {pages} {4206} (\bibinfo {year}
  {1995})}\BibitemShut {NoStop}%
\bibitem [{\citenamefont {Roberts}\ \emph {et~al.}(2010)\citenamefont
  {Roberts}, \citenamefont {Stoica}, \citenamefont {Li}, \citenamefont
  {Yardibi},\ and\ \citenamefont {Sadjadi}}]{Roberts2010}%
  \BibitemOpen
  \bibfield  {author} {\bibinfo {author} {\bibfnamefont {W.}~\bibnamefont
  {Roberts}}, \bibinfo {author} {\bibfnamefont {P.}~\bibnamefont {Stoica}},
  \bibinfo {author} {\bibfnamefont {J.}~\bibnamefont {Li}}, \bibinfo {author}
  {\bibfnamefont {T.}~\bibnamefont {Yardibi}}, \ and\ \bibinfo {author}
  {\bibfnamefont {F.~A.}\ \bibnamefont {Sadjadi}},\ }\href@noop {} {\bibfield
  {journal} {\bibinfo  {journal} {IEEE Journal of Selected Topics in Signal
  Processing}\ }\textbf {\bibinfo {volume} {4}},\ \bibinfo {pages} {5}
  (\bibinfo {year} {2010})}\BibitemShut {NoStop}%
\bibitem [{\citenamefont {Badon}\ \emph {et~al.}(2017)\citenamefont {Badon},
  \citenamefont {Boccara}, \citenamefont {Lerosey}, \citenamefont {Fink},\ and\
  \citenamefont {Aubry}}]{badon2017multiple}%
  \BibitemOpen
  \bibfield  {author} {\bibinfo {author} {\bibfnamefont {A.}~\bibnamefont
  {Badon}}, \bibinfo {author} {\bibfnamefont {A.~C.}\ \bibnamefont {Boccara}},
  \bibinfo {author} {\bibfnamefont {G.}~\bibnamefont {Lerosey}}, \bibinfo
  {author} {\bibfnamefont {M.}~\bibnamefont {Fink}}, \ and\ \bibinfo {author}
  {\bibfnamefont {A.}~\bibnamefont {Aubry}},\ }\href@noop {} {\bibfield
  {journal} {\bibinfo  {journal} {Opt. Exp.}\ }\textbf {\bibinfo {volume}
  {25}},\ \bibinfo {pages} {28914} (\bibinfo {year} {2017})}\BibitemShut
  {NoStop}%
\bibitem [{\citenamefont {Aubry}\ and\ \citenamefont
  {Derode}(2011)}]{Aubry2011}%
  \BibitemOpen
  \bibfield  {author} {\bibinfo {author} {\bibfnamefont {A.}~\bibnamefont
  {Aubry}}\ and\ \bibinfo {author} {\bibfnamefont {A.}~\bibnamefont {Derode}},\
  }\href@noop {} {\bibfield  {journal} {\bibinfo  {journal} {J. Acoust. Soc.
  Am.}\ }\textbf {\bibinfo {volume} {129}},\ \bibinfo {pages} {225} (\bibinfo
  {year} {2011})}\BibitemShut {NoStop}%
\bibitem [{\citenamefont {Aubry}\ \emph {et~al.}(2008)\citenamefont {Aubry},
  \citenamefont {Derode},\ and\ \citenamefont {Padilla}}]{Aubry2008}%
  \BibitemOpen
  \bibfield  {author} {\bibinfo {author} {\bibfnamefont {A.}~\bibnamefont
  {Aubry}}, \bibinfo {author} {\bibfnamefont {A.}~\bibnamefont {Derode}}, \
  and\ \bibinfo {author} {\bibfnamefont {F.}~\bibnamefont {Padilla}},\
  }\href@noop {} {\bibfield  {journal} {\bibinfo  {journal} {Appl. Phys.
  Lett.}\ }\textbf {\bibinfo {volume} {92}},\ \bibinfo {pages} {124101}
  (\bibinfo {year} {2008})}\BibitemShut {NoStop}%
\bibitem [{\citenamefont {Mohanty}\ \emph {et~al.}(2017)\citenamefont
  {Mohanty}, \citenamefont {Blackwell}, \citenamefont {Egan},\ and\
  \citenamefont {Muller}}]{Mohanty2017}%
  \BibitemOpen
  \bibfield  {author} {\bibinfo {author} {\bibfnamefont {K.}~\bibnamefont
  {Mohanty}}, \bibinfo {author} {\bibfnamefont {J.}~\bibnamefont {Blackwell}},
  \bibinfo {author} {\bibfnamefont {T.}~\bibnamefont {Egan}}, \ and\ \bibinfo
  {author} {\bibfnamefont {M.}~\bibnamefont {Muller}},\ }\href@noop {}
  {\bibfield  {journal} {\bibinfo  {journal} {Ultrasound Med. Biol.}\ }\textbf
  {\bibinfo {volume} {43}},\ \bibinfo {pages} {993} (\bibinfo {year}
  {2017})}\BibitemShut {NoStop}%
\bibitem [{\citenamefont {Priestley}(1988)}]{Priestley1988}%
  \BibitemOpen
  \bibfield  {author} {\bibinfo {author} {\bibfnamefont {M.}~\bibnamefont
  {Priestley}},\ }\href@noop {} {\emph {\bibinfo {title} {Spectral analysis and
  time series}}}\ (\bibinfo  {publisher} {Academic Press, London},\ \bibinfo
  {year} {1988})\BibitemShut {NoStop}%
\bibitem [{\citenamefont {Goodman}(2000)}]{Goodman2000}%
  \BibitemOpen
  \bibfield  {author} {\bibinfo {author} {\bibfnamefont {J.~W.}\ \bibnamefont
  {Goodman}},\ }\href@noop {} {\emph {\bibinfo {title} {Statistical Optics}}}\
  (\bibinfo  {publisher} {Wiley, New York},\ \bibinfo {year}
  {2000})\BibitemShut {NoStop}%
\bibitem [{\citenamefont {Aubry}\ and\ \citenamefont
  {Derode}(2010)}]{Aubry2010}%
  \BibitemOpen
  \bibfield  {author} {\bibinfo {author} {\bibfnamefont {A.}~\bibnamefont
  {Aubry}}\ and\ \bibinfo {author} {\bibfnamefont {A.}~\bibnamefont {Derode}},\
  }\href@noop {} {\bibfield  {journal} {\bibinfo  {journal} {Waves Random
  Complex Media}\ }\textbf {\bibinfo {volume} {20}} (\bibinfo {year}
  {2010})}\BibitemShut {NoStop}%
\bibitem [{\citenamefont {Mar\u{c}enko}\ and\ \citenamefont
  {Pastur}(1967)}]{marcenko}%
  \BibitemOpen
  \bibfield  {author} {\bibinfo {author} {\bibfnamefont {M.}~\bibnamefont
  {Mar\u{c}enko}}\ and\ \bibinfo {author} {\bibfnamefont {L.}~\bibnamefont
  {Pastur}},\ }\href@noop {} {\bibfield  {journal} {\bibinfo  {journal} {Math.
  USSR-Sbornik}\ }\textbf {\bibinfo {volume} {1}},\ \bibinfo {pages} {457}
  (\bibinfo {year} {1967})}\BibitemShut {NoStop}%
\end{thebibliography}

%

\end{document}